\documentclass[format=manuscript, review=false, screen=true]{acmart}
\usepackage{amssymb}
\usepackage{amsmath}

\usepackage{color, colortbl}
\usepackage{xcolor}
\definecolor{darkblue}{rgb}{0,0,.5}
\definecolor{Brown}{cmyk}{0,0.81,1,0.60}
\definecolor{OliveGreen}{cmyk}{0.64,0,0.95,0.40}
\definecolor{CadetBlue}{cmyk}{0.62,0.57,0.23,0}
\definecolor{darkgray}{rgb}{0.5, 0.5, 0.5}
\definecolor{lightlightgray}{gray}{0.9}
\definecolor{lightgreen}{rgb}{0.67, 1, 0.32}
\definecolor{EsiOrange}{HTML}{F39600}
\definecolor{EsiBlue}{HTML}{006d9a}
\definecolor{EsiBlue2}{HTML}{006d9a}
\definecolor{EsiGreen}{HTML}{9ac013}
\definecolor{AimesGreen}{HTML}{42A83B}
\definecolor{AimesBlue}{HTML}{31A5F3}
\definecolor{bubblegum}{rgb}{0.99, 0.76, 0.8}
\definecolor{palegreen}{rgb}{0.6, 0.98, 0.6}
\definecolor{blond}{rgb}{0.98, 0.94, 0.75}

\usepackage[math]{cellspace}
\usepackage{footnote}
\usepackage{enumitem}

\usepackage{listings}
\DeclareCaptionFormat{listing}{\colorbox{AimesBlue!40}{\parbox{\textwidth}{\hspace{15pt}#1#2#3}}}
\usepackage{tocloft}

\usepackage{booktabs} % For formal tables

\usepackage{placeins} %fixing tables in appendix

\usepackage{pdfpages} %including PDFs like response letter

\copyrightyear{2018}
\setcopyright{acmlicensed}
\acmDOI{0000001.0000001}

\received{July 2017}
\received[revised]{January 2018 (minor)}
\received[accepted]{February 2018}

\begin{document}
% Title portion. Note the short title for running heads
\title[New GPGPU Code Transformation Framework Applied to Weather Prediction]{New High Performance GPGPU Code Transformation Framework Applied to Large Production Weather Prediction Code}
\author{Michel M\"uller}
\orcid{1234-5678-9012-3456}
\affiliation{%
  \institution{Tokyo Institute of Technology}
%   \streetaddress{104 Jamestown Rd}
%   \city{Williamsburg}
%   \state{VA}
%   \postcode{23185}
%   \country{USA}
}
\author{Takayuki Aoki}
\affiliation{%
  \institution{Tokyo Institute of Technology}
%   \department{School of Engineering}
%   \city{Charlottesville}
%   \state{VA}
%   \postcode{22903}
%   \country{USA}
}
% \author{Ting Yan}
% \affiliation{%
%   \institution{Eaton Innovation Center}
%   \city{Prague}
%   \country{Czech Republic}}
% \author{Tian He}
% \affiliation{%
%   \institution{University of Minnesota}
%   \country{USA}}
% \author{Chengdu Huang}
% \author{John A. Stankovic}
% \author{Tarek F. Abdelzaher}
% \affiliation{%
%   \institution{University of Virginia}
%   \department{School of Engineering}
%   \city{Charlottesville}
%   \state{VA}
%   \postcode{22903}
%   \country{USA}
% }

\begin{abstract}
We introduce ``Hybrid Fortran'', a new approach that allows a high performance GPGPU port for structured grid Fortran codes. This technique only requires minimal changes for a CPU targeted codebase, which is a significant advancement in terms of productivity. It has been successfully applied to both dynamical core and physical processes of ASUCA, a Japanese mesoscale weather prediction model with more than 150k lines of code. By means of a minimal weather application that resembles ASUCA's code structure, Hybrid Fortran is compared to both a performance model as well as today's commonly used method, OpenACC. As a result, the Hybrid Fortran implementation is shown to deliver the same or better performance than OpenACC and its performance agrees with the model both on CPU and GPU. In a full scale production run, using an ASUCA grid with 1581 x 1301 x 58 cells and real world weather data in 2km resolution, 24 NVIDIA Tesla P100 running the Hybrid Fortran based GPU port are shown to replace more than 50 18-core Intel Xeon Broadwell E5-2695 v4 running the reference implementation - an achievement comparable to more invasive GPGPU rewrites of other weather models.
\end{abstract}

%
% The code below should be generated by the tool at
% http://dl.acm.org/ccs.cfm
% Please copy and paste the code instead of the example below.

%
% End generated code
%


\thanks{
This research has been partly supported by the Japan Science and Technology Agency (JST) Core Research of Evolutional Science and Technology (CREST) research program ``Highly Productive, High Performance Application Frameworks for Post Peta-scale Computing'', partly by ``KAKENHI'', Grant-in-Aid for Scientific Research (S) 26220002 from the Ministry of Education, Culture, Sports, Science and Technology (MEXT) of Japan, partly by ``Joint Usage/Research Center for Interdisciplinary Large-scale Information Infrastructures (JHPCN)'' and ``High Performance Computing Infrastructure (HPCI)'' in Japan and partly by the Advanced Computation and I/O Methods for Earth-System Simulations (AIMES) project running under the multinational priority program ``Software for Exascale Computing'' (SPPEXA)
}

\maketitle

% The default list of authors is too long for headers}
\renewcommand{\shortauthors}{M. M\"uller and T. Aoki}

% remove this (not part of acm template)
% \newpage
% (TOC to be removed before submission)
% \tableofcontents
% \newpage

\section{Introduction} \label{sec:intro}

With the growth in single threaded performance on CPUs slowing down \cite{SingleThreadCPUPerf}, HPC applications today are increasingly required to make effective use of on-chip parallelism. In 2005 Microsoft's H. Sutter called this paradigm shift ``the free lunch is over'': rather than relying on a steady and exponential increase of computational performance to automatically benefit already existing software, it has become necessary for applications to specifically target, among other aspects, on-chip parallelism \cite{sutter2005free}.

\medskip

As a consequence of this paradigm shift, co-processors commonly referred to as ``accelerators'' have become increasingly common in supercomputers since these architectures are generally designed to offer as much computational throughput as possible for a given number of transistors. Top500 currently lists five of the top ten systems to make use either NVIDIA Tesla or Intel Xeon Phi based accelerators \cite{Top500-2016-11}. Offloading computationally intensive tasks to co-processors has thus become an increasingly common task in the field of HPC.

\medskip

This need for architecture specific programming has, however, created a divide between domain scientists who mainly care about application logic, and the supercomputers their applications run on. Due to this, implementations optimized for hardware have often been left to engineers. Both on the application - as well as on the hardware side - there is often a high rate of change, which creates maintainability issues. As an example, Oak Ridge National Laboratory's Norman et al. write about a recent accelerator implementation of the U.S. DOE's ``Accelerated Model for Climate and Energy'': "The original CUDA Fortran\footnote{see Appendix \ref{sec:gpgpu-glossary} for a definition of CUDA and CUDA Fortran} refactoring gave good performance, but it turned out that the codebase was still somewhat in flux. This meant that changes needed to be propagated into the CUDA code, which was quite difficult to support"  \cite{norman2017exascale}. See also Section \ref{sec:existing} for a further discussion of this related work.

\medskip

The work this paper is based on has initially been motivated by an evaluation of operative GPU support for the Japanese production weather prediction model ``ASUCA'', as discussed in Section \ref{par:asuca}. ASUCA is the main mesoscale weather prediction model used in operation and developed at the Japan Meteorological Agency \cite{ishida2010development}. ASUCA uses a structured grid and is implemented in Fortran. In order to gain accelerator support, the aforementioned maintainability issues are a main point of contention. ASUCA is constantly being pushed further by the application owners, and it is thus of fundamental importance to maintain ease-of-use and high productivity for the domain scientists when working with the code, both on CPU and GPU. A single codebase for both architectures is therefore paramount.

\medskip

In this work we therefore investigate methods to re-target structured grid applications in Fortran with low programming effort for an efficient GPU implementation while at the same time keeping CPU compatibility. This implies the following requirements for potential solutions:

\begin{enumerate}
    \item \label{hf-req:hybrid} The same user code should be applicable to both CPU and GPU architectures. In current research the focus is on NVIDIA GPUs and x86 CPU architectures, but the switch to others should be possible and streamlined.
    \item \label{hf-req:close} This user code (from here on called ``Hybrid ASUCA'') should stay as close as possible to the original codebase to ease the transition. Since the original code is written in Fortran, which does have some industry support for accelerators, the language should thus be kept if possible.
    % \item\label{hf-req:perf-port} The implementation should be designed to allow performance portability between CPU and GPU.
\end{enumerate}

Requirement \ref{hf-req:close} rules out applying domain-specific languages (such as stencil DSLs) to the computational code, to avoid a complete rewrite. In an evaluation of existing directive based methods (see Section \ref{sec:design-perf-port}) we conclude that industry standards alone are insufficient to fulfill the above requirements, which has motivated the invention of a new method for porting CPU-targeted structured grid applications in Fortran to GPU.

\medskip

In Section \ref{sec:hybridFortran} we thus propose ``Hybrid Fortran'', a source-to-source transformation and language extension that supports different targets. It uses Modern Fortran together with higher level parallelization and data specification abstractions as input, and outputs either OpenMP Fortran, CUDA Fortran or OpenACC Fortran sources. Crucially, this tool introduces a new assisted splitting method for large kernels, a necessity in order to enable GPU compatibility of physical processes within ASUCA without requiring a rewrite. We use a reduced weather application (introduced in Section \ref{sec:minimal-application}) as a basis to compare the features of Hybrid Fortran with the industry standard OpenACC. A performance model (discussed in Section \ref{sec:perfAnalysis}) is introduced in order to evaluate the performance of Hybrid Fortran and OpenACC for this use case (see Section \ref{sec:performance-sample}).

\medskip

% OpenACC has been evaluated thoroughly (see Section \ref{sec:SampleApplication}), but various stability issues with PGI's implementation (which at the time of this writing is the only available one with a near complete feature set except for the proprietary Cray implementation
% %ft: Cite
% ) lead to employing CUDA Fortran and a transpiler instead. It is however still used as a third possible backend implementation, mainly in order to make use of its reduction features (kernel reductions are only supported in the OpenMP and OpenACC based backends
% %ft: reference
% ). This set of tools and the underlying language extension is from here on referred to as ``Hybrid Fortran''.
% \medskip

By applying Hybrid Fortran to both dynamical core and physical processes of ASUCA a GPGPU implementation of this weather model has been completed, resulting in only 5\% increase in code size, even though the code now supports both CPU and GPU (see Section \ref{sec:hybrid-asuca}). To our knowledge this is the first time a complete production weather prediction model has been ported to GPGPU using a unified programming paradigm. This new implementation performs with a speedup of up to 4.9x comparing one GPU to one multi-core CPU socket (as discussed in sections \ref{sec:performance-asuca-small} and \ref{sec:performance-asuca}). The magnitude of these results are in line with previous GPU ports of ASUCA and other weather models using more invasive rewrites (as discussed in the related work section below).

\medskip

% The goal of this work is to evaluate and provide the means to create software that
% \begin{enumerate}
% \item can perform well on different hardware architectures with varying levels of parallelism,
% \item has good maintainability characteristics, e.g. adheres to the DRY principle \cite{hunt1999pragmatic},
% \item enables familiar coding practices,
% \item requires the least amount of portation effort for large existing codebases.
% \end{enumerate}

% All tools presented here are compatible to extend the parallelization to multiple sockets, GPUs and nodes, however the focus of this work are software tools for on-chip parallelisms.

% The background of this work is ongoing research to develop a complete GPU portation strategy for Japan's next generation weather model ``ASUCA'' \cite{ASUCA2010}. This work outlines the existing stack of software tools (Section \ref{sec:existing}), introduces and analyses a sample application (Section \ref{sec:SampleApplication}), shows the problems when trying to account for above requirements (sections \ref{sec:storage_order}, \ref{sec:VectorSize}, \ref{sec:multipleParallelizations}), proposes a new tool that bridges these gaps (Section \ref{sec:hybridFortran}), and, finally, presents the performance results for six different applications with this new tool (Section \ref{sec:performance}). These findings are applicable to all data parallel HPC applications written in Fortran that are required to support both GPUs and CPUs in a unified codebase.

Finally we draw conclusions and point out future work in sections \ref{sec:conclusions} and \ref{sec:future}, respectively.
\subsection{Related Work}\label{sec:existing}

The following GPGPU ports of atmospheric models are relevant for the discussion of our work.
% Section \ref{sec:performance-asuca-small} will furthermore compare the GPU performance of our approach for ASUCA, with that of other rectangular-grid GPGPU weather model ports presented here.

\subsubsection{ASUCA GPU Port by Shimokawabe et al.}\label{sec:existing-shimokawabe}

In 2010, Shimokawabe et al. completed a GPU-only CUDA C port of the dynamical core and part of the physical core of ASUCA. %A performance model for the five most relevant kernels was established as part of this work.
14 TFLOP/s were achieved in single precision using 528 Tesla S1070 GPUs \cite{shimokawabe201080}.

\medskip

In 2011, 24 GFLOP/s was achieved in double precision for ASUCA dynamical core on a single NVIDIA Fermi based Tesla M2050 GPU on TSUBAME 2.0, a 6-fold speedup when compared to the system's Xeon X5670 CPUs running the original Fortran code. This work also included extensive work regarding the optimization of inter-node communication and an overlapping method in order to run communication and computation in parallel. By applying weak scaling (i.e. increasing the grid size together with the number of GPUs), up to 145 TFLOP/s were achieved in single precision on 3990 Tesla M2050 GPUs \cite{shimokawabe2011145}.

\medskip

In 2014, the previous work was generalized by employing a stencil library and DSL based on C++ templates that generates either CUDA kernels or CPU code. Employing previously researched optimization methods to this more generalized method, up to 209.6 TFLOP/s were achieved for ASUCA in single precision on 4108 Tesla K20x GPUs \cite{shimokawabe:2014}.

\subsubsection{COSMO}\label{sec:existing-cosmo}

In 2014, MeteoSwiss' Fuhrer et al. proposed a mixed approach for COSMO, another structured grid regional weather and climate model used in operation in several European nations. In this mixed approach, a C++ stencil library and DSL called ``STELLA'' is used for the dynamical core, while a ``less disruptive'' OpenACC based method is applied to the physical processes in order to retain most of the original Fortran codebase \cite{fuhrer2014towards}. In COSMO's case, the physical processes already expose fine grained parallelism and kernel fusion was employed in some instances in order to optimize for less kernel launches by Lapillonne et al. \cite{lapillonne2014using}. Thanks to the code refactoring a speedup of 1.5x was achieved on CPU. By employing GPU code generation an additional speedup of 2.9x was achieved (comparing Tesla K20x to 8-core Xeon E5-2670 on Piz Daint) \cite{fuhrer2014towards}. STELLA has subsequently been generalized for use cases beyond COSMO and rectangular grids under the new name ``GridTools'' \cite{GridTools}.

\subsubsection{Weather Research and Forecasting Model (WRF)}\label{sec:existing-wrf}

The U.S. National Center for Atmospheric Research's ``Weather Research and Forecasting model'' (WRF) is the most widely used weather model worldwide \cite{MichalakesV08}. It too uses a structured grid. WRF consists of multiple hundreds of thousands of lines of Fortran code. GPU ports known to us have been only partial, leading to heavy pressure on the memory bus between CPU and GPU and thus a lower speedup compared to full GPU ports of other weather models \cite{vanderbauwhede2013investigation}. The following partial GPU ports have been achieved among others:

\begin{itemize}
    \item in 2009, Michalakes and Vachharajani ported the ``WSM5'' cloud microphysics to CUDA C, achieving an overall speedup of 1.25x for a benchmark workload \cite{MichalakesV08},
    \item in 2010, Ruetsch et al. ported the longwave radiation physics to CUDA Fortran \cite{ruetsch2010gpu},
    \item in 2012, Mielikainen et al. ported the Goddard shortwave radiation scheme to CUDA C \cite{mielikainen2012gpu},
    \item in 2013, Vanderbauwhede and Takemi ported the advection scheme to OpenCL and integrated it together with the Fortran based WRF. Doing so, an overall speedup of 2x was achieved \cite{vanderbauwhede2013investigation}.
\end{itemize}

\subsubsection{Non-hydrostatic Icosahedral Model (NIM)}

In 2010 the dynamical core of the Non-hydrostatic Icosahedral Model (NIM) developed at the U.S. National Oceanic and Atmospheric Administration (NOAA) was ported to GPU by Govett et al. by using a Fortran-to-CUDA-C transpiler called "F2C-ACC" \cite{govettGPU}. In 2014 this work was compared to the industry standard ``OpenACC'' on Titan using Tesla K20x GPUs, showing 77\% higher performance with F2C-ACC compared to Cray's OpenACC Fortran implementation and 220\% higher performance compared to PGI's OpenACC. Overall however the GPU implementation was not able to achieve a speedup compared to CPU due to the missing physical processes and the thus high GPU-CPU data transfer overhead \cite{GovettDirective}. Since then, performance of both hardware and software have improved significantly however, resulting in a speedup of 2.5x when comparing NVIDIA GPUs to CPUs and 2.0x when comparing Xeon Phi to the same CPU in 2017. Govett et al. have also reported that OpenACC's performance issues and bugs have been fixed and F2C-ACC has thus been fully replaced with the industry standard \cite{govett2017parallelization}.

\subsubsection{Accelerated Model for Climate and Energy (ACME)}\label{sec:existing-acme}

Norman et al. have investigated methods to create a hybrid GPU/CPU codebase for the U.S. Department of Energy's ``Accelerated Model for Climate and Energy'' (ACME). A previous CUDA Fortran based port was dropped due to issues with having to maintain two separate codebases for CPU and GPU. For this reason, OpenACC was evaluated as a replacement, but limitations with resource pressure and inlining when porting large kernels to GPU were found. It was concluded that such large kernels need to be split up to be feasible on GPU with OpenACC, which aligns with our findings about ASUCA's physical processes (see Section \ref{par:asuca:granularity}). Since CPUs perform better with fused loops for caching reasons, for a pure OpenACC based solution it was found that two separate code versions are required for CPU and GPU, resulting in similar maintainability issues as with a pure CUDA Fortran port \cite{norman2017exascale}.
\section{Background and Motivation} \label{par:asuca}

This section discusses the background and motivation of this work. As such it introduces ASUCA (the main target application for the work presented in this paper), the goals of an accelerator port, a discussion of existing methods for porting such applications to accelerators, as well as a problem statement.

\subsection{The ASUCA Weather Prediction Model} \label{sec:asuca-model}

\begin{figure}[htb]
  \centering
  \includegraphics[width=.7\linewidth]{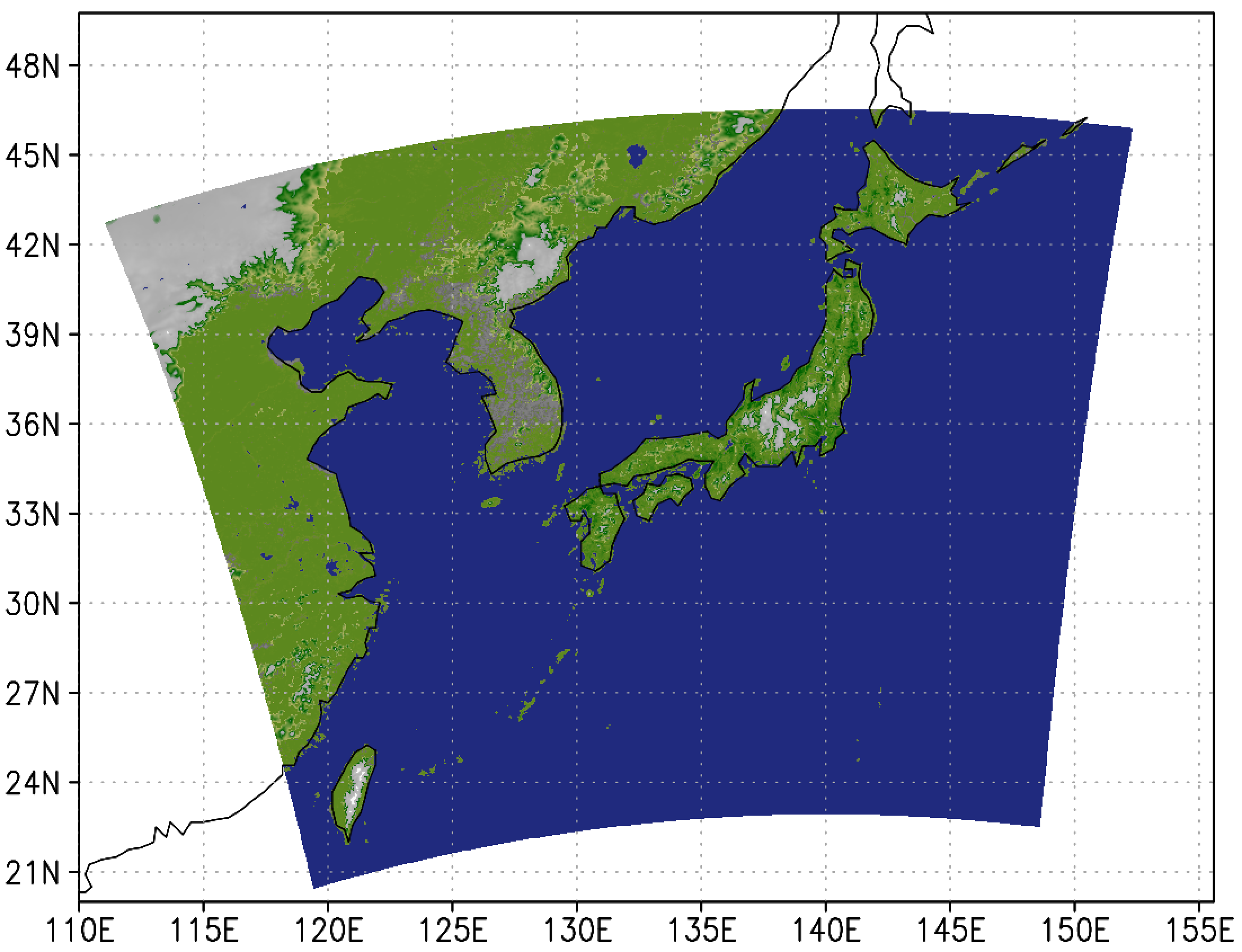}
  \caption{ASUCA's model simulation boundaries}
  \label{fig:asuca}
\end{figure}

ASUCA is a mesoscale weather prediction model developed by the Japan Meteorological Agency (JMA) \cite{ishida2010development}. It is used in production since 2014 as one of the main operational weather forecast models in Japan and covers the area depicted in Figure \ref{fig:asuca} in two kilometer resolution, generating nine-hour-forecasts every hour \cite{sakamotodevelopment}. ASUCA uses an Arakawa C-grid (one type of rectangular grid) \cite{shimokawabe:2014}. Time discretization is implemented by employing a third order Runge-Kutta scheme, enabling long time steps \cite{wicker2002time}. Sound and gravity waves are treated separately using a second order Runge-Kutta scheme to enable a higher time resolution, employing the HEVI scheme (Horizontally explicit - vertically implicit). Vertical advection of water substances are calculated using a separate time step for each column, based on the Courant-Friedrichs-Lewy convergence condition \cite{ishida2010development}. ASUCA employs a generalized coordinate system that is used in conjunction with Lambert conformal conic- as well as latitude/longitude projections \cite{shimokawabe:2014}. It is discretized using a structured grid on a finite volume method in three spatial dimensions: \verb|I| and \verb|J| as interchangeable horizontal dimensions and \verb|K| as the separately treated (largely sequentially implemented) vertical dimension.

\medskip

ASUCA has been implemented in Fortran, employing multidimensional arrays as its main data structure due to their simplicity and performant implementation on hardware. In ASUCA these data objects tend to be directly stored in modules (i.e. only making rare use of derived types) and imported where necessary through \verb|use| statements. The program structure can roughly be divided into two categories: Physical processes (later depicted under \verb|pp interface| as well as \verb|phys. adjust| in Figure \ref{fig:asuca-structure}, Section \ref{sec:hybrid-asuca}) and dynamical core (all other modules depicted in Figure \ref{fig:asuca-structure}, Section \ref{sec:hybrid-asuca}). This distinction is common in weather models such as WRF \cite{wrf_on_phi} and COSMO \cite{cummings:review}.

\medskip

ASUCA is parallelized for multi-core and multi-node in the horizontal domain (i.e. \verb|I| and \verb|J| dimensions). For the multi-node parallelization, given a full grid of $nx \cdot ny \cdot nz$ size, with $nx$ and $ny$ being the horizontal boundaries farthest from the origin and $nz$ being the upper vertical boundary, ASUCA's grid is decomposed into blocks and scattered across the available nodes, with each node computing a block of $nx' \cdot ny' \cdot nz$ size with $nx' <= nx$ and $ny' <= ny$, thus requiring halo communication for each of the employed Runge-Kutta schemes. For the on-chip parallelization, OpenMP parallel loop directives are used. The vertical domain is executed sequentially due to loop carried dependencies in the physical processes.

\medskip

ASUCA's runtime is dominated by the dynamical core, which as a stencil code is heavily bounded by memory bandwidth \cite{shimokawabe201080}  \cite{dursun2009core}. It therefore stands to reason to investigate hardware architectures with a high memory bandwidth as potential implementation targets.

% \medskip

% Shimokawabe et al. have already clearly demonstrated the feasibility for an ASUCA GPU implementation in terms of performance (see Section \ref{sec:existing-shimokawabe}). Hybrid ASUCA thus has an additional goal: Allowing the existing Fortran based user code to be reusable as much as possible (see also the requirements as outlined in Section \ref{sec:intro}). Furthermore, CPU compatibility is to be kept since operationally JMA still rely on CPU based cluster for the foreseeable future. Figures \ref{fig:asuca-structure} and \ref{fig:asuca-numbers} show the current state of completion of this project.

\subsection{GPGPU Computing} \label{sec:gpgpu}

This section discusses the motivation for GPGPU support in a weather prediction codebase. See also Appendix \ref{sec:gpgpu-glossary} for a glossary of GPGPU specific terms.

\begin{table}[htpb]
  \centering
  \footnotesize
  \caption{Comparison of the latest generation Intel CPUs and NVIDIA GPUs used in HPC}
  \label{table:gpu-cpu-comparison}
	\begin{tabular}{|Sc|Sc|Sc|}
	\hline
	Characteristic & CPU & GPU \tabularnewline
	\hline
	Number of parallelisms in & \cellcolor{bubblegum} 2: multi-core, Vector & \cellcolor{palegreen} 1: CUDA core \tabularnewline
	programming model & \cellcolor{bubblegum} & \cellcolor{palegreen} \tabularnewline
	\hline
	Floating-point units (FPUs) & \cellcolor{bubblegum} $\sim 100$ & \cellcolor{palegreen} $\sim 3000$ \tabularnewline
	(core count multiplied with vector length) & \cellcolor{bubblegum} & \cellcolor{palegreen} \tabularnewline
	\hline
	Single-threaded performance & \cellcolor{palegreen} high & \cellcolor{bubblegum} low \tabularnewline
	\hline
	Context switching overhead & \cellcolor{bubblegum} high & \cellcolor{palegreen} low \tabularnewline
	\hline
	Worst case performance loss & \cellcolor{bubblegum} $\sim80x$ & \cellcolor{bubblegum} $\sim64x$ \tabularnewline
	with branching & \cellcolor{bubblegum} (pipeline length times vector size) & \cellcolor{bubblegum} (number of cores per SMX) \tabularnewline
	\hline
	Double- (DP) to single-precision (SP)  & \cellcolor{palegreen} $\sim0.5x$ & \cellcolor{palegreen} $\sim0.5x$ \tabularnewline
	FLOP/s ratio & \cellcolor{palegreen} & \cellcolor{palegreen} \tabularnewline
	\hline
	Cache per FPU & \cellcolor{palegreen} 8 KB (L1), 32 KB (L2), $\sim$ 300 KB (L3) & \cellcolor{bubblegum} $\sim$ 0.4 KB (L1), $\sim$ 1 KB (L2) \tabularnewline
	\hline
	DP Registers per core & \cellcolor{bubblegum} $\sim80$ & \cellcolor{palegreen} $\sim1000$ \tabularnewline
	\hline
	Memory bandwidth for sequential access & \cellcolor{bubblegum} $\sim80$ GB/s & \cellcolor{palegreen} $\sim500$ GB/s \tabularnewline
	\hline
	Memory bandwidth for random access & \cellcolor{bubblegum} $\sim0.1$ GB/s & \cellcolor{palegreen} $\sim0.9$ GB/s \tabularnewline
	\hline
	\end{tabular}
\end{table}

Table \ref{table:gpu-cpu-comparison} sums up the architectural differences that are relevant to the use case at hand. Specific performance numbers from 18-core Broadwell CPUs and Tesla P100 are used here (see also Appendix \ref{sec:hardware-metrics}).

\medskip

GPUs are designed to support the highest number of parallel threads possible for a given number of transistors, while still assigning each thread enough resources to support common use cases in scientific and graphics computing. This results in GPU programs operating on a much higher number of threads in parallel compared to CPU (supported by the high core count, high number of registers per core and high memory bandwidth), however the sequential computational performance is lower due to the threads operating on a shorter pipeline with no additional vectorization and with less cache per thread.

\medskip

In terms of memory bandwidth, GPUs typically have a 5-7x advantage over CPUs, which makes GPUs interesting for bandwidth bounded applications such as stencil applications. It should be noted however that the CPU's larger caches per floating-point unit can lead to the CPU being faster for the bandwidth bounded case if and only if cached accesses dominate over main memory accesses in the runtime profile. Experience shows that this is usually not the case however, instead for most use cases the CPU's more prominent caches lead to a somewhat lesser speedup for the GPU implementation compared to the bandwidth difference between GPU and CPU.

\medskip

A unified programming model for the on-chip parallelism (compared to vectorization and multi-core, which need to be treated separately on CPU) as well as the low context switching cost for fine grained parallelizations (thanks to keeping inactive thread context stored in the large register files where possible) are further advantages of GPUs.

\medskip

The GPU's main disadvantage is a high occupancy requirement (discussed in Appendix \ref{sec:gpgpu-glossary}), which results in a minimum problem size per GPU as well as the need for being more economical with each thread's private resources in order to achieve a high performance. This has the following main effects on how GPGPU applications are constructed and run:

\begin{itemize}
    \item The resource constraints generally lead to a higher parallelization granularity being optimal, i.e. the amount of code that can belong to a single GPU kernel is limited. Disregarding this leads to local resources (particularly registers) overflowing, resulting in excessive main memory accesses.
    \item There is a limit to how few threads a GPU can be given without starving for work and subsequently giving poor performance. This, in turn, enforces a lower limit on time-to-solution (i.e. an upper limit to the number of GPUs a given problem scales to). This limitation is common in all high-throughput processor architectures, to various degrees.
\end{itemize}

However, even for a fixed given time to solution, assuming a large enough problem size per microchip, GPUs are usually more energy efficient thanks to their high computational and memory bandwidth ``density'' in terms of rack space (as shown by table \ref{table:gpu-cpu-comparison}).

\medskip

To acquire an understanding of the feasibility of GPU ports for memory bandwidth bounded applications, we can define the necessary and sufficient condition for a speedup on GPU as ${t_H} > {t_D} + {t_{comm}}$, where $t_H$ is the computation time for the host implementation, $t_D$ is the computation time on the device and $t_{comm}$ is the time spent for communication between host and device. Assuming memory bandwidth of sequentially accessed values to be the bottleneck of an application\footnote{This assumption holds for typical stencil algorithms as the threads within a warp operate in lockstep (all threads in the warp execute the same instruction except for data index offsets), which leads to coalesced memory accesses as long as the data is ordered in memory as threads are ordered in the CUDA grid.}, we substitute ${t_H} = n_i\frac{{m \cdot b}}{{B{W_H}}}$, ${t_D} = n_i\frac{{m \cdot b}}{{B{W_D}}}$, and ${t_{comm}} = \frac{{{m_{HtoD}} \cdot b}}{{B{W_{HtoD}}}}$, where $m$ is the number of values to read and write from/to memory per point update, $b$ is the byte length of each value, $n_i$ is the number of iterations over the data, ${m_{HtoD}}$ is the number of values that need to be transferred from/to the device after $n_i$ iterations and ${B{W_H}}$, ${B{W_D}}$ and ${B{W_{HtoD}}}$ are the bandwidths available on host, device and between host and device, respectively. After substitution the speedup condition becomes

\begin{equation}\label{eq:speedup_req_mem_bound}
{n_{i}}\frac{m}{{{m_{HtoD}}}} > \frac{{\frac{{B{W_H}}}{{B{W_{HtoD}}}}}}{{(1 - \frac{{B{W_H}}}{{B{W_D}}})}}.
\end{equation}
\smallskip

The left-hand side of this inequality is dictated by the application while the right-hand side is purely hardware dependent. We can therefore determine that ${n_{i}}\frac{m}{{{m_{HtoD}}}}$ must be larger than $5.4$, $8.3$ and $5.2$ on the GPU supercomputers TSUBAME 2.5, Reedbush-H and Piz Daint, respectively (using the performance numbers listed in appendix \ref{sec:hardware-metrics}).

\subsection{Existing Hybrid CPU/GPU Parallelization Methods} \label{sec:existing-methods}

As discussed in Section \ref{sec:intro}, the main goals of this work is to find a solution for accelerating the ASUCA weather prediction model with GPUs with a unified code and without requiring a rewrite. An overview over existing software tools that allow unified GPU/CPU codes is provided in this section. Table \ref{table:existingTools} lists these options. For ``STELLA'' and "F2C-ACC" see also the discussion in Section \ref{sec:existing}.

\begin{table*}[hbtp]
  \centering
  \footnotesize
  \caption{Existing tools for unified CPU/GPU code}
  \label{table:existingTools}
	\begin{tabular}{|c|c|c|c|c|}
	\hline
	 & Language & Usage & Implementations & Notable Applications\tabularnewline
	\hline
	OpenACC\cite{openACCSpec} & C,C++,Fortran & Directives & PGI \cite{PGIAcc}, & COSMO\tabularnewline
    & & & Cray \cite{CrayAcc} & Physical Processes\cite{cummings:review}\cite{fuhrer2014towards}\cite{lapillonne2014using}\tabularnewline
	& & & & NIM\cite{GovettDirective}\tabularnewline
	\hline
	OpenMP 4.0+\cite{openmp4} & C,C++,Fortran & Directives & GCC 4.9.1+ ,  & Currently no major \tabularnewline
	 &  &  & Clang\cite{clang:OpenMP4},  & known applications \tabularnewline
	 &  &  & ROSE\cite{openmp4:early} & \tabularnewline
	\hline
	F2C-ACC\cite{govett:f2c-acc} & Fortran & Directives & NOAA & NIM\cite{govettGPU}\cite{GovettDirective}\tabularnewline
	\hline
	STELLA\cite{fuhrer2014towards} & C++ & Stencil Library / DSL & SCS/CSCS & COSMO \tabularnewline
	& & & & Dynamical Processes\cite{fuhrer2014towards}\cite{cummings:review} \tabularnewline
	\hline
	Physis\cite{maruyama} & C, Fortran & Stencil Library / DSL & RIKEN & Sample applications provided \tabularnewline
	 &  &  &  & in the github repository\tabularnewline
	\hline
	Shimokawabe & C++ & Stencil Library / DSL & Tokyo Tech & ASUCA \tabularnewline
	et al. Stencil &  &  &  & Dynamical Processes\cite{shimokawabe:2014} \tabularnewline
	Framework\cite{shimokawabe:2014} & & & & \tabularnewline
	\hline
	Kokkos\cite{edwards2014kokkos} & C++ & Unified parallelization & Sandia & Albany/FELIX\cite{tezaurtowards} \tabularnewline
	& &  / Mem. Layout & & \tabularnewline
	& & Transformation & & \tabularnewline
	\hline
	\end{tabular}
\end{table*}

\medskip

The following criteria are important for the consideration of these candidates:

\begin{enumerate}
\item Due to their invasive nature when applying such techniques to ASUCA, Non-Fortran based solutions as well as domain-specific languages for stencils (stencil DSLs) have been ruled out from this evaluation. One should note that Shimokawabe et al. have already completed an investigation concerning the application of a C++ template based DSL to ASUCA \cite{shimokawabe:2014}.
% \item While allowing a higher level of abstraction and thus reduced code size, DSL based approaches lead to a high number of code changes in existing computational user codes. Since this is again problematic with regards to requirement \ref{hf-req:close}, such techniques have been ruled out (i.e. Physis).
\item Since this work focuses on production weather prediction, the maturity of the pursued tools is important. Despite OpenMP allowing to target accelerators from 4.0 on, its support has not yet reached a mature level as of 2016, according to NVIDIA engineers \cite{OpenMP45GPU}. For this reason it has been ruled out for a further evaluation for Hybrid ASUCA at this point in time.
\item F2C-ACC and OpenACC share many common features, with OpenACC being the much more widely supported choice. Furthermore, the maintainers of F2C-ACC at NOAA have recently dropped it in favor of OpenACC \cite{govett2017parallelization}.
\end{enumerate}

By this process of elimination we arrive at OpenACC as the only viable option in terms of already existing solutions for the problem at hand. Its limitations, and finally the decision against using it, will be discussed in the following Section \ref{sec:design-perf-port}.

\subsection{Design Considerations Regarding Code Reusability and Portability} \label{sec:design-perf-port}

In order to design the method to allow for a high reusability of existing code structure, the considerations outlined in this section have been crucial.

\subsubsection{Architecture-Dependant Parallelization Granularity} \label{par:asuca:granularity}

While ASUCA's dynamical core is already implemented with a high parallelization granularity in the original code and is thus already ``GPU friendly'' (as discussed in Section \ref{sec:gpgpu}), the physical processes are originally implemented with a very large kernels and thus low granularity. Since computations can be executed on each \verb|K|-column separately, without affecting neighboring columns until the next run of the dynamical core, most of the physical processes are implemented using a single kernel with one thread per column, in order to increase cache locality and decrease the amount of context switching and thread synchronization \cite{kwiatkowski2001evaluation} \cite{douglas2000cache}. This amounts to approximately 25k lines of code in a single parallel kernel. For a GPU implementation this is problematic due to the following reasons (as discussed in Section \ref{sec:gpgpu} and found independently by Norman et al. \cite{norman2017exascale}):

\begin{enumerate}
  \item Deep call graphs under a single kernel are difficult to maintain since kernel code relies on compiler inlining, which breaks easily and only supports a subset of language constructs.
  \item On GPUs the memory, cache and register resources per thread are much more limited, in favor of supporting a much higher number of parallel threads.
\end{enumerate}

% \footnote{The minimal weather application's physics module (see Appendix \ref{sec:sources-physical}) provides an example of a somewhat coarse-grained parallelism, however it does not apply sufficient register- or cache pressure or to show a noticeable impact on the performance on GPU.}

\medskip

For a GPGPU port it is therefore necessary to split this large kernel in order to achieve finer grained tasks for each thread. Doing so manually would however require a complete rewrite of the physical processes code. Furthermore it would make a performance portable solution impossible since the CPU benefits from a high cache locality of the column based code.

\smallskip

An automated or assisted approach for kernel fission is therefore highly desirable in order to allow for GPU acceleration while keeping cache locality and low overhead for the CPU case. OpenACC does not offer a coding facility that allows to abstract the parallelization granularity - instead, two separate code versions are required, once with the parallelization applied to the call graph leafs and once applied to a coarse grained level. In ASUCA's case this would mean a doubling of most of code used for the physical processes. We note that Norman et al., as discussed in Section \ref{sec:existing-acme}, have arrived at the same conclusion \cite{norman2017exascale}.

% In order to achieve a high performance on both CPU and GPU it is therefore necessary to allow the parallelization to be applied at different levels in the call graph, depending on what hardware architecture is targeted. On CPU the physical processes depicted under \verb|pp interface| in Figure \ref{fig:asuca-structure} are to keep their coarse grained parallelization, while on GPU, parallelization is to be applied at a finer-grained level in the call graph. Kernel fission and/or kernel fusion is therefore required in order to achieve a user code that is performance portable to CPU and GPU. Furthermore, kernel fission is to be preferred since this would allow the user code to be close to the reference codebase with a large kernel, rather than having to rewrite the code using many tight kernels that the framework then fuses for CPU \cite{norman2017exascale}.

% \subsubsection{Course-Grained versus Fine-Grained Parallelization}\label{sec:multipleParallelizations}

\subsubsection{Architecture-Dependent Privatization} \label{par:asuca:privatization}

Due to the large amount of calculations being performed locally per thread in the reference physical processes as discussed in Section \ref{par:asuca:granularity}, there are also many data objects created locally. Kernel fission requires the sharing of such data objects between threads. If kernel fission is to be employed and cache locality on CPU is to be kept it is necessary to define the privatization depending on what architecture is targeted. OpenACC allows for automatic privatization by means of explicit clauses, however this only works reliably within the same routine as the parallel loop is defined. For a rewrite with finer grained parallelization granularity across a deep call graph it would still be necessary to rewrite most of the data objects for a higher number of domains (in case of ASUCA's physics 3D grid data instead of 1D column data).

\subsubsection{Architecture-Dependent Storage Order} \label{par:asuca:storage}

In a largely memory bandwidth application it is paramount to implement a storage order that matches the target hardware's cache architecture since cache misses directly influence memory pressure \cite{dursun2009core}. Different hardware architectures with different cache architectures have different ideal storage orders. On CPU, the innermost loop should always be given stride-1 memory access in order to increase cache locality. Due to the characteristics of physical processes discussed in Section \ref{par:asuca:granularity} the innermost loop is always operating on the \verb|K|-domain, therefore the natural storage order choice is \verb|KIJ| or \verb|KJI| \cite{shimokawabe201080}. Meanwhile on GPUs the domain that is mapped to the first thread index (i.e. one of the parallel domains) should be given stride-1 access to enable coalesced memory access \cite{harris2007optimizing}, which requires either \verb|IJK| or \verb|JIK| orderings in case of ASUCA \cite{shimokawabe201080}. Storage order abstraction is therefore necessary to achieve performance portability. Section \ref{sec:storage_order} shows this as well, based on the findings from a model application.

\smallskip

Storage order therefore strongly impacts performance portability. In OpenACC Fortran there is no provided method to specify multidimensional Fortran arrays in a way that the storage order is efficient on both GPU and CPU without requiring a rewrite of all array accesses and specifications.

% \begin{enumerate}
% \item the user code is unified across GPU and CPU,
% \item the storage order is efficient on both GPU and CPU,
% \item array accesses and specifications do not need to be rewritten.
% \end{enumerate}

\subsection{Problem Statement}\label{sec:problem-description}

In Section \ref{sec:design-perf-port} we have outlined what we consider the most important porting aspects regarding efficiency on GPU for hybrid codes, i.e. applications with the capability to run on either CPU or GPU. In existing methods used for NWP these issues have been solved using a combination of the following approaches:

\begin{itemize}
    \item Maximum relearning: Through a complete reimplementation in a meta-programming capable language (so far for HPC this means C++), memory layout abstraction and parallelization on different target platforms has been solved. Examples for this approach are Fuhrer et al. 2014 \cite{fuhrer2014towards} and Shimokawabe et al. 2014 \cite{shimokawabe:2014}.
    \item Efficiency loss: Reducing the scope of the problem by only porting the parts of the code that are already fine-grained (and thus suited for GPU) solves the issues described in Sections \ref{par:asuca:granularity} and \ref{par:asuca:privatization}. This however leads to a significant overhead in GPU-CPU data exchange since the remaining code runs on CPU, thus requires data synchronization for every time step. This approach has for example been taken by Govett et al. \cite{govettGPU} \cite{GovettDirective} and by various projects involving GPU acceleration of WRF as outlined in Section \ref{sec:existing-wrf}.
    \item Code divergence: If the entire logic is to be kept in a style of Fortran language that is familiar to domain scientists (by making use of directive-based approaches such as OpenMP and/or OpenACC), it ultimately leads to code duplication and divergence since no directive-based approaches have support for code granularity transformations or memory layout transformations. Norman et al. 2017 show an example of this approach \cite{norman2017exascale}.
\end{itemize}

Our goal for ASUCA was to find a solution that has none of the aforementioned drawbacks, i.e. a full GPU port that remains CPU compatible, requires minimal code divergence and is able to reuse as much of the existing code as possible.

\section{New Approach}\label{sec:hybridFortran}

A new approach called ``Hybrid Fortran'' has been developed in order to solve the problem described in Section \ref{sec:problem-description}. Hybrid Fortran employs a source transformation with the following characteristics:

\begin{enumerate}
\item \label{enum:automation} porting to this framework from existing CPU code requires a low amount of effort, i.e. existing computational code targeted at CPUs can be used as-is where possible,
\item as a consequence (see also Sections \ref{par:asuca:granularity} and \ref{par:asuca:storage}), storage order and parallelization granularity are to be made variable, such that CPU targeted user code can be transformed to match the architecture specific granularity and storage order,
% \item \label{enum:reorder} a method to reorder the storage order of multidimensional arrays is provided,
% \item \label{enum:reorder_centralised} the specification for the reordering are as centralized as possible, i.e. changing computational code lines is not required,
% \item \label{enum:multi_parallelization} a higher level abstraction is provided in order to specify both coarse-grained and fine-grained parallelization in a unified codebase,
% \item \label{enum:levels} unlike inlining routines into kernels, \ref{enum:multi_parallelization} can be applied across arbitrarily many subroutine calls,
% \item \label{enum:privatization} in order to enable \ref{enum:multi_parallelization}, data privatizations can be applied selectively, depending on the target architecture,
\item \label{enum:implementation} the parallelization technology (here OpenMP, OpenACC or CUDA) is transparent to the framework user such that with changing industry standards the backend implementation can easily be modified,
\item \label{enum:abstraction} whenever architecture specific programming is needed, Hybrid Fortran uses an abstraction in order to centralize these specific settings in a codebase-wide configuration rather than leaking it to the user code,
\item \label{enum:control} however, control is not be taken away from the framework user, i.e. the implementation should be predictable and adjustable in order to allow for a highly productive environment to create performant implementations,
\item and finally, Hybrid Fortran is suitable for both dynamical processes (stencil patterns) as well as column-only physical parametrizations.
\end{enumerate}

\medskip

In this section we first introduce a model application used for the further discussion (Section \ref{sec:minimal-application}). We then apply OpenACC parallelization for GPU and OpenMP parallelization for CPU and compare it to a Hybrid Fortran implementation from a user-centric viewpoint in Section \ref{sec:hf-user-centric}, followed by an implementation-centric discussion in Section \ref{sec:hf-implementation-centric}. Finally, the current limitations of this approach are discussed in Section \ref{hf-limitations}.

\subsection{Reduced Weather Application} \label{sec:minimal-application}
Since ASUCA is too large to present and analyze here in its entirety (see also Figure \ref{fig:asuca-numbers} in Section \ref{sec:hybrid-asuca}), this section introduces a reduced, yet scientifically meaningful, weather application as a vehicle to show relevant code and performance characteristics. Equation \ref{eq:minimal} shows the main equation governing this application: the heat diffusion equation with $T$ being the temperature defined in \verb|IJK|-space, $\kappa$ being the thermal diffusivity and $f$ being the rate of heat input through radiation.

\begin{align}\label{eq:minimal}
\frac{\partial T}{\partial t}= \kappa \nabla^2 T + f
\end{align}
\smallskip

\begin{figure}[htpb]
  \centering
  \includegraphics[width=0.8\linewidth]{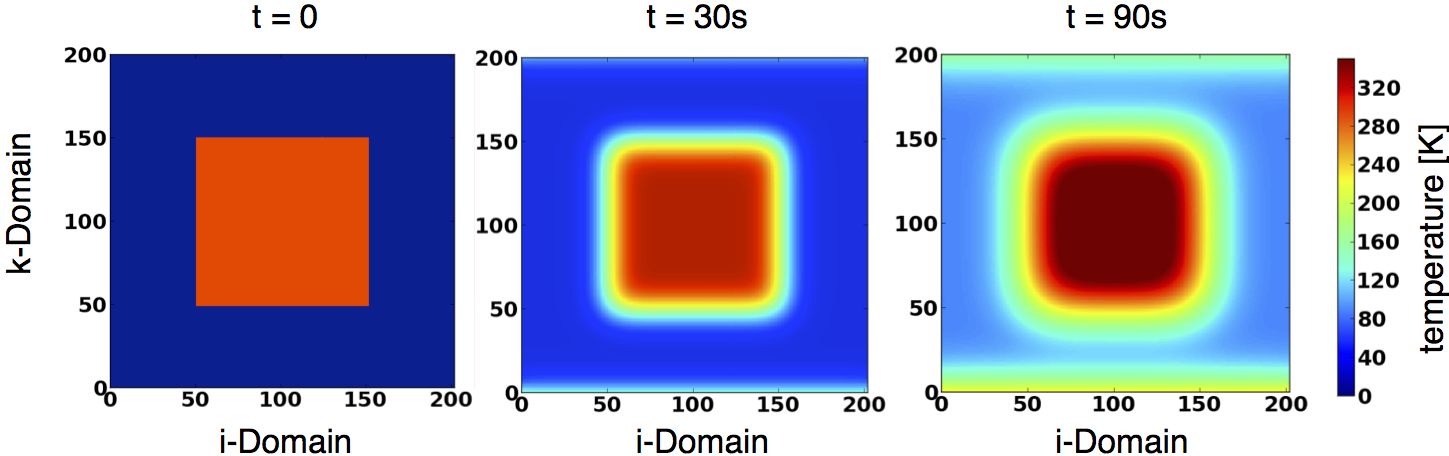}
  \caption{Output at $j=100$.}
  \label{figure:simple_weather_data}
\end{figure}

This equation is solved numerically by using the explicit Euler method in 3D. In addition, the $k = 0$ and $k = nz$ boundaries are modeled as infinite heat wells with temperatures $330\mathrm{K}$ and $200\mathrm{K}$ respectively. Radiation $f$ is assumed to be uniform and statically applied across the spatial domain. The \verb|KI| and \verb|KJ| boundaries are modeled cyclically. We initialize the temperature function to $0\mathrm{K}$ with a $300\mathrm{K}$ box from one quarter to three quarters of each spatial dimension and set a delta of $0.1\mathrm{s}$ between time steps. Figure \ref{figure:simple_weather_data} shows slices at $j=100$ for the resulting data with time steps at $t=0\mathrm{s}$, $t=30\mathrm{s}$ and $t=90\mathrm{s}$. This shows the heat diffusion, the radiation effect and the surface ($k=0$) and planetary boundary heat exchange ($k=200$).

\begin{figure}[htpb]
  \centering
  \includegraphics[width=1.0\linewidth]{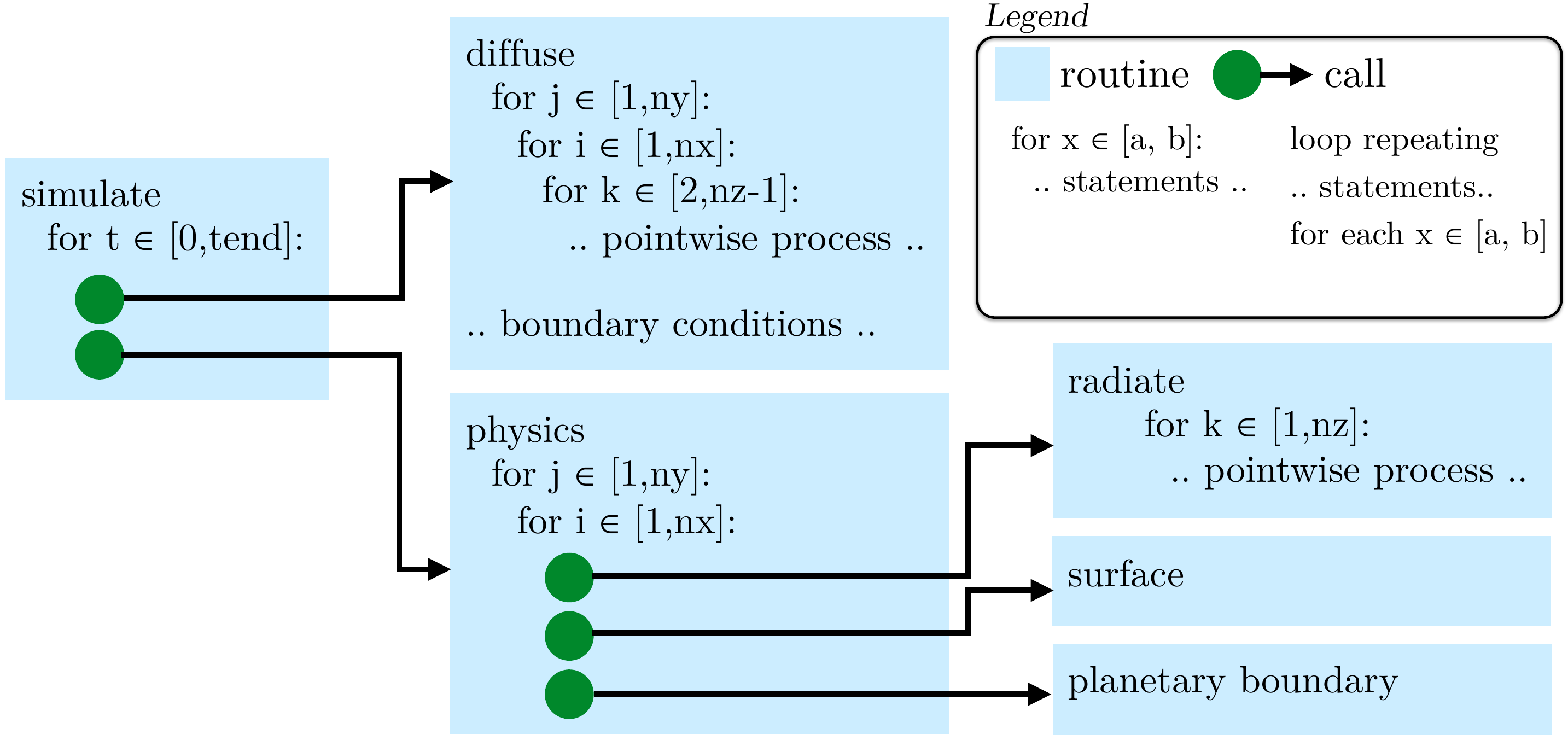}
  \caption{Code structure of the reduced weather application.}
  \label{figure:simple_weather_structure}
\end{figure}

\medskip

In order to adhere to ASUCA's code structure the application is split into a dynamical core (which here only consists of the diffusion $\frac{\partial T}{\partial t}= \kappa \nabla^2 T$) and physical processes with no interaction in the horizontal (radiation, surface- and planetary boundary heat exchange). Analogous to ASUCA's physical processes interface (see Section \ref{par:asuca:granularity}), such processes are parallelized (here in \verb|run_physics|) with each parallel thread executing all physical processes for a single \verb|K|-column. The dynamical core consists of four stencil code kernels - the inner \verb|IJK| region as well as the boundary regions for the \verb|IJ|, \verb|KI| and \verb|KJ| boundaries. Figure \ref{figure:simple_weather_structure} shows this code structure including the computationally relevant call graph, loops and data domains. The entire code is listed in Appendix \ref{sec:sources}.

\subsection{Design of the Hybrid Fortran Language Extension} \label{sec:hf-user-centric}

In the following section, by using the previously introduced reduced weather application as an example, we discuss the application of Hybrid Fortran with regards to the design goals stated in the beginning of section \ref{sec:hybridFortran}.

\subsubsection{Basic Parallelization} \label{sec:hf-basics}

An example of a parallelizeable code is the following extract of the diffusion algorithm (see Appendix \ref{sec:sources-diffusion} for a full listing):

\begin{lstlisting}[name=diffuse, label=listing:diffuse-reduced-sequential, caption={Diffusion, implemented sequentially.}]
    subroutine diffuse(energy_u, energy)
       real(8),intent(out):: energy_u(0:nx+1,0:ny+1,nz)
       real(8),intent(in):: energy(0:nx+1,0:ny+1,nz)
       ! ... further specifications and initializations ...
       do j = 1,ny
  	   do i = 1,nx
       do k = 2, nz-1
          energy_u(i,j,k) = energy(i,j,k) &
    &     * (1.0d0 - 6 * diffusion_velocity) + diffusion_velocity &
    &     * ( energy(i-1,j,k) + energy(i+1,j,k) + ... ) &
       end do
       end do
       end do
       ! ...
\end{lstlisting}

There are no loop carried dependencies for the sole modified data object \verb|energy_u|, thus this inner loop region is inherently parallelizable. One can parallelize this code for CPU and GPU with OpenMP and OpenACC by adding directives as follows\footnote{It is notable that OpenACC's kernel directive has symbol sharing defaults that are sufficient for the described use case, i.e. the default behaviour is equivalent to using default(firstprivate) and declaring all shared arrays explicitly. It will parallelize the stencil kernel based on heuristics as well as a dependency analysis (i.e. recognizing loops as parallelizeable when there are no loop carried dependencies).}:

\begin{lstlisting}[name=diffuse, label=listing:diffuse-reduced-par, caption={Diffusion, naive parallel implementation on CPU and GPU.}]
       !$omp parallel default(firstprivate) shared(energy, energy_u)
       !$acc kernels
       do j = 1,ny
       do i = 1,nx
       do k = 2, nz-1
       ! ... loop body ...
       end do
       end do
       end do
\end{lstlisting}

\smallskip

In contrast, in Hybrid Fortran the loops to be parallelized are replaced by the directive \verb|@parallelRegion| as follows\footnote{Here we only parallelize in I and J as otherwise the kernel thread runtime becomes sub-optimally small, which creates scheduler overhead.}\footnote{Without further specification, Hybrid Fortran assumes arrays to start at 1, which can be changed by adding additional clauses to the parallel region directive.}:

\begin{lstlisting}[name=diffuse, label=listing:diffuse-reduced-hf, caption={Diffusion, implemented in parallel with Hybrid Fortran.}]
    @parallelRegion{domName(i,j), domSize(nx,ny)}
    do k = 2, nz-1
       ! ... loop body ...
    end do
    @end parallelRegion
\end{lstlisting}

By replacing parallelizable loops with this explicit code construct the following is gained:

\begin{enumerate}
 \item It becomes immediately clear that loops are semantically different from parallel regions. Loops are always sequential constructs and thus allow any dependencies on previous values, while in a parallel regions the order of execution is not defined.
 \item For the GPU implementation there is a clear boundary between device code and host code (see also the glossary in Appendix \ref{sec:gpgpu-glossary}). All code inside the parallel region directive gets implemented as device code, while all other code remains on the host. Thus, Hybrid Fortran follows a more explicit approach for the parallelization, which more easily allows the programmer to have a mental model of how an algorithm is implemented on the device.
%  \item The source code overall becomes shorter, i.e. separate directives for CPU and GPU parallelizations as well as the \verb|do|/\verb|end do| constructs are not required.
\end{enumerate}

\smallskip

Through only the information provided in this one directive, Hybrid Fortran is able to generate OpenMP parallel loops for the CPU- as well as CUDA Fortran or OpenACC kernels for GPU-parallelization. The generated code for GPU is listed in Appendix \ref{sec:sources-diffusion-generated} while the generated CPU parallelization corresponds to Listing \ref{listing:diffuse-reduced-par}. It is noteworthy that Hybrid Fortran automatically adds the required sharing clauses. For that purpose it is assumed by default that all arrays need to be shared while all scalars can be treated as thread private\footnote{Reductions thus require an explicit clause for the parallel region directive, analogous to OpenMP reductions, see also Appendix \ref{sub:parallelRegionDirective}.}. Data object specification- as well as usage within parallel regions is parsed in order to enable useful default sharing behaviour. This combination of design choices makes implementing data parallel code very simple for the framework user, i.e. only the domain boundaries of the parallelization need to be specified.

\subsubsection{Device Data Region} \label{sec:hf-data-region}

The GPU's advantage in terms of memory bandwidth can only be used effectively if the copying of data via the comparatively slow PCI Express bus can be minimized.
%\footnote{see also inequality \ref{eq:speedup_req_mem_bound} in Section \ref{sec:perfModel} that describes the relationship between the data region size and the expected speedup.}
To ensure this, it is necessary to gather device state information for each data object in each routine, i.e. whether an existing device memory version can be utilized or whether the data object first needs to be synchronized with their host versions.

\smallskip

In order to improve on this with OpenACC, data directives are added for the time integration loop (Appendix \ref{sec:sources-time}) as follows:
% \footnote{For completeness sake we have used two exit data directives here, even though the second one can never be reached. In the general case (e.g. break statements in the while loop) this is necessary however. We note that the need to specify all the arrays in all entry- and exit points is a weakness here, as it does not adhere to the DRY\cite{hunt1999pragmatic} principle.}

%--------------------------------------------------!
%                                       ---------->!
%                                      max length  !
\begin{lstlisting}[name=dataDirective, label=listing:dataDirective, caption={The data directives.}]
!$acc enter data copyin(energy), &
!$acc& $copyin(energy_u, energy_surf, energy_pbl) &
do while (.true.)
  ! .. time step code ..
  time = time + timestep
  if (time > end_time) then
!$acc exit data delete(energy, energy_u), &
!$acc& delete(energy_surf, energy_pbl)
  	return
  end if
 end do
!$acc exit data delete(energy, energy_u), &
!$acc& delete(energy_surf, energy_pbl)
\end{lstlisting}

In addition, \verb|present| clauses need to be added to all OpenACC kernel directives for all the data objects that are already present thanks to this newly introduced data region.

\smallskip

By contrast, in Hybrid Fortran the device state of data objects is defined declaratively using \verb|@domainDependant| directives that are added after the specification parts of each relevant subroutine:

\begin{lstlisting}[name=diffuse, label=listing:diffuse-hf-data-region, caption={Specification of device present data}]
    subroutine diffuse(energy_u, energy)
       real(8),intent(out):: energy_u(0:nx+1,0:ny+1,nz)
       real(8),intent(in):: energy(0:nx+1,0:ny+1,nz)
       ! ... further specifications
       @domainDependant{attribute(autoDom, present)}
       energy_u, energy
       @end domainDependant
       ! ...
\end{lstlisting}

Through the \verb|present| attribute, Hybrid Fortran is instructed to treat data as device present. The \verb|autoDom| attribute furthermore instructs Hybrid Fortran to use the dimensions from the user specification (which becomes relevant with the features discussed in Section \ref{sec:hf-privatization}). By applying analogous \verb|@domainDependant| directives with \verb|transferHere| attribute at the data region boundaries (e.g. at the subroutine that implements the time integration loop), Hybrid Fortran is instructed to generate the data copies at those boundaries (instead of the default behavior, which is to transfer at every kernel invocation). Again, the user specified \verb|intent| is used for each data object to determine the type of memory transfer to be implemented, which minimizes the potential for bugs in comparison to OpenACC's explicit \verb|copyIn|, \verb|copyOut| and \verb|copy| clauses.

\subsubsection{Architecture-Dependent Parallelization Granularity} \label{sec:hf-granularity}

As discussed in Section \ref{par:asuca:granularity}, one of the main problems to solve in order to gain performance portability and to allow code reusability is to support parallelizations with variable granularity to co-exist within the same user code. For this purpose, Hybrid Fortran allows a parallel region to be applied to only a partial set of hardware architectures.

\smallskip

For the physical processes in the example we originally have the following code (see the full listings in Appendix \ref{sec:sources-physical}, \ref{sec:sources-radiation} and \ref{sec:sources-exchange}):

\begin{lstlisting}[name=diffuse, label=listing:physical-sequential, caption={Physical processes, implemented sequentially}]
   subroutine run_physics(...)
      ! ... specifications ...
      do j = 0,ny+1
      do i = 0,nx+1
         call radiate(energy(i,j,:))
         call exchange_heat_with_boundary( energy(i,j,:), energy_surf(i,j), 1 )
         call exchange_heat_with_boundary( energy(i,j,:), energy_pbl(i,j), nz )
      end do
      end do
   end subroutine

   subroutine radiate(energy)
      real(8), intent(inout), dimension(nz) :: energy
      ! ... more specifications and initializations ...
      do k=1,nz
         energy(k) = energy(k) + radiation_intensity
      end do
   end subroutine

   ! ... exchange_heat_with_boundary definition in 1D
\end{lstlisting}

Hybrid Fortran allows to write this as follows:

\begin{lstlisting}[name=diffuse, label=listing:physical-hf, caption={Physical processes, implemented with Hybrid Fortran}]
   subroutine run_physics(...)
       ! ... specifications ...
       @parallelRegion{ &
    &     appliesTo(CPU), domName(i,j), domSize(0:nx+1,0:ny+1), &
    &     startAt(0,0), endAt(nx+1,ny+1) &
    &  }
       call radiate(energy(i,j,:))
       call exchange_heat_with_boundary( energy(i,j,:), energy_surf(i,j), 1 )
       call exchange_heat_with_boundary( energy(i,j,:), energy_pbl(i,j), nz )
       @end parallelRegion
   end subroutine

  subroutine radiate(energy)
     real(8), intent(inout), dimension(nz) :: energy
     ! ... more specifications and initializations ...
     @parallelRegion{ &
    &     appliesTo(GPU), domName(i,j), domSize(0:nx+1,0:ny+1), &
    &     startAt(0,0), endAt(nx+1,ny+1) &
     }
     do k=1,nz
        energy(k) = energy(k) + radiation_intensity
     end do
     @end parallelRegion
  end subroutine

  ! ... exchange_heat_with_boundary definition in 1D, analogous to radiation ...
\end{lstlisting}

\begin{figure}[htpb]
  \centering
  \includegraphics[width=1.0\linewidth]{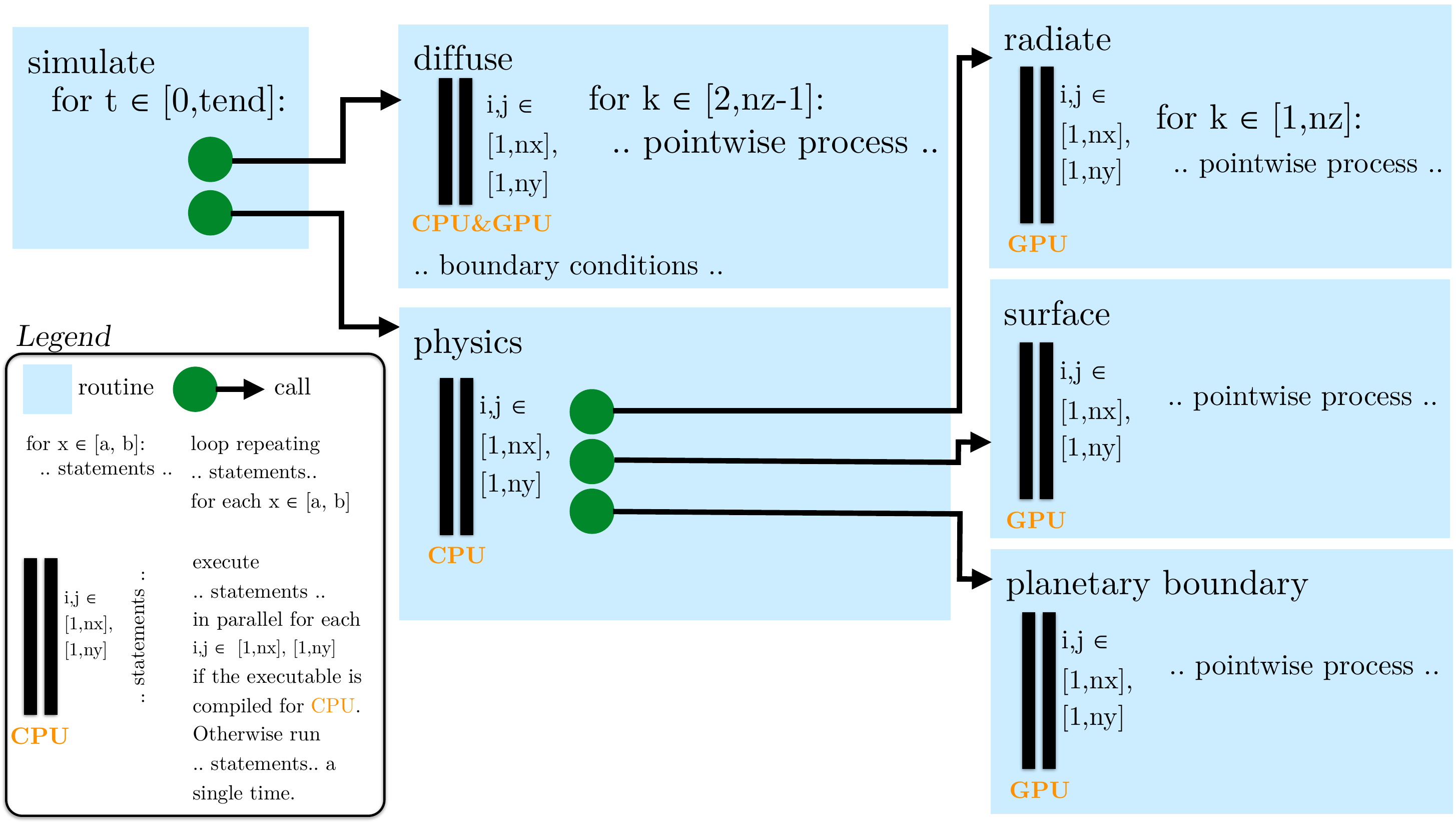}
  \caption{Code structure of the reduced weather application, adapted for Hybrid Fortran.}
  \label{figure:simple_weather_structure_hf}
\end{figure}

This shows the following:

\begin{enumerate}
 \item Using an \verb|appliesTo| clause (e.g. for the physical processes), parallel regions can be applied partially to specific architectures, that is, for architectures not listed in this clause the region body will be run only once per call and the parallelization can be applied at a higher granularity (here at the radiation, surface and planetary boundary processes). See also Figure \ref{figure:simple_weather_structure_hf} for the resulting code structure and compare to Figure \ref{figure:simple_weather_structure} in Section \ref{sec:minimal-application}.
%  \item In the GPU case, \verb|run_physics| becomes a sequence of 3D kernel invocations, while in the CPU case it is itself a kernel, calling 1D-subroutines.
 \item The user code within parallel region bodies can remain untouched compared to the sequential CPU version. Hybrid Fortran automatically extends data accesses and specifications, here from 1D to 3D for \verb|radiate| (see also Section \ref{sec:hf-privatization}).
\end{enumerate}

\begin{figure}[htpb]
  \centering
  \includegraphics[width=0.8\linewidth]{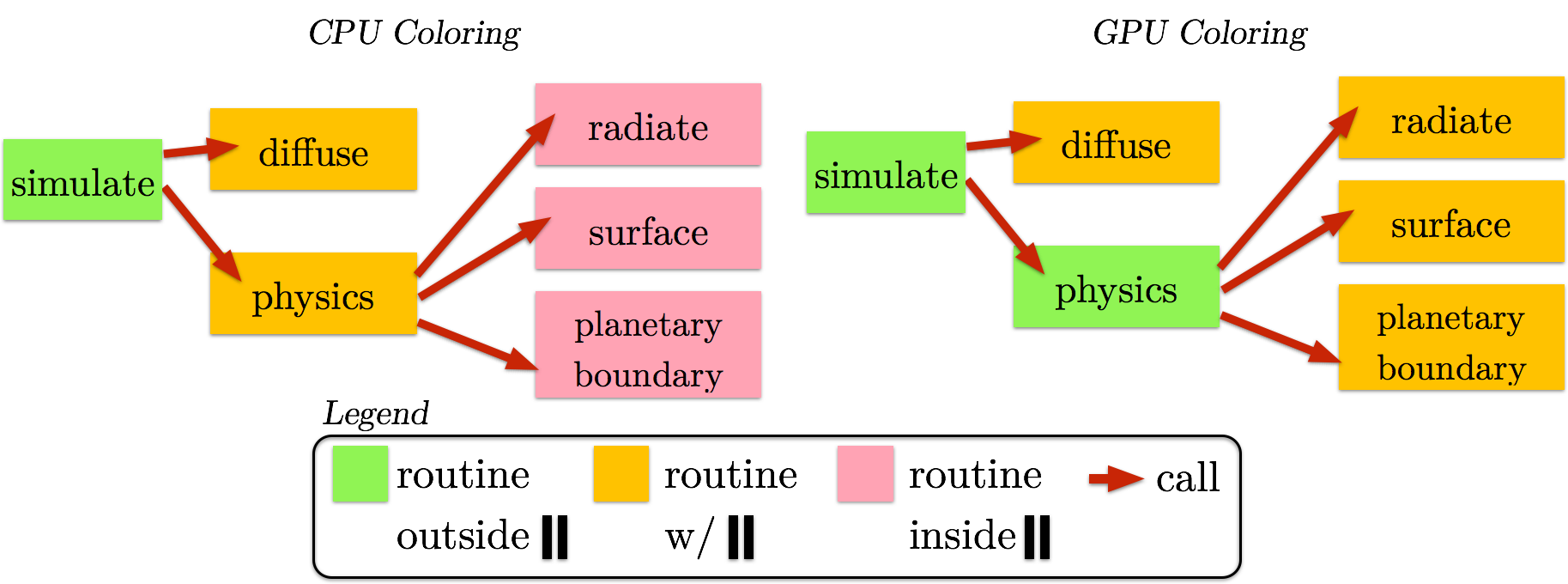}
  \caption{Architecture specific call graph colorings.}
  \label{figure:coloring}
\end{figure}

\smallskip

In order to be able to correctly generate the code for varying parallelization granularity, the transformation process requires positioning information for each routine towards relevant parallel regions. Each routine either directly contains a parallel region, has parallel regions within its call graph, or is itself called within a parallel region. This information can be visualized as a coloring of the call graph, with a separate coloring for each target architecture. Parallel region \verb|appliesTo| clauses are the deciding factor for this coloring. Figure \ref{figure:coloring} shows this for the reduced weather application. This global positioning information is required by the code transformation in order to implement the correct dimensionality (see also Section \ref{sec:hf-privatization}) as well as the correct type of code in case of GPUs being the target, i.e. host-, kernel- or device routine code (see also the relevant definitions in Appendix \ref{sec:gpgpu-glossary}).

\subsubsection{Architecture-Dependent Privatization} \label{sec:hf-privatization}

Since Hybrid Fortran allows compile-time defined parallelization granularity, as discussed in Section \ref{sec:hf-granularity}, it is necessary for the code transformation to change the dimensionality of data objects in order to expose their data parallelism at a finer-grained level. This requires hints from the framework user. The following listing shows the \verb|@domainDependant| directives required in order to extend the data objects from 1D to 3D (adding the missing parallel dimensions). This is used in column-only physical parametrizations (radiation and boundary layer in case of ASUCA) where the original CPU code is parallelized in a single place using OpenMP-loops over the horizontal domain. Applying this technique allows the user to simply wrap the existing code at a fine-grained level with GPU-targeted \verb|@parallelRegion| constructs in order to delegate the task of exposing data parallelism to the framework. 

\begin{lstlisting}[name=radiate, label=listing:radiation-hf, caption={Extending data object dimensions with Hybrid Fortran}]
  subroutine radiate(energy)
     real(8), intent(inout), dimension(nz) :: energy
     ! ... more specifications ...
     @domainDependant{attribute(autoDom, present), domName(i,j), domSize(0:nx+1,0:ny+1)}
     energy
     @end domainDependant

     @parallelRegion{ &
  &     appliesTo(GPU), domName(i,j), domSize(0:nx+1,0:ny+1), &
  &     startAt(0,0), endAt(nx+1,ny+1) &
  &  }
     do k=1,nz
        energy(k) = energy(k) + radiation_intensity
     end do
     @end parallelRegion
  end subroutine
\end{lstlisting}

% Using \verb|domName| and \verb|domSize| clauses in the \verb|@domainDependant| directive together with the \verb|autoDom| attribute allows the user to specify the additional dimensions that are required when the subroutine has parallel regions in ``inside'' or ``within'' position.
% %This position is determined according to the call graph coloring shown in Figure \ref{figure:coloring}.
% The dimensions specified with \verb|domName| and \verb|domSize| are required to match the ones used in the \verb|@parallelRegion| directives.
% % As an example the \verb|dimension| specification of \verb|energy| is rewritten to \verb|dimension(DOM(nx,ny,nz))| here. \verb|DOM| is a macro wrapping used to define the storage order - see also Section \ref{sec:hf-storage-order}.

\subsubsection{Storage Order} \label{sec:hf-storage-order}

As will be shown in Section \ref{sec:storage_order}, variable and architecture-dependent storage order is necessary for performance portability as well as code reuse. Hybrid Fortran therefore automatically adds default macro wrappings for multidimensional arrays. This allows a specification of the storage order depending on the target architecture in a centralized location. Unlike with other code hybridization methods supporting Fortran, this enables variable storage order without the need to rewrite any user code. In case more flexibility is needed (such us varying storage order in different parts of an application), Hybrid Fortran allows an explicit definition of the macro names in the \verb|@domainDependant| directive.

\medskip

As an example, consider again Listing \ref{listing:physical-hf}, which shows the Hybrid Fortran version of the reduced weather application's physics code. Through the transformation processes, array accesses and specifications are rewritten as follows:

\begin{lstlisting}[name=storageOrderMacros, label=listing:storageOrderMacros, caption={Storage order and coarse-grained parallelization after automated transformation for CPU}]
do j = 0,ny+1
  do i = 0,nx+1
    call radiate(energy(AT(i,j,:)))
    !...
  end do
end do
\end{lstlisting}

\medskip

In an separate source file that is automatically included in all sources, the macro \verb|AT| is specified as follows:

\begin{lstlisting}[name=storageOrderMacros, label=listing:storageOrderMacrosDEFINITION, caption={Storage order definition}]
	#if (GPU)
		#define AT(i,j,k) i, j, k
	#else
		#define AT(i,j,k) k, j, i
	#endif
\end{lstlisting}

% The macro \verb|DOM| is simply an alias for \verb|AT|, kept separately in order to allow for deviating definitions in the future.

% \medskip

% While Hybrid Fortran generates this code automatically for all array accesses and specifications, in OpenACC all array accesses would need to be rewritten in order to achieve the same.

\subsubsection{Block Size} \label{sec:hf-block}

% PGI Accelerator however uses the sub-optimal block size $128 \times 1$ in this case. To avoid this it is necessary to prefix each parallel loop inside \verb|kernels| with an \verb|!$acc loop independent vector(SIZE)| directive, with \verb|SIZE| being 32 for loops over \verb|i| and 16 for loops over \verb|j| domain.

Block size (see also Appendix \ref{sec:gpgpu-glossary}) can have a significant effect on GPU performance. Hybrid Fortran by default chooses a block size of \verb|32x16x1|, which is a good default for data parallel atmospheric applications (see also Section \ref{sec:VectorSize}). This default can be changed through macros in a configuration source that gets automatically included in all other sources. In cases where kernels have varying optimal block sizes, Hybrid Fortran allows the specification of a macro name suffix in the parallel region directive.

\smallskip

In contrast, PGI Accelerator's OpenACC implementation by default uses a $128 \times 1$ block size configuration, which usually leads to a subpar performance due to a low occupancy (see also Appendix \ref{sec:gpgpu-glossary}). This therefore needs to be changed explicitly for each loop invocation by using \verb|vector| clauses in \verb|$acc loop| directives added to all parallel loops. Furthermore it is often advisable to use \verb|!$acc loop seq| for loops over \verb|k| in order to force the compiler not to assign the first thread index to \verb|k| (which is sub-optimal for the chosen storage order).

\smallskip

By using macros, Hybrid Fortran avoids leaking this architecture dependant information to the user code in favor of a centrally defined configuration.

\subsection{Source Transformation Method}  \label{sec:hf-implementation-centric}

\begin{figure}[htpb]
  \centering
  \includegraphics[width=1.0\linewidth]{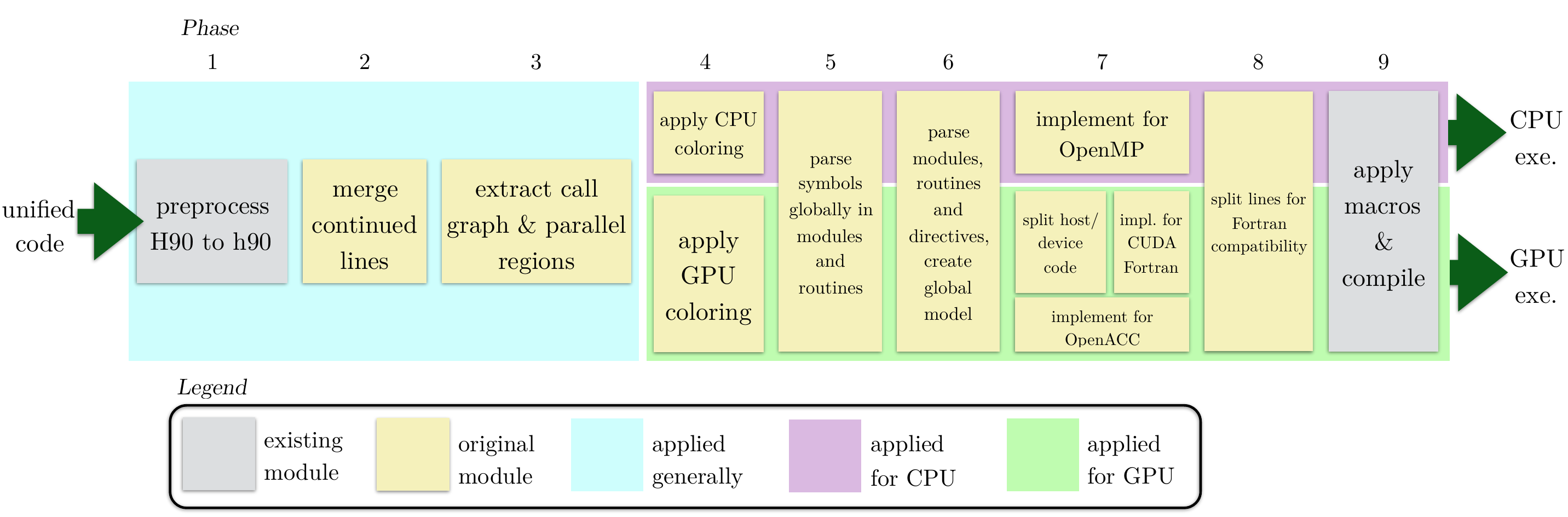}
  \caption{Steps involved in the code transformation method.}
  \label{figure:hfMethod}
\end{figure}

In this section we introduce the code transformation, involved in implementing the framework logic discussed. Figure \ref{figure:hfMethod} shows this process in nine distinct phases, with phases four through nine being applied separately depending on the target architecture. This process runs transparently from the user point of view, i.e. it is applied automatically by the means of a common Makefile provided together with the framework. Each of the numbered paragraphs in this section corresponds to one transformation phase.

\subsubsection{Macro Processing for Unified Sources}

In addition to the automatically applied transformation process that implements the behavior described in Section \ref{sec:hf-user-centric}, a further layer of meta-programming is supported by applying a macro processing step to Hybrid Fortran sources\footnote{Hybrid Fortran's build system recognizes sources from four distinct formats: f90 (Modern Fortran), F90 (Modern Fortran with Macros), h90 (Hybrid Fortran) and H90 (Hybrid Fortran with Macros). Depending on the source format, the build process starts at phase one (H90), two (h90) or nine (f90 / F90).}, including directives. We use the GNU compiler tool-chain for this purpose (\verb|gcc -E|).

\subsubsection{Merging of Continued Lines}

In order to simplify the parsing in subsequent phases, Fortran continuation lines are merged.

\subsubsection{Call Graph and Parallel Region Parsing}

Facilitating phase 4, the application's call graph and parallel region directives are parsed globally.

\subsubsection{Call Graph Coloring} \label{sec:phase-coloring}

A call graph coloring is created, as shown previously in Figure \ref{figure:coloring}.

\subsubsection{Global Symbol Parsing}

In order to gather symbol information, the application is parsed twice in this phase: A first time in order to gather module data object specifications and a second time to link this information to the routines where such module data objects get imported, together with the information of data objects defined in each routine scope. This allows Hybrid Fortran to globally determine the domain information for data objects.

\subsubsection{Global Model Generation}

All previous information is compiled into a global application model, with model classes representing the modules, routines and code regions. Code regions are modelled using subclasses for specification-, sequential-, call- and parallel regions. For each code line this model contains a list of the data objects used within the line and their known meta information (gathered previously from local module scope, imported scope, routine scope as well as added information from Hybrid Fortran directives). Each routine model is also assigned an implementation class (see phase 7).

\subsubsection{Implementation} \label{sec:phase-implementation}

One of the stated goals of Hybrid Fortran is to create a higher level tool-chain that is reusable for various hardware- and software architectures. Thus, only a single transformation phase is dependent on the backend implementation, here called the ``implementation phase''. For this purpose, a Python implementation class hierarchy is used in order to facilitate the reuse or specialization of a set of implementation methods for the varying backends. Previously gathered information is used by instance methods of these implementation classes to create the sources from the previously built model. More specifically, using implementation classes for either standard Fortran, OpenMP Fortran, CUDA Fortran or OpenACC Fortran, the following transformations are applied:

\begin{itemize}
    \item data object specifications and accesses are transformed according to object type, implementation class and routine coloring,
    \item data copies from and to the device are generated,
    \item device- and host code boilerplate is generated. Since this requires separate routines for CUDA (one for the host code and one for each kernel in the user code), parallel region bodies are split off into their own generated routines beforehand.
    % (see also listings \ref{listing:diffuse-reduced-cuda-host} and \ref{listing:diffuse-reduced-cuda-device}).
\end{itemize}

CUDA Fortran and OpenACC GPU implementations can be mixed in the same project. This is mainly done in order to make use of OpenACC's reduction feature (which is not supported for the CUDA Fortran backend). Switching between the different backend architectures is done by specifying a default implementation in the configuration and wrapping routines in a \verb|@scheme| directive where a different one needs to be applied. In order to facilitate interoperatibility, Hybrid Fortran uses CUDA Fortran allocated device pointers for all GPU implementations.

\subsubsection{Line Splitting}

In order to make the generated code compatible with standard Fortran compilers (which usually enforce a maximum line length), the generated lines are split at appropriate positions containing white space until the maximum line length is satisfied.

\subsubsection{Final Macro Processing and Compilation}

The GNU preprocessor is applied again in order to process the generated macros (e.g. for storage reordering, see also Section \ref{sec:hf-storage-order}). Subsequently, a user specified compiler and linker is used in order to create the CPU and GPU executables. For GPU, the generated code is compatible with PGI Accelerator in case of the CUDA Fortran implementation, and PGI Accelerator or Cray Fortran compiler in case of the OpenACC implementation.

\subsection{Limitations} \label{hf-limitations}

Hybrid Fortran has at this point in time mainly been tested to implement structured grid applications. We expect that unstructured grid applications would reveal currently unsupported use cases.

\medskip

Additionally, some limitations apply to the code within GPU-applied parallel region bodies: Such code must not contain recursion, call other kernel routines, use data objects with \verb|DATA| or \verb|SAVE| attribute, contain I/O statements (except \verb|print|) or contain array or slice expressions such as \verb|A(:,:,:) = 0.0| or \verb|C = A + B| (i.e. all operations in GPU parallel region bodies are required to be scalar).

\section{Performance Model} \label{sec:perfAnalysis}

In this section a performance model is constructed for the application introduced in Section \ref{sec:minimal-application}. For this purpose we simplify the problem as follows:

\begin{enumerate}
\item By assuming an optimized storage order (see also Section \ref{sec:storage_order}) the bandwidth bounded model for sequentially accessed memory can be applied to the inner region of \verb|diffusion| as well as \verb|radiation|.
\item Memory accesses from cache are assumed to not affect execution time at all (i.e. such accesses are assumed to be hidden behind host- or device memory accesses in the pipeline).
\item By assuming sufficiently large spatial dimensions we omit all other program parts from further modelling since their runtime effect should be negligible.
\item We assume the compiler to optimize all operations that can be hoisted out of loops.
\end{enumerate}

\medskip

Per loop iteration, \verb|diffusion|\footnote{See also the source in Appendix \ref{sec:sources-diffusion}}, being a seven point stencil code, requires a maximum of eight memory accesses per point update as well as six add and two multiply instructions. The arithmetic intensity according to an adapted roofline model\cite{roofline} is thus

\begin{equation}
\frac{c}{{{b \cdot m}}} = \frac{8}{8 \cdot 8} \begin{bmatrix} \frac{{FLOP}}{B} \end{bmatrix} = 0.125 \begin{bmatrix} \frac{{FLOP}}{B} \end{bmatrix}, 
\end{equation}
\smallskip

where $m$ is the number of values to read and write from/to memory per point update, $b$ is the byte length of each value and $c$ is the number of floating point cycles spent per point update\footnote{In contrast to the roofline model we use floating point cycles rather than operations here in order to have a more generalized model that can be applied to multi-cycle floating point operations such as division or exponentiation}.

\medskip

The minimum arithmetic intensity for compute boundedness is $5.9 \lbrack \frac{{FLOP}}{B} \rbrack$ and $7.8 \lbrack \frac{{FLOP}}{B} \rbrack$ for Tesla K20x and P100, respectively (using the performance numbers from Appendix \ref{sec:hardware-metrics}). Thus, \verb|diffuse| is clearly memory bandwidth bounded. This holds true even if six of the eight memory accesses are cached (see also the stencil access pattern in Appendix \ref{sec:sources-diffusion}). Due to the lower random access memory bandwidth, the boundary region \verb|JK| is clearly also memory bandwidth bounded. The same holds for the radiation process, since it requires two memory accesses and one FLOP per iteration\footnote{One should note however that this does not necessarily reflect typical physical processes in atmospheric models - such processes tend to consist of a mix of compute bounded and memory bandwidth bounded algorithms.}. We therefore conclude that the reduced weather application is memory bandwidth bounded for all processes.

\medskip

The reduced weather application is thus highly sensitive to the cache performance of the target architecture. Applying the above parameters to the speedup condition (equation \ref{eq:speedup_req_mem_bound}), a speedup on GPU is feasible if the output is not required at every time step.

\medskip

With some simplification in the domain boundaries (by assuming sufficiently large domains for halos to be neglectable) the model GPU execution time becomes

\begin{equation}
\frac{\delta t_{output}}{\delta t_{timestep}} (n_x \cdot n_y \cdot n_z \cdot ( \frac{b \cdot m_{sa}}{B{W_D}} + \frac{b \cdot m_{HtoD}}{B{W_{HtoD}}} ) + n_y \cdot n_z \cdot \frac{m_{ra}}{RA_{D}}),
\end{equation}
\smallskip

and the model CPU execution becomes

\begin{equation}
\frac{\delta t_{output}}{\delta t_{timestep}} (n_x \cdot n_y \cdot n_z \cdot \frac{b \cdot m_{sa}}{B{W_H}} + \cdot n_y \cdot n_z \cdot \frac{m_{ra}}{RA_{H}}),
\end{equation}
\smallskip

where $b$ is the byte length of each value, $m_{sa}$ is denoting the number of sequential memory accesses in the inner region of \verb|diffuse| and \verb|radiation| and $m_{ra}$ is denoting the number of random access updates in the boundary region \verb|JK|, $n_x$, $n_y$ and $n_z$ are denoting the grid sizes in x-, y- and z-direction, respectively, ${B{W_H}}$, ${B{W_D}}$ and ${B{W_{HtoD}}}$ are the bandwidths available on host, device and between host and device, respectively, and ${RA_{H}}$ and ${RA_{D}}$ are the number of random memory accesses per second (measured in $GUP/s$) \cite{luszczek2006hpc}. It holds that $m_{sa}=10$ without caching, $m_{sa}=4$ with caching and $m_{ra}=4$. See also Section \ref{sec:performance-sample} for a comparison of the measured performance with this model.
\section{Performance Results and Discussion}\label{sec:performance}

This section discusses performance results obtained using the Hybrid Fortran approach. In Section \ref{sec:performance-sample} the reduced weather application (introduced in Section \ref{sec:minimal-application}) is used to compare the Hybrid Fortran user code implementation to an OpenACC implementation and the performance model (introduced in Section \ref{sec:perfAnalysis}) with respect to the overall performance as well as individual performance related implementation aspects. Hybrid ASUCA is presented as a new implementation in Section \ref{sec:hybrid-asuca}. Sections \ref{sec:performance-asuca-small} and \ref{sec:performance-asuca} subsequently compare its GPU performance to the reference implementation (CPU-only), for small and large scale grids, respectively.

\subsection{Comparing Hybrid Fortran with OpenACC and Model}\label{sec:performance-sample}

This section facilitates the previously discussed reduced weather application in order to compare Hybrid Fortran with OpenACC and a performance model. We focus here on the desired portability features discussed in Section \ref{sec:design-perf-port}. For the performance measurements in this section the configuration as listed in Appendix \ref{sec:config-reduced} has been used.

\subsubsection{Performance Portable Storage Order}\label{sec:storage_order}

\medskip

Table \ref{table:storageOrderAndExecutionTime} shows the impact, storage order has on execution time. In the case of the currently discussed application, choosing a sub-optimal storage order impacts CPU execution time negatively by 35\%, while on GPU the slowdown is 7.7x. This verifies the necessity of a flexible storage order for applications with similar data structures as ASUCA, as discussed in Section \ref{par:asuca:storage}.

\begin{savenotes}
\begin{table}[htpb]
  \centering
  \footnotesize
  \caption{Influence of Storage Order on Execution Time, $nx=ny=nz=128$}
  \label{table:storageOrderAndExecutionTime}
	\begin{tabular}{|Sc|Sc|Sc|}
	\hline
	 & IJK Order & KIJ Order\tabularnewline
	\hline
	CPU Single Core & 1.73s & 1.28s\tabularnewline
	\hline
	GPU (OpenACC) & 0.10s & 0.77s \tabularnewline
	(Fastest Implementation) &  & \tabularnewline
	\hline
	\end{tabular}
\end{table}
\end{savenotes}

\subsubsection{``Naive'' Parallelization}\label{sec:naiveParallelization}

\begin{table}[htpb]
  \centering
  \footnotesize
  \caption{Execution Time with ``Naive'' Parallelization}
  \label{table:executionTimeNaiveParallelization}
	\begin{tabular}{|Sc|Sc|Sc|}
	\hline
	$nx \cdot ny \cdot nz$ & $128^3$ & $256^3$\tabularnewline
	\hline
	CPU Single Core Measurement & 1.28s & 8.20s\tabularnewline
	\hline
	CPU Single Core Model w/ cache & 0.74s & 5.70s\tabularnewline
	\hline
	CPU Single Core Model w/o cache & 1.77s & 13.91s\tabularnewline
	\hline
	CPU 6 Core Measurement & 0.40s & 4.25s\tabularnewline
	\hline
	CPU 6 Core Model w/ cache & 0.38 & 2.84s\tabularnewline
	\hline
	CPU 6 Core Model w/o cache & 0.87 & 6.77s\tabularnewline
	\hline
	GPU Measurement & 163.13s & n/a\tabularnewline
	\hline
% 	GPU Model w/o cache & 0.087s & 0.66s\tabularnewline
% 	\hline
	\end{tabular}
\end{table}

In order to establish a baseline performance we apply a basic parallelization to the reduced weather application with OpenACC and OpenMP directives as discussed in Listing \ref{listing:diffuse-reduced-par}, Section \ref{sec:hf-basics}. Storage order is made variable across the whole application by using macros for accessing and specifying multi-dimensional arrays. We are using I-J-K order for GPU and K-I-J order for the CPU implementation. Table \ref{table:executionTimeNaiveParallelization} shows the resulting CPU performance from this parallelization.

\medskip

We conclude that the measured CPU performance is already well within the models with perfect and no cache. The GPU performance is however very slow - more than 400x slower than the six core CPU version, which is not in agreement with the models we have constructed. This is to be expected however, since no data region has yet been defined, without which the data is being copied over the slow PCI express bus for every kernel invocation. Further discussion for the reduced weather application therefore focuses on GPU performance.

\subsubsection{Data Region}\label{sec:dataRegion}

\begin{table}[htpb]
  \centering
  \footnotesize
  \caption{Execution Time with Data Region}
  \label{table:executionTimeDataRegion}
	\begin{tabular}{|Sc|Sc|Sc|}
	\hline
	$nx \cdot ny \cdot nz$ & $128^3$ & $256^3$\tabularnewline
	\hline
	GPU Measurement & 0.095s & 0.63s\tabularnewline
	\hline
	GPU Model w/ cache & 0.027s & 0.19s\tabularnewline
	\hline
	GPU Model w/o cache & 0.087s & 0.66s\tabularnewline
	\hline
	\end{tabular}
\end{table}

The amount of I/O to and from the GPU can be greatly reduced by using a data region in order to avoid host-to-device data copies at every kernel invocation, as discussed in Section \ref{sec:hf-data-region}. Table \ref{table:executionTimeDataRegion} shows the resulting performance with OpenACC.
% Section \ref{sec:perfComparisonSimpleWeather} will also show Hybrid Fortran performance results for a version with a declared data region.
% As demonstrated here, implementing a data region and choosing a large enough problem size\footnote{in order to increase the GPU occupancy, see also the description in Appendix \ref{sec:gpgpu-glossary}} is sufficient to yield a reasonable performance for this application, however there is still room for improvement as the following section will show.

\subsubsection{Block Size}\label{sec:VectorSize}

\begin{table}[htpb]
  \centering
  \footnotesize
  \caption{Block Size Impact on Execution Time}
  \label{table:executionTimeVectorSize}
	\begin{tabular}{|Sc|Sc|Sc|Sc|}
	\hline
	$nx \cdot ny \cdot nz$ & $128^3$ & $256^3$ & $256^2 \cdot 512$\tabularnewline
	\hline
	GPU Automatic Block size & 0.095s & 0.63s & 2.99s\tabularnewline
	\hline
	GPU 32 x 16 Block size & 0.10s & 0.55s & 2.16s\tabularnewline
	\hline
	GPU Model w/ cache & 0.027s & 0.19s & 0.69s\tabularnewline
	\hline
	GPU Model w/o cache & 0.087s & 0.66 & 2.60s\tabularnewline
	\hline
	\end{tabular}
\end{table}

As Table \ref{table:executionTimeVectorSize} shows, up to 38\% performance improvement is possible over the version discussed in Section \ref{sec:dataRegion} for the OpenACC implementation by changing the default block size using vector clauses for each parallel loop (for a definition of block size, see also the GPGPU terms glossary in Appendix \ref{sec:gpgpu-glossary}). PGI accelerator uses a $128 \times 1$ block size for its OpenACC implementation by default while $32 \times 16$ has been experimentally determined to be the best block size for this application except for a small degradation for the smallest measured grid size. See also Section \ref{sec:perfComparisonSimpleWeather} for a performance comparison with Hybrid Fortran (which chooses this block size by default).

\subsubsection{Performance Comparison}\label{sec:perfComparisonSimpleWeather}

\begin{figure}[htpb]
  \centering
  \includegraphics[width=0.8\linewidth]{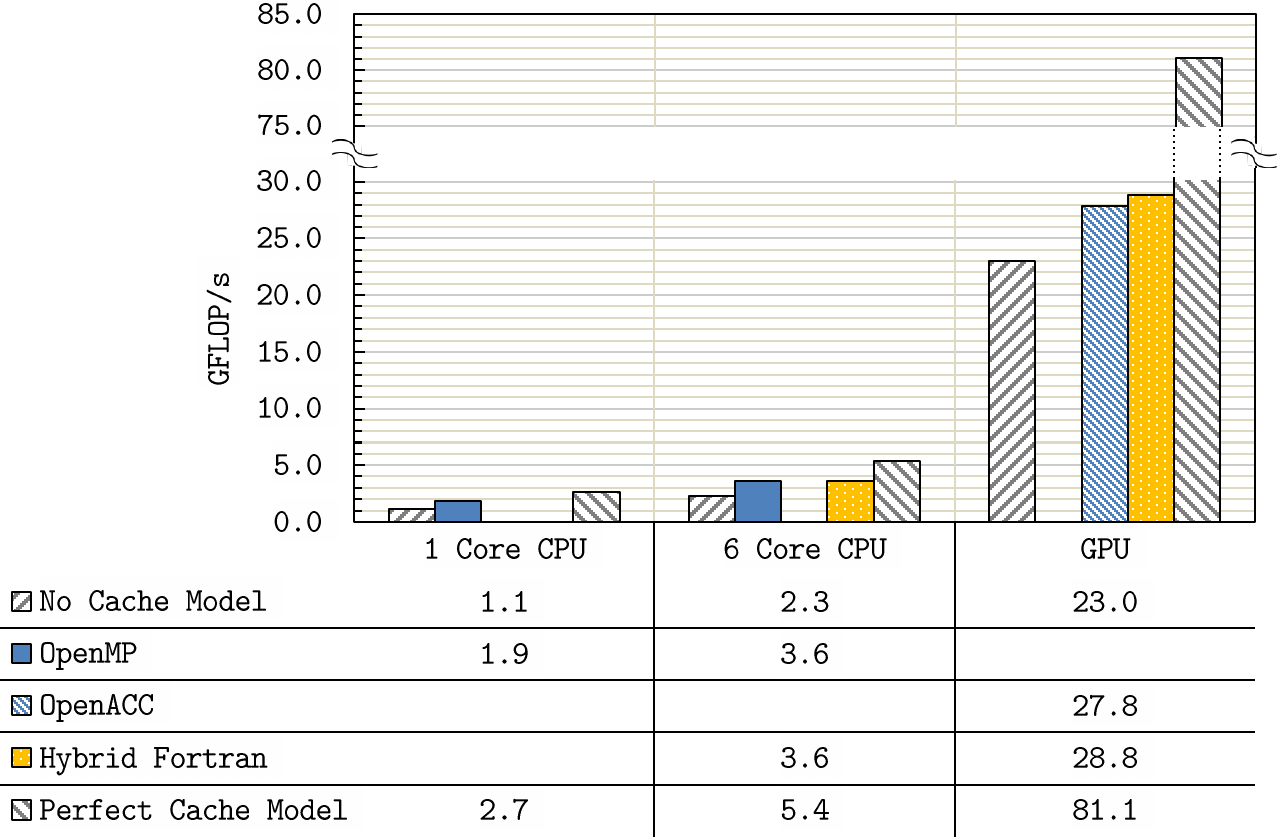}
  \caption{Comparing performance with the reduced weather application for handwritten vs. Hybrid Fortran generated vs. model on $256^3$ Grid.}
  \label{figure:performanceSimpleWeather}
\end{figure}

Figure \ref{figure:performanceSimpleWeather} shows the performance results for the reduced weather application with the Hybrid Fortran based implementation (discussed in Section \ref{sec:hf-user-centric}) in comparison with the models introduced in Section \ref{sec:perfAnalysis} as well as the OpenMP/OpenACC based approach discussed in Section \ref{sec:hf-user-centric}. For each of the implementations the highest performing version (32 x 16 block size with data region) has been selected for this comparison. This shows that measured CPU performance aligns well in between the two models (with and without cache) while for the TSUBAME 2.5 Kepler based GPU architecture, L1 cache appears to be less effective for this application and the measured performance is closer to the model without cache. It is noteworthy that the Hybrid Fortran generated code performs as well or better than the equivalent OpenACC code for both target architectures in this example.

\medskip

Overall we observe a speedup of 8x for the Hybrid Fortran version vs. 6-core CPU, compared to the speedup of 7.7x for the OpenACC version. The Hybrid Fortran implementation's speedup is thus very close to the bandwidth increase 8.2x between the two architectures (see Appendix \ref{sec:hardware-metrics}).

\medskip

% By defining

% \begin{equation}\label{eq:efficiency}
% \text{efficiency} = \frac{\text{speedup}}{\frac{BW_{D}}{BW_{H}}}
% \end{equation}
% \smallskip

% we arrive at an efficiency of 97\% for the Hybrid Fortran implementation versus 93\% of the OpenACC implementation.

On CPU the Hybrid Fortran implementation performs the same as the manually coded OpenMP implementation, which is unsurprising given that the CPU code generated by Hybrid Fortran is practically identical.

\subsection{Hybrid ASUCA Implementation} \label{sec:hybrid-asuca}

\begin{figure}[htb]
  \centering
  \includegraphics[width=0.8\linewidth]{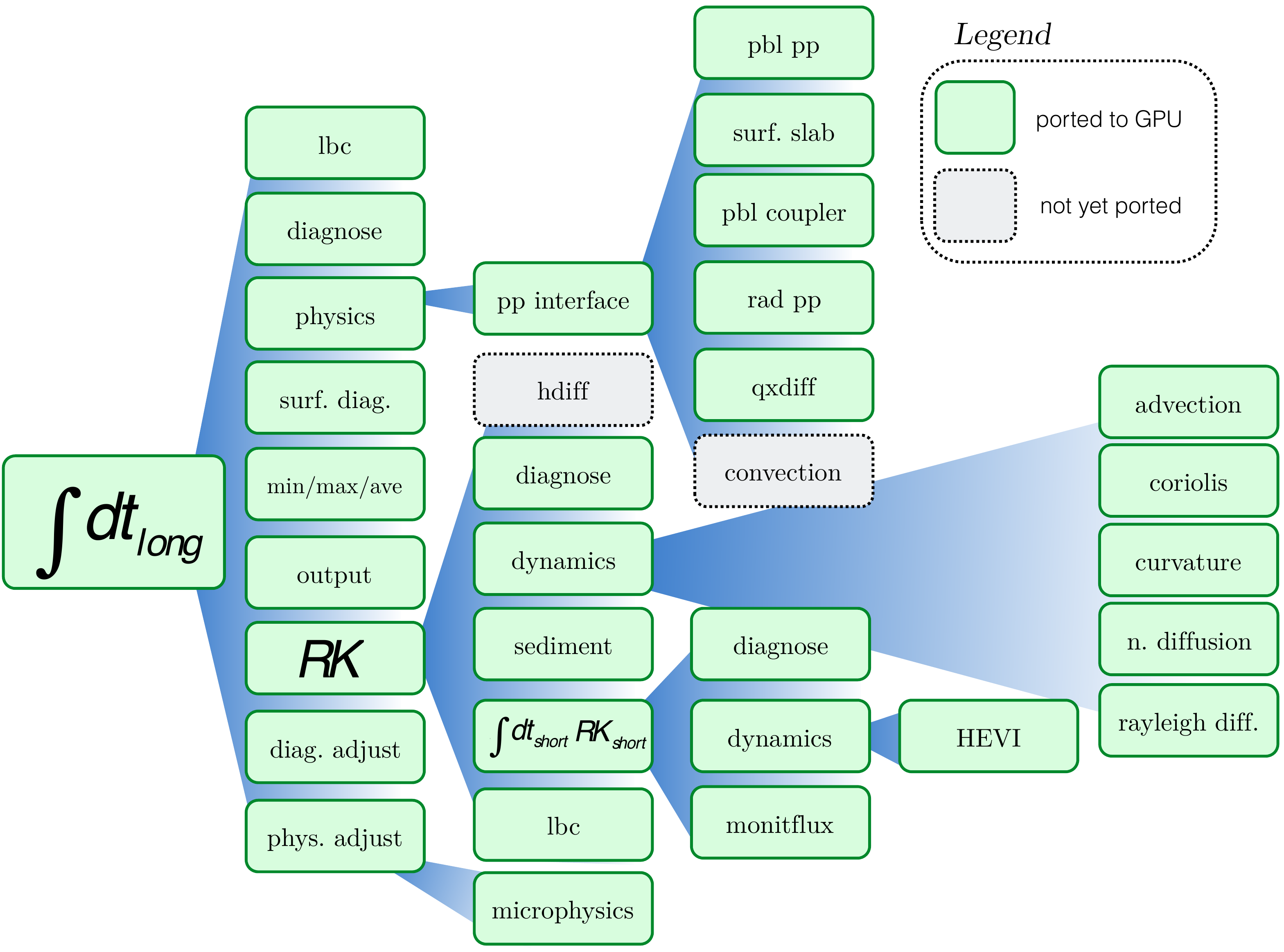}
  \caption{Simplified call graph of ASUCA and status of Hybrid Fortran based implementation}
  \label{fig:asuca-structure}
\end{figure}

\begin{figure}[htb]
  \centering
  \includegraphics[width=1.0\linewidth]{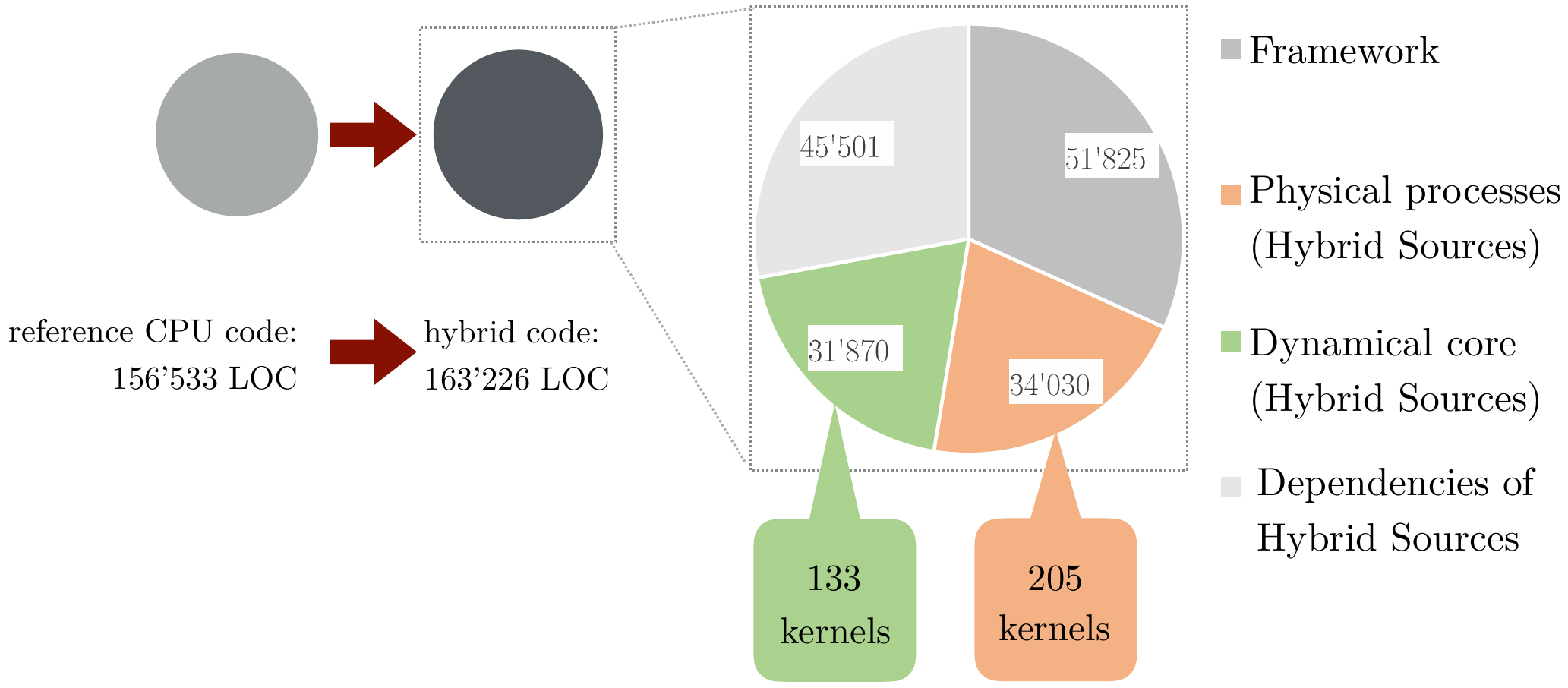}
  \caption{Hybrid ASUCA in numbers}
  \label{fig:asuca-numbers}
\end{figure}

As Figure \ref{fig:asuca-structure} shows, by employing Hybrid Fortran directives to both the dynamical core and the physical processes of the existing CPU-targeted ASUCA user code, nearly all modules required for an operative weather prediction have been ported to GPU while retaining CPU compatibility. All ported modules have been validated by comparing the output of all relevant variables to the original implementation and ensuring the normalized root mean square error to be smaller than \verb|1E-10| after 10 time steps in the coarsest time discretization, both on CPU and GPU.

\medskip

Figure \ref{fig:asuca-numbers} shows the impact this new implementation has on the code size, as well as the overall distribution of code lines in the application. Additionally, in \cite{MUELLER2017HYBRID} we have studied the productivity effects in more detail, which we describe here briefly: ASUCA's codebase encompasses more than 150k lines of code. By using our method, more than 85\% of the original codebase was left unchanged and the overall code size has been extended by less than 5\%, even though the new codebase targets two very different hardware architectures instead of one. This shows very promising productivity, maintainability and ease-of-adoption of the proposed method in an operational setting. Using an application model we have established a detailed estimate of the required code changes for a comparable OpenACC implementation. The result: OpenACC would require more than 73\% of additional changes compared to the changes required using Hybrid Fortran. We reckon that the same would be the case when using OpenMP directives targeting GPU, since to our knowledge it only offers a subset of OpenACC's GPU features to this date.

\medskip

Of the approximately of 163k lines of code, around 19.5\% are used for the dynamical core and 20.8\% are used for the physical processes, mainly to implement 133 and 205 kernels\footnote{We arrive at this number when counting the kernels from the point of view of the GPU - through kernel fusion this number is considerably lower for the CPU.}, respectively.

\medskip

By applying this unified programming model to both dynamical core and physical processes of ASUCA, host-to-device communication has been eliminated with the only exceptions being setup, halo communication as well as file output.

\subsection{Hybrid ASUCA Kernel Performance}\label{sec:performance-asuca-small}

\begin{figure}[htb]
  \centering
  \includegraphics[width=0.8\linewidth]{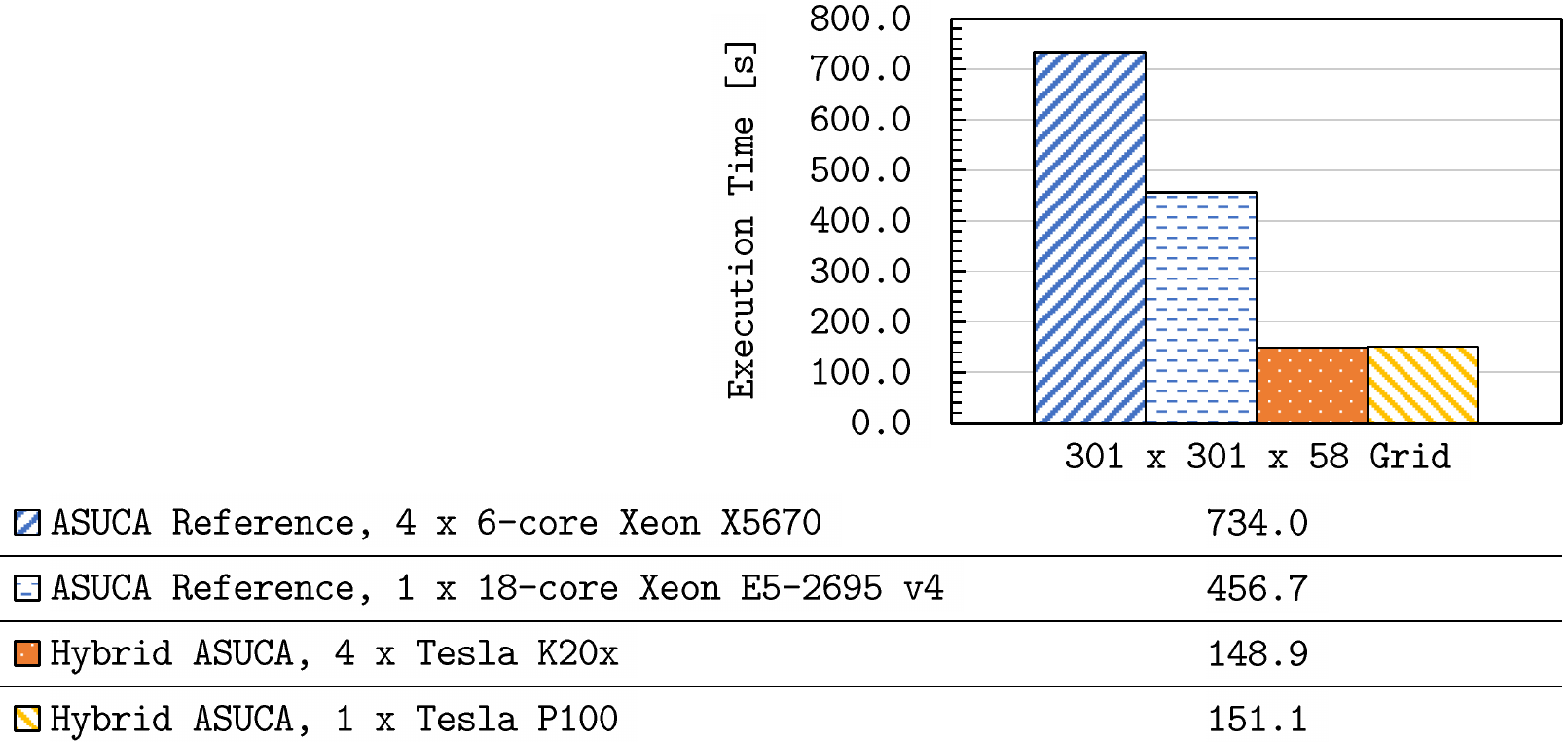}
  \caption{Execution time measurements for 75 long time steps of ASUCA executed in four different configurations}
  \label{fig:asuca-301x301-exectime}
\end{figure}

In this section, performance results for this new implementation are discussed for a 301 x 301 x 58 grid that is small enough for single GPU or single socket execution with the latest architecture (Tesla P100 on Reedbush-H), yet still allows a useful performance analysis in terms of occupancy (see Appendix \ref{sec:gpgpu-glossary}). This allows to draw conclusions for the kernel performance as opposed to the communication overhead (which impacts performance more strongly, the more nodes are used for the same grid size, i.e. when applying strong scaling as will be shown in Section \ref{sec:performance-asuca}). Additionally, an older system (TSUBAME 2.5 with Tesla K20x) is used for comparison purposes. Since the two compared GPUs differ in their available device memory (16 GB for P100 vs. 6 GB for K20x) we compare four K20x GPUs to one P100. See Appendix \ref{sec:config-small} for the software configuration used for this test. Hybrid ASUCA is also compared to previous GPGPU ports of weather models as discussed in Section \ref{sec:existing}.

% \medskip

% The problem size that can be executed on a single GPU is limited by the available device memory. Meanwhile, in order to achieve an optimal occupancy (see Appendix \ref{sec:gpgpu-glossary}), the problem size should be as large as possible. Since the two compared GPUs differ in their available device memory (16 GB for P100 vs. 6 GB for K20x) we compare four K20x GPUs to one P100.

\medskip

Figure \ref{fig:asuca-301x301-exectime} shows the execution times for this configuration on four hardware/software configurations as listed. A speedup of 4.9x has been achieved by the port on Kepler GPU vs. Westmere 6-core Xeon X5670. On newer hardware however (Pascal vs. Broadwell) the speedup has so far been a more modest 3.1x, which is partly explained by the lower memory bandwidth difference between the two comparisons and the increased caching performance on Broadwell- versus Westmere CPU architecture (as caching generally has a lower impact on GPU compared to CPU).

\FloatBarrier
\subsection{Hybrid ASUCA Strong Scaling GPU Results}\label{sec:performance-asuca}

\begin{figure}[htb]
  \centering
  \includegraphics[width=.7\linewidth]{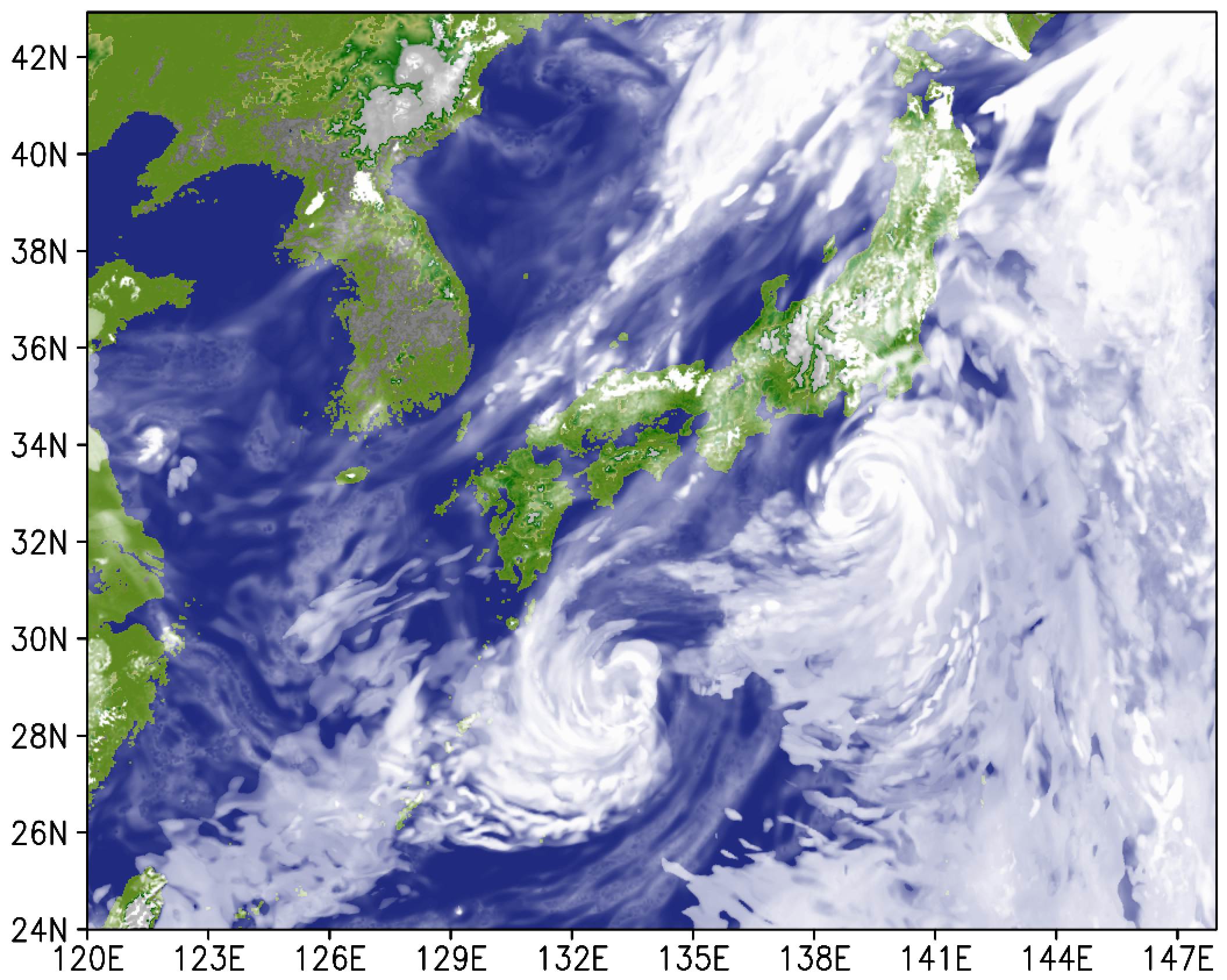}
  \caption{Total cloud cover result with ASUCA using a 2km resolution grid with a real world weather situation}
  \label{fig:asuca-cloud-cover}
\end{figure}

\begin{figure}[htb]
  \centering
  \includegraphics[width=0.6\linewidth]{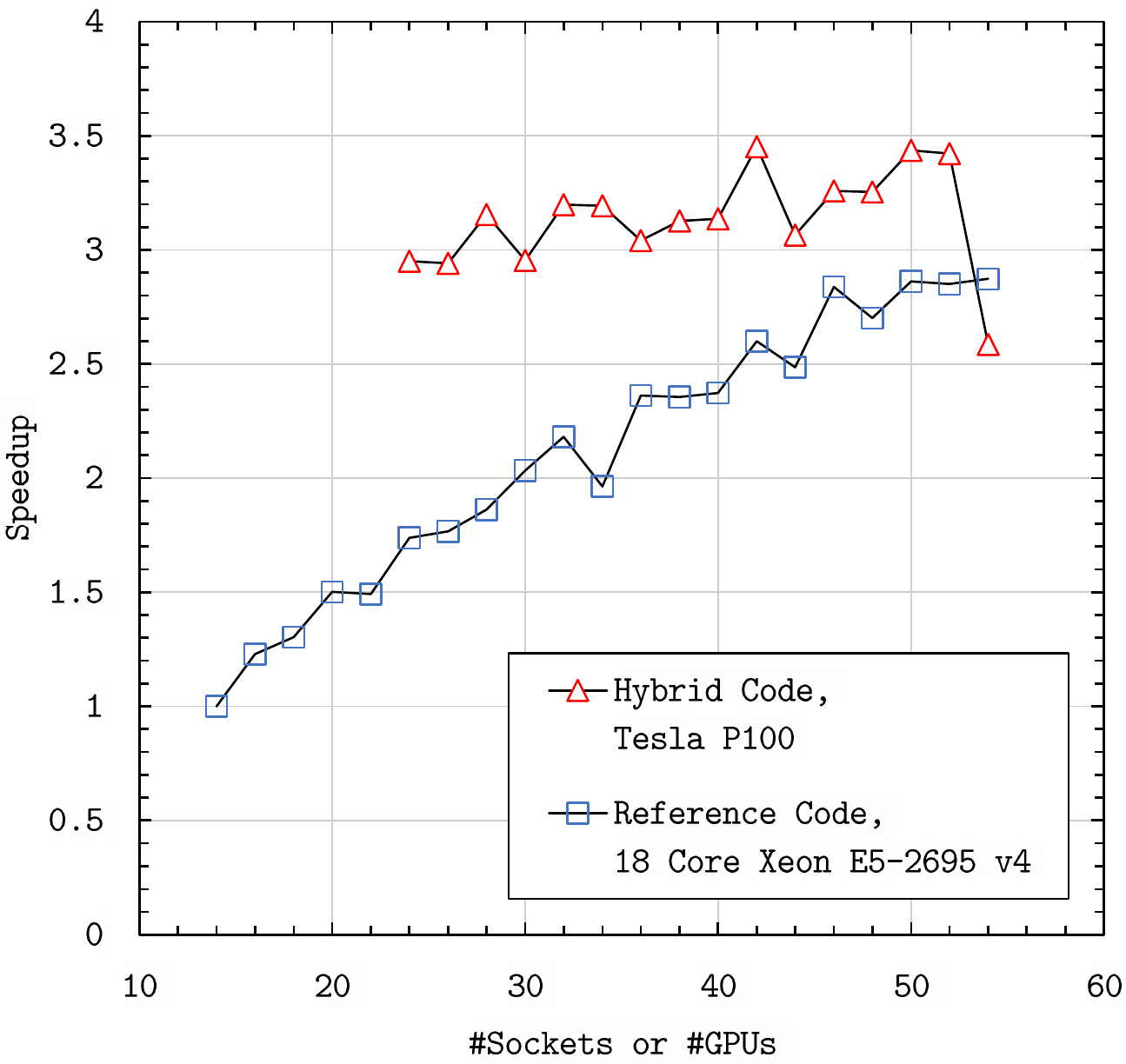}
  \caption{Strong scaling speedup on 1581 x 1301 x 58 ASUCA Grid.}
  \label{fig:asuca-speedup}
\end{figure}

Using a full scale production grid, Hybrid ASUCA has been tested on the new Tokyo University cluster ``Reedbush H'' \cite{reedbush}. This cluster has two 18-core Xeon E5-2695 v4 CPUs per node as well as two NVIDIA Tesla P100 GPUs per node. At least seven nodes (14 sockets) or 24 GPUs are required for this test due to the grid's memory requirements. A visualization of the resulting cloud cover from this simulation is depicted in Figure \ref{fig:asuca-cloud-cover}. 

\medskip

Figure \ref{fig:asuca-speedup} shows that 24 GPUs can replace more than 50 18-core CPU sockets. When comparing the same number of GPUs and CPU sockets, the GPU is up to 76\% faster. See Appendix \ref{sec:config-large} for the software configuration used in this test.

%a speedup of up to 3.5x on GPU has been achieved for this production run, on a 1581 x 1301 x 58 grid with 2km horizontal resolution using real world sample input data, when comparing the fastest time to solution on GPU with the minimum number of CPU sockets required to complete the simulation due to host memory requirements. For a given performance it shows that 24 GPUs can replace more than 50 18-core CPU sockets. When comparing the same number of GPUs and CPU sockets, the GPU is up to 76\% faster. See Appendix \ref{sec:config-large} for the software configuration used in this test.

\begin{figure}[htb]
  \centering
  \includegraphics[width=0.6\linewidth]{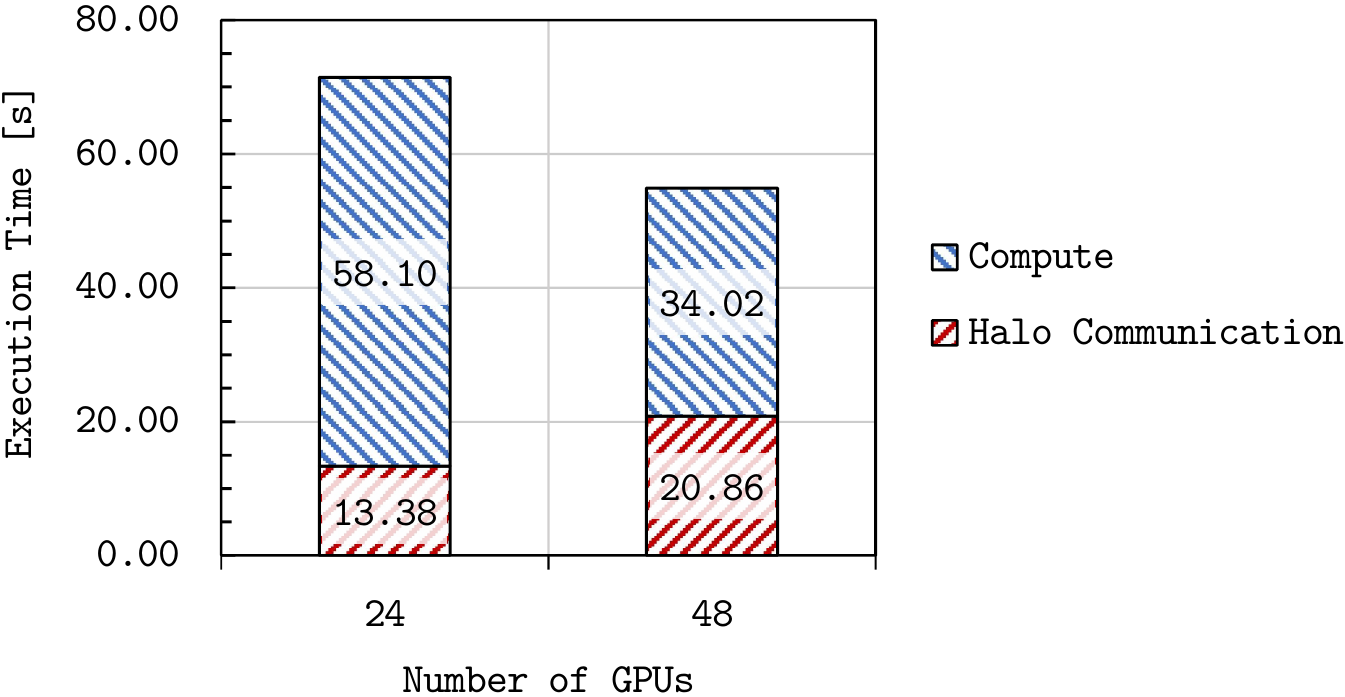}
  \caption{Impact of communication for strong scaling on 1581 x 1301 x 58 ASUCA Grid.}
  \label{fig:asuca-comm-impact}
\end{figure}

\medskip

The following factors influence scalability and will need to be improved in order to achieve better strong scaling:

\begin{enumerate}
\item MPI communications code has been used as-is, with no further optimizations applied with respect to the targeted cluster. When testing the impact of communication on performance on TSUBAME 3.0 using the minimally required 24 GPUs, as shown in Figure \ref{fig:asuca-comm-impact}, communication requires approximately 13.4s or 18.7\% of the overall runtime of 71.5s (to compute a 600 seconds simulation of the full regional grid in 2km resolution). When doubling the number of GPUs this increases to 20.9s while the compute time decreases from 58.1s to 34.0s, thus the communication then takes 38\% of the runtime. Overlapping communication and computations has been shown to be effective in enabling better scaling by Shimokawabe et al. \cite{shimokawabe201080}, thus this approach will be the first step to improve performance at larger scales.
%The communications overhead is larger for GPU than for CPU since the GPU also requires communication between host and device, in addition to the network.
%This can be improved by employing a communication/computation overlapping scheme (such as demonstrated by Shimokawabe et al. \cite{shimokawabe201080}) as well as ``GPUDirect'', a technology allowing direct network access from GPUs (which requires hardware support from the cluster).
\item Since GPUs require a large enough problem size per chip in order to have a sufficient number of threads to fill all schedulers, strong scaling is limited when the problem size per GPU becomes too small.
% \item The domain decomposition can be optimized for some of the results, as can be seen by some of the performance degradation with some of the different aspect ratios. GPU and CPU have different characteristics, which would require tuning of the decomposition for each of the reported number of processes and for each of the two architectures.
\item In our ASUCA code version, as in the given reference, we do not have a distributed file I/O system as used in production. Due to the large amount of data, even though for this test the output is only run at the beginning and end of the simulation, it still has a strong impact on the overall execution time, resulting in part of the discrepancy between for the speedups between the 301 x 301 and 1581 x 1301 grid sizes.
\end{enumerate}

\begin{figure}[htb]
  \centering
  \includegraphics[width=0.6\linewidth]{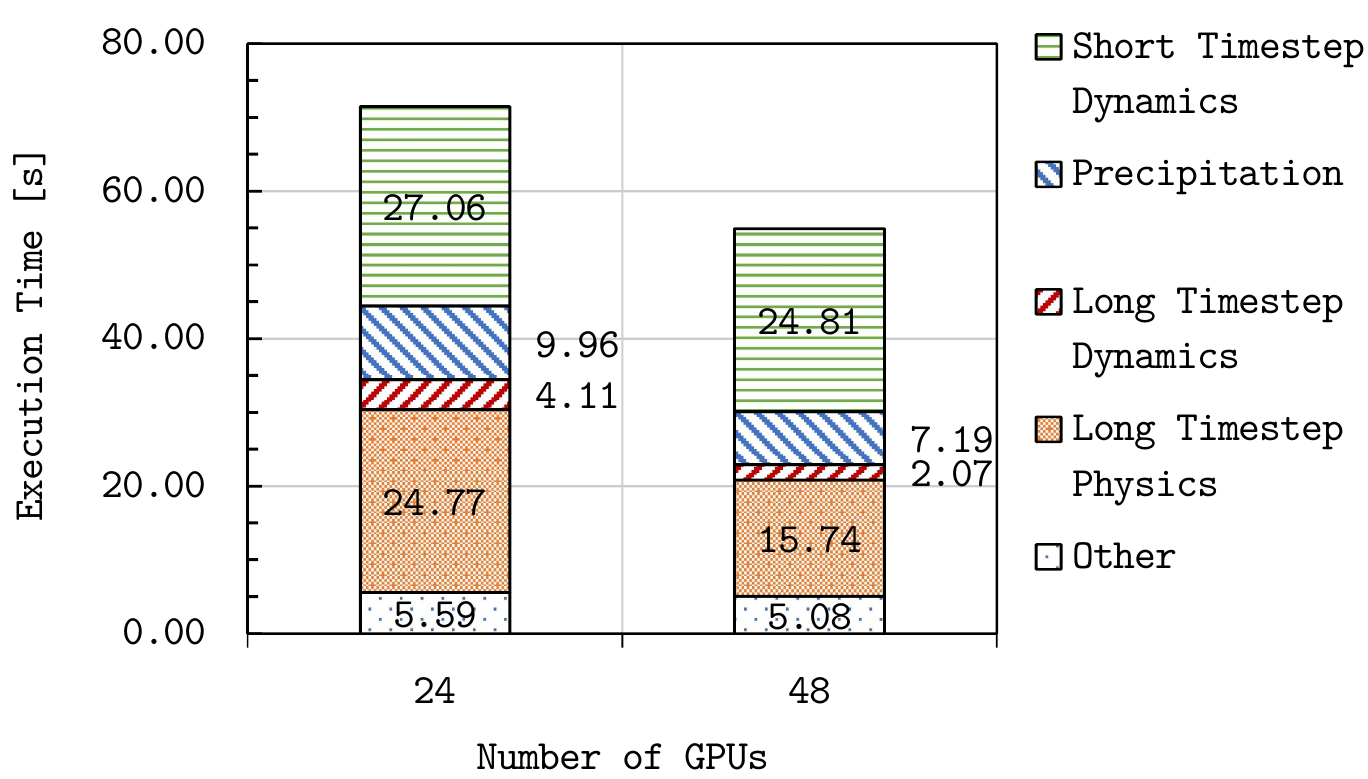}
  \caption{Impact of modules for strong scaling on 1581 x 1301 x 58 ASUCA Grid.}
  \label{fig:asuca-module-impact}
\end{figure}

Figure \ref{fig:asuca-module-impact} categorizes the performance impact of the different modules of ASUCA on performance, including communication. The simulation of fast moving sound- and gravity waves has the highest impact, followed by radiation- and boundary layer physics. Since the physics calculations do not require communication, the impact of fast moving dynamics increases with the number of nodes, rendering it the most important optimization target for larger scale simulations. For a listing of the configurations used to gather the data for figures \ref{fig:asuca-comm-impact} and \ref{fig:asuca-module-impact}, please refer to Appendix \ref{sec:config-large-tsubame3}.

\FloatBarrier
\subsection{Comparison to Previous Work} \label{phd:cha4:sec:shimokawabe-comparison}

In Section \ref{sec:existing-shimokawabe}, Dr. Shimokawabe's work on applying GPUs to ASUCA was briefly discussed. In this section we compare that implementation to Hybrid ASUCA.

\begin{table}[htpb]
  \centering
  \footnotesize
  \caption{Comparison of Hybrid ASUCA to GPU implementation by Shimokawabe et al.}
  \label{table:shimokawabe-comparison}
	\begin{tabular}{|c|c|c|}
	\hline
	Characteristic & \textbf{Hybrid ASUCA} & \textbf{Shimokawabe}\tabularnewline
	\hline
	\textbf{Main focus} & Productivity & Efficiency at large scale \tabularnewline
	& &  (\# of GPUs) \tabularnewline
	\hline
	\textbf{User language} & Fortran & C++ \tabularnewline
	\hline
	\textbf{Parallelization} & DSL & DSL \tabularnewline
	\hline
	\textbf{Memory layout} & Transformation & DSL \tabularnewline
	\hline
	\textbf{Dynamical core} & Port of Ishida et al. 2010 \cite{ishida2010development}  & Port of Ishida et al. 2010 \cite{ishida2010development} \tabularnewline
	\hline
	\textbf{Cloud microphysics} & Warm rain (Kessler-type) & Warm rain (Kessler-type) \tabularnewline
	\hline
	\textbf{Radiation} & MSM0705 based  & n/a \tabularnewline
	\hline
	\textbf{Boundary layer physics} & MSM0705 based  & n/a \tabularnewline
	\hline
	\textbf{Strong scaling} & 42 P100 & 512 K20x \tabularnewline
	\textbf{(\# of GPUs)} & (equiv. to 159 K20x) &  \tabularnewline
	\hline
	\textbf{Strong scaling} & n/a & 4108 K20x \tabularnewline
	\textbf{(\# of GPUs)} & & \tabularnewline
	\hline
	\end{tabular}
\end{table}

\medskip

Table \ref{table:shimokawabe-comparison} shows the differences in focus between Hybrid ASUCA and Shimokawabe et al. \cite{shimokawabe:2014}. Dr. Shimokawabe's work was targeted at achieving computations on the largest scale available, i.e. a near-full utilization of TSUBAME 2.5 for a weak scaling result as well as using 512 GPUs for strong scaling. For comparison, Hybrid ASUCA was only scaled using strong scaling, reaching a performance peak at 42 GPUs, using a device model that is equivalent in theoretical performance to 159 GPUs used by Shimokawabe (P100 vs. K20x). On the other hand, with Hybrid ASUCA we were first able to port a complete model including operation-ready physical processes using a unified framework and we have demonstrated that only a minor rewrite is necessary for a previously CPU-targeted code to achieve full GPU compatibility (while Shimokawabe et al. relied on a full rewrite using a different user language and did not demonstrate radiation- as well as boundary layer physics).

\FloatBarrier
\section{Conclusions}\label{sec:conclusions}

In this work, ``Hybrid Fortran'' has been introduced as a new approach to port structured grid Fortran applications to GPU. It provides abstractions over architecture specific programming while allowing existing CPU targeted user code to be reused as much as possible (see also Section \ref{sec:hf-user-centric}). This approach allows using a mixture of OpenMP, OpenACC, and CUDA Fortran in the backend with a unified programming interface (see also Section \ref{sec:hf-implementation-centric}). Storage order and parallelization granularity are made variable. Existing CPU targeted user code is transformed to match the architecture specific granularity and storage order.

\medskip

A new implementation of ASUCA, the main mesoscale weather model used in weather prediction operations by the national Japanese weather agency, has been created by applying Hybrid Fortran to both dynamical core and physical processes. The resulting speedup on GPU (up to 4.9x on single GPU compared to a single Xeon socket) is comparable to architecture specific rewrites of similar weather models (see also Section \ref{sec:performance-asuca-small}). To our knowledge this is the first time a complete production weather prediction model has been ported to GPU using a single unified programming paradigm. The resulting user code is less than 5\% larger compared to the reference CPU code, which is only made possible by sharing all of the user code between the two architectures (see also Figure \ref{fig:asuca-numbers}). A speedup of up to 3.5x on GPU was achieved for a full scale production run (1581 x 1301 x 58 grid with 2km horizontal resolution using real world sample input data, see also Section \ref{sec:performance-asuca}).

% In order to facilitate the above, Hybrid Fortran implements the design goals discussed in the introduction of Section \ref{sec:hybridFortran}:

% \begin{enumerate}
% \item Parallel loops are replaced with \verb|@parallelRegion| directives that allow to be applied partially by target architecture. This results in a compile-time defined parallelization granularity (design goal \ref{enum:multi_parallelization}) by giving the user a facility to express the fusion of kernels depending the architecture.
% \item Stencil code can be kept in the same number of dimensions as used in CPU-only code, thus inserting privatizations for the parallel domains where needed. This results in a compile-time defined privatization (design goals \ref{enum:privatization} and \ref{enum:levels}). This is especially relevant for the physical processes that are, in ASUCA's case, almost exclusively programmed as column based operations, while the GPU requires these to be expressed as 3D grid operations.
% \item Storage order is abstracted and defined using a central configuration per codebase. This allows an architecture-dependant storage order (design goals \ref{enum:reorder} and \ref{enum:reorder_centralised}).
% \end{enumerate}

\FloatBarrier
\section{Future Work}\label{sec:future}

Future efforts will focus on three aspects: performance, productivity and scope.

\medskip

In order to improve the performance on GPGPU, the following avenues will be explored:

\begin{enumerate}
\item Communication will be optimized in order to improve larger scale runs on more than 30 nodes. Network topology based optimization, as well as the overlapping of communication and computation will be the main focus for these optimizations.
\item GPU kernels will be more finely tuned as far as a unified user code allows. One possible avenue is the usage of launch bounds, a CUDA feature that supplies better hints to the compiler with regard to register usage (however it requires re-compilations for varying grid sizes).
\item Automated cache blocking transformations that are applicable to GPU (shared memory, L1, texture cache) are an important path to explore. Such transformations also have the potential to auto-tune the code for different CPU architectures.
\end{enumerate}

\medskip

Further productivity gains are possible by automating more of the mechanical aspects of porting to Hybrid Fortran. More specifically, replacing the device data state attributes per routine and per data object with a generalized data region facility (and keeping track of all the involved data objects automatically) will greatly reduce the work needed to port to Hybrid Fortran. This in turn can make Hybrid Fortran viable for a GPGPU port of the WRF model (which is one order of magnitude larger in code size compared to ASUCA).

\medskip

In terms of scope, applying Hybrid Fortran to unstructured grids will be explored.

% As related research, the weather model ASUCA will be ported using a complete Hybrid Fortran implementation. Performance results for large scale runs on the TSUBAME supercomputer will be published. This will extend the work already done by Shimokawabe et al. in relation to GPU portations of ASUCA (based on CUDA C)\cite{Shimokawabe2010}\cite{Shimokawabe2011}. In order to further simplify porting large applications to GPU, \verb|@domainDependant| directives will be made optional by using static data flow analysis. F2C-ACC\cite{govett:f2c-acc} and OpenMP4 accelerator\cite{openmp4} implementations will be made available for Hybrid Fortran in order to have more performance comparisons. Optimizations in relation to GPU occupancy are planned by applying automatic kernel fission and fusion using a scheme developed by Wahib et al.\cite{wahib}

% Appendix
\newpage
\appendix
\section{Hybrid Fortran Language Extension} \label{sub:hf-documentation}

This appendix section gives an overview of the main language extensions provided by Hybrid Fortran. For a complete listing, refer to the documentation in the repository\footnote{Hybrid Fortran master: \url{https://github.com/muellermichel/Hybrid-Fortran}}.

\subsection{Parallel Region Construct} \label{sub:parallelRegionDirective}

Listing \ref{listing:parallelRegion} shows the parallel region construct. It allows the framework to define parallelization as well as the thread granularity at compile-time.

\begin{lstlisting}[name=parallelRegion, label=listing:parallelRegion, caption={Parallel region directive syntax.}]
@parallelRegion{ATTRIBUTE_NAME1(MEMBER1, MEMBER2, ...), ...}
! code to be executed in parallel !
@end parallelRegion
\end{lstlisting}

The following attributes are supported for this directive:

\begin{description}
 \item [appliesTo] Specify one or more of the following attribute members in order to set this parallel region to apply to either the CPU code version, the GPU version or both. Omitting this attribute has the same effect as specifying all supported architectures.
  \begin{enumerate}
   \item CPU
   \item GPU
  \end{enumerate}
 \item [domName] (Required) Specify one or more domain names over which the code can be executed in parallel. These domain names are being used as iterator names for the respective loops or CUDA kernels.
 \item [domSize] (Required) Set of the domain dimensions in the same order as their respective domain names specified using the \verb|domName| attribute. It is required that $|domName| = |domSize|$.
 \item [startAt] The lower boundary for each domain at which to start computations. Omitting this attribute will set all boundaries to \verb|1|. It is required that $|startAt| = |domName|$.
 \item [endAt] The upper boundary for each domain at which to end computations. Omitting this attribute will set all boundaries to \verb|domSize| for each domain. It is required that $|startAt| = |domName|$.
 \item [reduction] Works in the same way as OpenMP reduction directives. This is only supported with the OpenACC- and OpenMP backends. For example \verb|reduction(+:result)| sums up the results over all threads.
 \item [template] Defines a postfix that is to be applied to different attributes that are loaded using the preprocessor. Currently this only affects CUDA Blocksizes: These are loaded using \verb|CUDA_BLOCKSIZE_X_[Template-Name]| from the preprocessor (\verb|storage_order.F90| is the suggested place to define these). The goal of this attribute is to hoist hardware dependent attributes outside of the code in order to more easily facilitate performance portability.
\end{description}

\subsection{Domain Dependant Construct} \label{sub:domainDependantDirective}

Listing \ref{listing:domainDependant} shows the domain dependant construct. It is used to give hints to the framework how data objects are to be adjusted in the transformation process. This allows passing on the responsibility of rewriting certain aspects of data specifications and accesses to the framework. It should be noted that the framework only operates on local information available to each subroutine. As an example, whether a data object has already been copied to the GPU is not being analyzed. Domain Dependant constructs are required to be specified between the specification and the implementation part of a Fortran subroutine.

\begin{lstlisting}[name=domainDependant, label=listing:domainDependant, caption={Domain dependant construct syntax.}]
@domainDependant{ATTRIBUTE_NAME1(MEMBER1, MEMBER2, ...), ...}
! data objects that share the attributes !
! defined above to be defined here, separated  !
! by commas                                    !
...
@end domainDependant
\end{lstlisting}

The following attributes are supported for this construct:

\begin{description}
  \item [domName] Set of all domain names in which the data object needs to be privatized. This is required to be a superset of the domains that are being declared as the data object's dimensions in the specification part of the subroutine (except if the \verb|autoDom| attribute flag is used, see below). More specifically, the domain names specified here must be the set of domains from the specification part plus the parallel domains (as specified using the parallel region construct, see section \ref{sub:parallelRegionDirective}) for which privatization is needed.
% $domName_{Domain Dependant}|_{subroutine level} = domName_{Parallel Region}|_$.
 \item [domSize] Set of the domain dimensions in the same order as their respective domain names specified using the \verb|domName| attribute. It is required that $|domName| = |domSize|$.
 \item [accPP \label{descr:accPP}] Preprocessor macro name that takes $|domSize|$ arguments and outputs them comma separated in the current storage order for data object accesses. This macro must be defined in the file \verb|storage_order.F90|. In case the \verb|autoDom| attribute is being used, the \verb|accPP| specification is not necessary - \verb|AT|, \verb|AT4|, \verb|AT5| (...) are assumed as defined storage order macro names, depending on the number of array dimensions.
 \item [domPP \label{descr:domPP} ] Preprocessor macro name that takes $|domSize|$ arguments and outputs them comma separated in the current storage order for the data object declaration. This macro must be defined in the file \verb|storage_order.F90|. In case the \verb|autoDom| attribute is being used, the \verb|domPP| specification is not necessary - \verb|DOM|, \verb|DOM4|, \verb|DOM5| (...) are assumed as defined storage order macro names, depending on the number of array dimensions. This preprocessor macro is usually identical to the one defined in \verb|accPP|.
 \item [attribute] Attribute flags for these data objects. Currently the following flags are supported:
  \begin{description}
   \item [present] In case this flag is specified, the framework assumes array data to be already present on the device memory for GPU compilation and the data will not be transferred.
   \item [transferHere] In case this flag is specified, all domain dependants with \verb|intent| specified as \verb|in|, \verb|inout| or \verb|out| will be transferred to- and from the device (according to the intent). This flag may not be specified together with the \verb|present| flag as it has exactly opposite effects.
   \item [autoDom] In case this flag is specified, the framework will use the array dimensions that have been declared using standard Fortran syntax to determine the domains for each data object. In case the parallel domains are omitted from the Fortran specification (in order to allow different parallelization for CPU and GPU), they still need to be specified using \verb|domName| and \verb|domSize| for data objects that are to be privatized for each thread. The parallel domains will be inserted before any independent domains picked up through the declaration, depending on the subroutine's position towards the parallel region. In addition, using \verb|autoDom| will by default enable standard \verb|accPP| and \verb|domPP| settings, if not specified otherwise. Using this flag then greatly simplifies the \verb|@domainDependant| specification part, since the construct can be reused by data objects of different domains.
   \item [host] In subroutines or modules that are not related to a parallel region in their call graph, Hybrid Fortran would not know whether a data object resides on the host or the device. This attribute is thus used to specify this property unambiguously.
  \end{description}
\end{description}

It is important to consider that in case neither \verb|present| nor \verb|transferHere| flags are used, Hybrid Fortran will automatically transfer all domain dependants with their Fortran \verb|intent| specified (as \verb|in|, \verb|inout| or \verb|out|) from and/or to the device according to their intent in case of a routine containing a GPU parallel region.

\section{Hardware Performance} \label{sec:hardware-metrics}

\begin{table}[htpb]
  \centering
  \footnotesize
  \caption{Hardware performance numbers referred to in this work}
  \label{table:tsubamePerf}
	\begin{tabular}{|c|c|c|c|c|c|}
	\hline
	Term & Hardware & Sustained  & Unit & System & Comment\tabularnewline
	 &  & Performance & & & \tabularnewline
	\hline
	$P_H1C$ & Xeon X5670  & 9.50 & GFLOP/s & TSUBAME 2.5 & In double precision \cite{Xeon5670}\tabularnewline
	 & (1 Core) &  & & & \tabularnewline
	\hline
	$P_H$ & Xeon X5670  & 57 & GFLOP/s & TSUBAME 2.5 & In double precision \cite{Xeon5670} \tabularnewline
	 & (1 Socket) &  & & & \tabularnewline
	\hline
	$P_D$ & NVIDIA & 1030 & GFLOP/s & TSUBAME 2.5 & Linpack DP perf. \cite{k20linpack} \tabularnewline
	 & Tesla K20x &  & & & \tabularnewline
	\hline
	$P_D$ & NVIDIA & 3900 & GFLOP/s & Reedbush-H, & Linpack DP perf. \cite{p100linpack} \tabularnewline
	 & Tesla P100 & & & Piz Daint & \tabularnewline
	\hline
	$BW_{H1C}$ & Xeon X5670  & 9.80 & GB/s & TSUBAME 2.5 & according to Intel specifications\cite{Xeon5670} \tabularnewline
	 & (1 Core) &  & & & \tabularnewline
	\hline
	$BW_{H}$ & Xeon X5670  & 20.5 & GB/s & TSUBAME 2.5 & according to Intel specifications \cite{Xeon5670}, \tabularnewline
	 & (1 Socket) &  & & & confirmed by STREAM benchmark\tabularnewline
	\hline
	$BW_{H}$ & Xeon E5-2670 & 51.2 & GB/s & Piz Daint & according to Intel specifications\cite{XeonE5-2670} \tabularnewline
	 & (1 Socket) & & & & \tabularnewline
	\hline
	$BW_{H}$ & Xeon E5-2695 v4 & 76.8 & GB/s & Reedbush-H & according to Intel specifications \cite{XeonE5-2695v4}, \tabularnewline
	 & (1 Socket) &  & & & confirmed by STREAM benchmark\tabularnewline
	\hline
	$BW_{D}$ & NVIDIA  & 108.6 & GB/s & TSUBAME 2.0 & Reported by CUDA bandwidth test \tabularnewline
	 & Tesla M2050 &  & & & \tabularnewline
	\hline
	$BW_{D}$ & NVIDIA  & 169.4 & GB/s & TSUBAME 2.5, & Reported by CUDA bandwidth test \tabularnewline
	 & Tesla K20x & & & Piz Daint, & \tabularnewline
	\hline
	$BW_{D}$ & NVIDIA & 499.4 & GB/s & Reedbush-H, & Reported by CUDA bandwidth test \tabularnewline
	 & Tesla P100 &  & & Piz Daint, & on Reedbush-H \tabularnewline
	 & & & & TSUBAME 3.0 & \tabularnewline
	\hline
	$BW_{HtoD}$ & PCI Express 2.x & 4.32 & GB/s & TSUBAME 2.5 & Reported by CUDA bandwidth test \tabularnewline
	\hline
	$BW_{HtoD}$ & PCI Express 3.x & 10.96 & GB/s & Reedbush-H, & Reported by CUDA bandwidth test \tabularnewline
	& & & & Piz Daint & on Reedbush-H \tabularnewline
	\hline
	$RA_{H}$ & Xeon X5670  & 0.12 & GUP/s & TSUBAME 2.5 & Reported by \tabularnewline
	 & (1 Socket) &  & & & RandomAccess benchmark \tabularnewline
	\hline
	$RA_{D}$ & NVIDIA  & 0.88 & GUP/s & TSUBAME 2.5 & Reported by \tabularnewline
	 & Tesla K20x &  & & & RandomAccess benchmark \tabularnewline
	\hline
	\end{tabular}
\end{table}

Table \ref{table:tsubamePerf} lists the relevant performance metrics that are used for the performance models and efficiency discussions in this work. For a definition of the terms, see Section \ref{sec:perfAnalysis}. For the random access memory performance, the test included with the HPC Challenge\cite{luszczek2006hpc} benchmark suite has been used (employing the unit ``Giga Updates Per Second''). For the GPU, a CUDA port\footnote{\url{https://github.com/muellermichel/cuda_randomaccess}} has been used while for the CPU test an open source version by Underwood et al. has been employed\cite{underwood2006simple} (using $2^{20}$ table size and $1000$ update sets). The GPU bandwidth is specified as reported by the NVIDIA provided CUDA bandwidth test.

\section{GPGPU Specific Programming Terms} \label{sec:gpgpu-glossary}

The following glossary gives an overview for terms commonly used for GPGPU programming with modern NVIDIA architectures.

\begin{itemize}[leftmargin=25mm]
    \item [Block size] Size of a block of threads that is fetched and loaded onto one [SMX] simultaneously. Can be defined in 1D, 2D or 3D. The first dimension should usually chosen as a multiple of the warp size in single precision or as a multiple of a half warp size in double precision. Each thread in a block requires register-  and, if used, shared memory resources. Thus, there is a trade-off between the block size and optimizing for the required number of registers per thread. If too many registers are used, spilling occurs in which data is offloaded to the device memory.
    \item [CUDA] NVIDIA's proprietary GPGPU programming language, available for C, C++ and Fortran.
    \item [CUDA core] NVIDIA's marketing term for a combination of general purpose ALU and FPU on the GPU, each able to operate on one CUDA thread.
    \item [CUDA Fortran] A Fortran interface to the CUDA tool chain, supported by the Portland Group (PGI).
    \item [Device code] Code that exclusively runs on the GPU. Two types of device code exist: Kernel routines and device routines. The latter type is required for routines called within kernels. CUDA requires the programmer to explicitly set the type of each routine.
    \item [Grid size] Set of blocks that is loaded onto the GPU simultaneously. See also [Block size].
    \item [Host code] Code that exclusively runs on the CPU.
    \item [Launch bounds] Compile-time provided upper bounds for the problem size per block - this allows the CUDA compiler to be more conservative with the assigned resources per thread and thus optimize for a higher occupancy. This however requires the application to be recompiled for each new problem size to be executed.
    \item [Occupancy] The percentage of CUDA cores that are occupied on average during execution. Occupancy can get lowered to to a number of issues, such as device memory stalls, high register pressure, high shared memory pressure, misconfigured block size configurations or problem sizes that are too small for the architecture. GPU specific optimizations are usually targeted at increasing the occupancy since this directly affects the runtime.
    \item [OpenACC] An industry standard for directive based GPGPU computing, available for C, C++ and Fortran. Notable implementations are available from Cray and Portland Group (PGI).
    \item [Cache] Each streaming multiprocessor (see [SM / SMX]) offers the following caches: Shared Memory, texture memory and L1 cache. Shared memory is used through explicitly programmed memory operations, texture memory is used in a declarative way and L1 cache operates automatically for device memory accesses, analogous to CPU cache architectures. Additionally an L2 cache is shared among all multiprocessors on one GPGPU.
    \item [SIMT]\label{enum:simt} Single instruction, multiple threads execution model. An extension of the traditional Single instruction, multiple data (SIMD) model, in which branching within kernels is possible (with a loss of performance however, see also [Warp]).
    \item [SM / SMX] Streaming multiprocessor - NVIDIA's marketing term for a set of CUDA cores that share certain resources, such as an L1 and shared memory cache as well as a register file.
    \item [Thread] One thread in the [SIMT] model. In GPU hardware such threads do not operate independently however - see also [Warp]. Each thread occupies one CUDA core during execution.
    \item [Warp] A set of threads that are executed using a single scheduler. Threads within a warp operate in lockstep on either the same operation or \verb|nop| (in case of branching). In current and recent architectures the warp size has been 32. For double precision, only each second thread per warp is executed (i.e. the data length per CUDA core and warp scheduler is fixed).
\end{itemize}

\section{Reduced Weather Configuration for Performance Tests} \label{sec:config-reduced}

The following configuration has been used for the results discussed in Section \ref{sec:performance-sample}:

\begin{enumerate}
\item TSUBAME 2.5 has been used, i.e. Tesla K20x GPUs and Xeon X5670 CPUs.
\item For the CPU performance measurements, \verb|ifort| and \verb|icc| with \verb|-fast| option has been used.
\item Thread affinity has been set to \verb|compact| in order to ensure that the comparison is being done to exactly one of the two CPU socket.
\item For the GPU performance measurements, PGI Accelerator version 15.04 with \verb|-fast| option and compute capability 3.x has been used.
\item GPU results include memory copy time from and to device unless stated otherwise.
\item Hybrid Fortran uses the OpenACC based backend implementation.
\item Hybrid Fortran uses the L1-cache-preferred setting for GPUs.
% \item All measurements have been repeated at least three times, low variance has been checked and the reported measurement the mean over all measurements.
\item The number of time steps \verb|NT| has been set to 100. Output to disk is only performed once at the end of the simulation.
% \item The number of vertical levels \verb|NK| has been set to 70.
\item All code measured has been open-sourced in the Hybrid Fortran GitHub repository\footnote{Hybrid Fortran: \url{https://github.com/muellermichel/Hybrid-Fortran/tree/0.95}}.
\end{enumerate}

\section{ASUCA Configuration for Tests with 301 x 301 x 58 Grid} \label{sec:config-small}

The following configuration has been used for the results discussed in Section \ref{sec:performance-asuca-small}:

\begin{enumerate}
    \item 75 time steps
    \item All ported modules enabled as depicted in Figure \ref{fig:asuca-structure}
    \item Real model input data
    \item All floating point computations done in double precision
    \item Intel Fortran compiler with \verb|-fast -no-ipo| settings has been used for the reference code running on CPU. Versions 14.0.2 and 17.0.2 have been used for Westmere Xeon X5670 and Broadwell Xeon E5-2695 v4 respectively (inter-procedural optimization has resulted in compiler errors).
    \item PGI accelerator has been used for the GPU measurements, with optimizations disabled (\verb|-O0|) due to accuracy problems; Versions 16.9 and 16.10 have been used for K20x and P100 respectively.
\end{enumerate}

\section{ASUCA Configuration for Tests with 1581 x 1301 x 58 Grid on Reedbush-H} \label{sec:config-large}

The following configuration has been used for the results discussed in Section \ref{sec:performance-asuca}:

\begin{enumerate}
\item All ported modules have been enabled as depicted in Figure \ref{fig:asuca-structure}.
\item 75 time steps have been executed.
\item OpenMPI version 1.10.7 has been used.
\item The GPU implementation uses two MPI processes per node.
\item The CPU implementation uses one MPI process per node, with OpenMP core affinity settings to keep the threads in a compact manner on each socket (in order to optimize cache locality).
\item The horizontal domain (\verb|IJ|) has been decomposed according to Appendix \ref{sec:decomposition}.
\item Output to file has been limited to the beginning and the end of the simulation.
\item A single MPI process has been used only for problem setup and output on a separate node. This process is not included in the number of processes shown in the x-axis of the result as well as Appendix \ref{sec:decomposition}.
\item Setup time is not included in the measured computation time and thus the speedup shown in Figure \ref{fig:asuca-speedup}.
\item For the CPU implementation, Intel compiler version 17.0.2 has been used with \verb|-fast| and \verb|-no-ipo| flags (inter-procedural optimization has resulted in compiler errors).
\item For the GPU implementation, PGI Accelerator version 16.10 has been used with compiler optimizations disabled (an unresolved issue with incorrect computations has been identified in conjunction with compiler optimizations).
\end{enumerate}

\section{ASUCA Configuration for Tests with 1581 x 1301 x 58 Grid on TSUBAME 3.0} \label{sec:config-large-tsubame3}

The following configuration has been used for the module- and communication impact results discussed in Section \ref{sec:performance-asuca}:

\begin{enumerate}
\item All ported modules have been enabled as depicted in Figure \ref{fig:asuca-structure}.
\item 36 time steps have been executed, with $\Delta T = 16.\overline{6}s$, thus having a simulation of 10 minutes physical time.
\item OpenMPI version 2.1.2 has been used.
\item The GPU implementation uses two MPI processes (i.e. GPUs) per node.
\item The horizontal domain (\verb|IJ|) has been decomposed according to Appendix \ref{sec:decomposition}.
\item A single MPI process has been used only for problem setup and output on a separate node. This process is not included in the number of processes shown in the x-axis of the result as well as Appendix \ref{sec:decomposition}.
\item Both setup- and file output time is not included in the measured computation time and thus the speedup shown in the figures.
\item PGI Accelerator version 17.10 has been used with -O2 optimizations.
\item Memory error correction (ECC) was enabled on GPU.
\end{enumerate}

\section{ASUCA IJ domain decomposition} \label{sec:decomposition}

\FloatBarrier
\begin{table*}[ht]
  \centering
  \footnotesize
  \caption{MPI IxJ domain decomposition by number of processes}
  \label{table:asuca-decomposition}
	\begin{tabular}{|c|c|}
	\hline
	\# processes & \verb|IxJ| decomposition \tabularnewline
	& (CPU and GPU) \tabularnewline
	\hline
    7 & 1 x 7\tabularnewline
    8 & 2 x 4\tabularnewline
    9 & 1 x 9\tabularnewline
    10 & 2 x 5\tabularnewline
    11 & 1 x 11\tabularnewline
    12 & 3 x 4\tabularnewline
    13 & 1 x 13\tabularnewline
    14 & 7 x 2\tabularnewline
    15 & 3 x 5\tabularnewline
    16 & 4 x 4\tabularnewline
    17 & 1 x 17\tabularnewline
    18 & 3 x 6\tabularnewline
    19 & 1 x 19\tabularnewline
    20 & 2 x 10\tabularnewline
    21 & 3 x 7\tabularnewline
    22 & 2 x 11\tabularnewline
    23 & 23 x 1\tabularnewline
    24 & 2 x 12\tabularnewline
    25 & 5 x 5\tabularnewline
    26 & 2 x 13\tabularnewline
    27 & 3 x 9\tabularnewline
	\hline
	\end{tabular}
	\quad
	\begin{tabular}{|c|c|}
	\hline
	\# processes & \verb|IxJ| decomposition \tabularnewline
	& (GPU only) \tabularnewline
	\hline
    28 & 2 x 14\tabularnewline
    30 & 3 x 10\tabularnewline
    32 & 2 x 16\tabularnewline
    34 & 2 x 17\tabularnewline
    36 & 3 x 12\tabularnewline
    38 & 2 x 19\tabularnewline
    40 & 4 x 10\tabularnewline
    42 & 2 x 21\tabularnewline
    44 & 4 x 11\tabularnewline
    46 & 2 x 23\tabularnewline
    48 & 4 x 12\tabularnewline
    50 & 2 x 25\tabularnewline
    52 & 4 x 13\tabularnewline
	\hline
	\end{tabular}
\end{table*}
\FloatBarrier

\section{Sources} \label{sec:sources}

This section presents the sources for the reduced weather application\footnote{The complete source code for the reduced weather application can also be compiled and run from the Hybrid Fortran GitHub repository: \url{https://github.com/muellermichel/Hybrid-Fortran/tree/0.95/examples/simple_weather}}, as well as an extract from ASUCA's source code, as discussed in this work. Please note that ASUCA is a closed source application and can therefore not be presented in full.

\subsection{Reduced Weather: Diffusion} \label{sec:sources-diffusion}

%--------------------------------------------------!
%                                       ---------->!
%                                      max length  !
\begin{lstlisting}[name=diffuse, label=listing:diffuse, caption={Diffusion.}]
	subroutine diffuse( &
		& energy_u, energy)
    ! ..more specifications..
    real(8),intent(out):: energy_u(0:nx+1,0:ny+1,nz)
    real(8),intent(in):: energy(0:nx+1,0:ny+1,nz)

    ! ------- inner region IJK: ------------
    do j = 1,ny
  	do i = 1,nx
    do k = 2, nz-1
      energy_u(i,j,k) = energy(i,j,k) * &
    & (1.0d0 - 6 * diffusion_velocity) &
    & + diffusion_velocity * ( &
    &   energy(i-1,j,k) + energy(i+1,j,k) + &
    &   energy(i,j-1,k) + energy(i,j+1,k) + &
    &   energy(i,j,k-1) + energy(i,j,k+1) &
    & )
    end do
    end do
    end do

    ! -- boundary regions IJ, IK, JK: -----
    do j = 1,ny
  	do i = 1,nx
      energy_updated(i,j,1) = (1.0d0 - 5 * diffusion_velocity) &
    & * energy(i,j,1) + diffusion_velocity * ( &
    &   energy(i-1,j,1) + energy(i+1,j,1) + &
    &   energy(i,j-1,1) + energy(i,j+1,1) + &
    &   energy(i,j,2) &
    & )
      energy_updated(i,j,nz) = (1.0d0 - 5 * diffusion_velocity) * &
    &    energy(i,j,nz) + diffusion_velocity * ( &
    &    energy(i-1,j,nz) + energy(i+1,j,nz) + &
    &    energy(i,j-1,nz) + energy(i,j+1,nz) + &
    &    energy(i,j,nz-1) &
    & )
    end do
    end do

  	do i = 0,nx+1
  	do k = 1,nz
      energy_updated(i,0,k) = (1.0d0 - 2 * diffusion_velocity) &
    & * energy(i,0,k) + diffusion_velocity * ( &
    &    energy(i,1,k) + energy(i,ny+1,k) &
    & )
      energy_updated(i,ny+1,k) = (1.0d0 - 2 * diffusion_velocity) &
    & * energy(i,ny+1,k) + diffusion_velocity * ( &
    &    energy(i,ny,k) + energy(i,0,k) &
    & )
    end do
    end do

    do j = 0,ny+1
  	do k = 1,nz
      energy_updated(0,j,k) = (1.0d0 - 2 * diffusion_velocity) &
    & * energy(0,j,k) + diffusion_velocity * ( &
    &    energy(1,j,k) + energy(nx+1,j,k) &
    & )
      energy_updated(nx+1,j,k) = (1.0d0 - 2 * diffusion_velocity) &
    & * energy(nx+1,j,k) + diffusion_velocity * ( &
    &    energy(nx,j,k) + energy(0,j,k) &
    & )
    end do
    end do
	end subroutine
\end{lstlisting}

\subsection{Reduced Weather: Code Generated by Hybrid Fortran for Diffusion} \label{sec:sources-diffusion-generated}

\begin{lstlisting}[name=diffuse, label=listing:diffuse-reduced-cuda-host, caption={Diffusion code generated for CUDA Fortran}]
    attributes(global) subroutine hfk0_diffuse(...)
      ! ... use statements ...
      real(8), device :: energy_u(0:nx+1,0:ny+1,nz)
      real(8), device :: energy(0:nx+1,0:ny+1,nz)
      ! ... further specifications and initializations
      i = (blockidx%x - 1) * blockDim%x + threadidx%x + 1 - 1
      j = (blockidx%y - 1) * blockDim%y + threadidx%y + 1 - 1
      if (i .GT. nx .OR. j .GT. ny) then
         return
      end if
      ! .... computational code within parallel region ...
    end subroutine

    subroutine diffuse(thermal_energy_updated, thermal_energy)
      ! ... use statements ...
      real(8), intent(out) :: energy_u(0:nx+1,0:ny+1,nz)
      real(8), device :: energy_u_hfdev(0:nx+1,0:ny+1,nz)
      real(8), intent(in) :: energy(0:nx+1,0:ny+1,nz)
      real(8), device :: energy_hfdev(0:nx+1,0:ny+1,nz)
      ! ... further specifications ...
      energy_u_hfdev(:,:,:) = 0
      if ( nx+1 - 0 + 1 .gt. 0 .and. ny+1 - 0 + 1 .gt. 0 .and. nz - 0 .gt. 0 ) then
         energy_hfdev(:,:,:) = energy(:,:,:)
      end if
      ! ... further initializations ...
      cugridSizeX = ceiling(real(nx) / real(32))
      cugridSizeY = ceiling(real(ny) / real(16))
      cugridSizeZ = 1
      cugrid = dim3(cugridSizeX, cugridSizeY, cugridSizeZ)
      cublock = dim3(32, 16, 1)
      call hfk0_diffuse <<< cugrid, cublock >>>( &
          ! ... passing in required scalars
          energy_u_hfdev, energy_hfdev &
      & )
      ! ... error handling code ...
      ! ... repeat the above for all boundary regions ...
      if ( nx+1 - 0 + 1 .gt. 0 .and. ny+1 - 0 + 1 .gt. 0 .and. nz - 0 .gt. 0 ) then
          energy_u(:,:,:) = energy_u_hfdev(:,:,:)
      end if
    end subroutine
\end{lstlisting}

\subsection{Reduced Weather: Physical Processes} \label{sec:sources-physical}

%--------------------------------------------------!
%                                       ---------->!
%                                      max length  !
\begin{lstlisting}[name=run_physics, label=listing:run_physics, caption={Physical processes.}]
	subroutine run_physics(&
		& energy, energy_surf, energy_pbl)
    ! ..more specifications..
    real(8),intent(inout):: energy(0:nx+1,0:ny+1,nz)
    real(8),intent(in):: energy_surf(0:nx+1,0:ny+1)
    real(8),intent(in):: energy_pbl(0:nx+1,0:ny+1)

    do j = 0,ny+1
    	do i = 0,nx+1
		    call radiate(energy(i,j,:))
		    call exchange_heat_with_boundary( &
		&     energy(i,j,:), energy_surf(i,j), 1)
		    call exchange_heat_with_boundary( &
		&     energy(i,j,:), energy_pbl(i,j), nz)
    	end do
    end do
  end subroutine
\end{lstlisting}

\subsection{Reduced Weather: Radiation} \label{sec:sources-radiation}

%--------------------------------------------------!
%                                       ---------->!
%                                      max length  !
\begin{lstlisting}[name=radiate, label=listing:radiate, caption={Radiation process.}]
  subroutine radiate(energy)
    ! ..more specifications..
    real(8), intent(inout), dimension(nz) :: energy

    radiation_intensity = 0.1d0
    do k=1,nz
        energy(k) = energy(k) + radiation_intensity
    end do
  end subroutine
\end{lstlisting}

\subsection{Reduced Weather: Heat Exchange} \label{sec:sources-exchange}

%--------------------------------------------------!
%                                       ---------->!
%                                      max length  !
\begin{lstlisting}[name=surface, label=listing:surface, caption={Surface / planetary heat exchange.}]
  subroutine exchange_heat_with_boundary( &
  	& energy, boundary_energy, boundary_level)
    ! ..more specifications..
    real(8), intent(inout), dimension(nz) :: energy
    real(8), intent(in) :: boundary_energy
    integer(4), intent(in) :: boundary_level

    transfer_velocity = 0.01d0
    energy_transfer_to_boundary = &
      & transfer_velocity * &
      & (energy(boundary_level) - boundary_energy)
    energy(boundary_level) = energy(boundary_level) &
      & - energy_transfer_to_boundary
  end subroutine
\end{lstlisting}

\subsection{Reduced Weather: Time Integration} \label{sec:sources-time}

%--------------------------------------------------!
%                                       ---------->!
%                                      max length  !
\begin{lstlisting}[name=simulate, label=listing:simulate, caption={Main time integration loop.}]
 	module simple_weather
 	real(8),pointer:: energy(:,:,:)
  real(8),pointer:: energy_u(:,:,:)

  ! .. initialization, cleanup and output code ..

	subroutine simulate( &
		& start_time, end_time, timestep, output_timestep)
    ! ..more specifications..
    real(8),pointer:: energy_temp(:,:,:)

    time = start_time
    do while (.true.)
      if (modulo(time + 0.001d0, output_timestep) &
     &     < 0.01d0) then
        call write_data(energy, "energy", time)
      end if
      call run_physics( &
     & energy, energy_surf, energy_pbl)
      call diffuse(energy_u, energy)

      energy_temp => energy_u
      energy_u => energy
      energy => energy_temp

      time = time + timestep
      if (time > end_time) then
        return
      end if
    end do
  end subroutine
  end module
\end{lstlisting}

\begin{acks}
This work has been supported by the Japan Science and Technology Agency (JST) Core Research of Evolutional Science and Technology (CREST) research program ``Highly Productive, High Performance Application Frameworks for Post Peta-scale Computing'', by KAKENHI Grant-in-Aid for Scientific Research (S) 26220002 from the Ministry of Education, Culture, Sports, Science and Technology (MEXT) of Japan, by ``Joint Usage/Research Center'' for Interdisciplinary Large-scale Information Infrastructures (JHPCN)" and ``High Performance Computing Infrastructure (HPCI)'' as well as by the ``Advanced Computation and I/O Methods for Earth-System Simulations'' (AIMES) project running under the German-Japanese priority program ``Software for Exascale Computing'' (SPPEXA). The authors thank the Japan Meteorological Agency for their extensive support, Tokyo University and the Global Scientific Information and Computing Center at Tokyo Institute of Technology for the use of their supercomputers Reedbush-H and TSUBAME 2.5, as well as Dr. Christian Conti and Dr. Mohamed Attia for their careful proof-reading.
\end{acks}

% Bibliography
\bibliographystyle{ACM-Reference-Format}
\bibliography{content/references}

%%% -*-BibTeX-*-
%%% Do NOT edit. File created by BibTeX with style
%%% ACM-Reference-Format-Journals [18-Jan-2012].

\begin{thebibliography}{00}

%%% ====================================================================
%%% NOTE TO THE USER: you can override these defaults by providing
%%% customized versions of any of these macros before the \bibliography
%%% command.  Each of them MUST provide its own final punctuation,
%%% except for \shownote{}, \showDOI{}, and \showURL{}.  The latter two
%%% do not use final punctuation, in order to avoid confusing it with
%%% the Web address.
%%%
%%% To suppress output of a particular field, define its macro to expand
%%% to an empty string, or better, \unskip, like this:
%%%
%%% \newcommand{\showDOI}[1]{\unskip}   % LaTeX syntax
%%%
%%% \def \showDOI #1{\unskip}           % plain TeX syntax
%%%
%%% ====================================================================

\ifx \showCODEN    \undefined \def \showCODEN     #1{\unskip}     \fi
\ifx \showDOI      \undefined \def \showDOI       #1{#1}\fi
\ifx \showISBNx    \undefined \def \showISBNx     #1{\unskip}     \fi
\ifx \showISBNxiii \undefined \def \showISBNxiii  #1{\unskip}     \fi
\ifx \showISSN     \undefined \def \showISSN      #1{\unskip}     \fi
\ifx \showLCCN     \undefined \def \showLCCN      #1{\unskip}     \fi
\ifx \shownote     \undefined \def \shownote      #1{#1}          \fi
\ifx \showarticletitle \undefined \def \showarticletitle #1{#1}   \fi
\ifx \showURL      \undefined \def \showURL       {\relax}        \fi
% The following commands are used for tagged output and should be
% invisible to TeX
\providecommand\bibfield[2]{#2}
\providecommand\bibinfo[2]{#2}
\providecommand\natexlab[1]{#1}
\providecommand\showeprint[2][]{arXiv:#2}

\bibitem[\protect\citeauthoryear{Bokhanko}{Bokhanko}{2014}]%
        {clang:OpenMP4}
\bibfield{author}{\bibinfo{person}{Andrey Bokhanko}.}
  \bibinfo{year}{2014}\natexlab{}.
\newblock \bibinfo{title}{{OpenMP} 4 Support in {Clang} / {LLVM}}.
\newblock   (\bibinfo{year}{2014}).
\newblock
\newblock
\shownote{Retrieved 2017-12-22 from OpenMP.org:
  \url{http://mail.openmp.org/sc14/BoF_Intel_Andrey_Clang.pdf}.}


\bibitem[\protect\citeauthoryear{Cumming, Osuna, Gysi, Bianco, Lapillonne,
  Fuhrer, and Schulthess}{Cumming et~al\mbox{.}}{2013}]%
        {cummings:review}
\bibfield{author}{\bibinfo{person}{Ben Cumming}, \bibinfo{person}{Carlos
  Osuna}, \bibinfo{person}{Tobias Gysi}, \bibinfo{person}{Mauro Bianco},
  \bibinfo{person}{Xavier Lapillonne}, \bibinfo{person}{Oliver Fuhrer}, {and}
  \bibinfo{person}{Thomas~C Schulthess}.} \bibinfo{year}{2013}\natexlab{}.
\newblock \showarticletitle{A review of the challenges and results of
  refactoring the community climate code {COSMO} for hybrid {Cray} {HPC}
  systems}.
\newblock \bibinfo{journal}{{\em Proceedings of Cray User Group\/}}
  \bibinfo{volume}{2013} (\bibinfo{year}{2013}), \bibinfo{pages}{1--11}.
\newblock


\bibitem[\protect\citeauthoryear{Douglas, Hu, Kowarschik, R{\"u}de, and
  Wei{\ss}}{Douglas et~al\mbox{.}}{2000}]%
        {douglas2000cache}
\bibfield{author}{\bibinfo{person}{Craig~C Douglas}, \bibinfo{person}{Jonathan
  Hu}, \bibinfo{person}{Markus Kowarschik}, \bibinfo{person}{Ulrich R{\"u}de},
  {and} \bibinfo{person}{Christian Wei{\ss}}.} \bibinfo{year}{2000}\natexlab{}.
\newblock \showarticletitle{Cache optimization for structured and unstructured
  grid multigrid}.
\newblock \bibinfo{journal}{{\em Electronic Transactions on Numerical
  Analysis\/}}  \bibinfo{volume}{10} (\bibinfo{year}{2000}),
  \bibinfo{pages}{21--40}.
\newblock


\bibitem[\protect\citeauthoryear{Dursun, Nomura, Wang, Kunaseth, Peng, Seymour,
  Kalia, Nakano, and Vashishta}{Dursun et~al\mbox{.}}{2009}]%
        {dursun2009core}
\bibfield{author}{\bibinfo{person}{Hikmet Dursun}, \bibinfo{person}{Ken-ichi
  Nomura}, \bibinfo{person}{Weiqiang Wang}, \bibinfo{person}{Manaschai
  Kunaseth}, \bibinfo{person}{Liu Peng}, \bibinfo{person}{Richard Seymour},
  \bibinfo{person}{Rajiv~K Kalia}, \bibinfo{person}{Aiichiro Nakano}, {and}
  \bibinfo{person}{Priya Vashishta}.} \bibinfo{year}{2009}\natexlab{}.
\newblock \showarticletitle{In-Core Optimization of High-Order Stencil
  Computations}. In \bibinfo{booktitle}{{\em PDPTA}}.
  \bibinfo{publisher}{CSCE}, \bibinfo{pages}{533--538}.
\newblock


\bibitem[\protect\citeauthoryear{Edwards, Trott, and Sunderland}{Edwards
  et~al\mbox{.}}{2014}]%
        {edwards2014kokkos}
\bibfield{author}{\bibinfo{person}{H~Carter Edwards},
  \bibinfo{person}{Christian~R Trott}, {and} \bibinfo{person}{Daniel
  Sunderland}.} \bibinfo{year}{2014}\natexlab{}.
\newblock \showarticletitle{Kokkos: Enabling manycore performance portability
  through polymorphic memory access patterns}.
\newblock \bibinfo{journal}{{\it J. Parallel and Distrib. Comput.}}
  \bibinfo{volume}{74}, \bibinfo{number}{12} (\bibinfo{year}{2014}),
  \bibinfo{pages}{3202--3216}.
\newblock


\bibitem[\protect\citeauthoryear{Fuhrer}{Fuhrer}{2014}]%
        {GridTools}
\bibfield{author}{\bibinfo{person}{Oliver Fuhrer}.}
  \bibinfo{year}{2014}\natexlab{}.
\newblock \bibinfo{title}{{Grid Tools: Towards a library for hardware oblivious
  implementation of stencil based codes}}.
\newblock \bibinfo{howpublished}{Retrieved 2017-12-22 from
  \url{http://www.pasc-ch.org/projects/2013-2016/grid-tools}}.
  (\bibinfo{year}{2014}).
\newblock


\bibitem[\protect\citeauthoryear{Fuhrer, Osuna, Lapillonne, Gysi, Cumming,
  Bianco, Arteaga, and Schulthess}{Fuhrer et~al\mbox{.}}{2014}]%
        {fuhrer2014towards}
\bibfield{author}{\bibinfo{person}{Oliver Fuhrer}, \bibinfo{person}{Carlos
  Osuna}, \bibinfo{person}{Xavier Lapillonne}, \bibinfo{person}{Tobias Gysi},
  \bibinfo{person}{Ben Cumming}, \bibinfo{person}{Mauro Bianco},
  \bibinfo{person}{Andrea Arteaga}, {and} \bibinfo{person}{Thomas~Christoph
  Schulthess}.} \bibinfo{year}{2014}\natexlab{}.
\newblock \showarticletitle{{Towards a performance portable, architecture
  agnostic implementation strategy for weather and climate models}}.
\newblock \bibinfo{journal}{{\em Supercomputing frontiers and innovations\/}}
  \bibinfo{volume}{1}, \bibinfo{number}{1} (\bibinfo{year}{2014}),
  \bibinfo{pages}{45--62}.
\newblock


\bibitem[\protect\citeauthoryear{Govett}{Govett}{2012}]%
        {govett:f2c-acc}
\bibfield{author}{\bibinfo{person}{Mark Govett}.}
  \bibinfo{year}{2012}\natexlab{}.
\newblock \bibinfo{title}{{F2C-ACC} Users Guide, Version 4.2}.
\newblock   (\bibinfo{year}{2012}).
\newblock
\newblock
\shownote{Retrieved 2017-12-22 from NOAA: \url{http://www.esrl.noaa.gov/gsd/
  ab/ac/Accelerators.html}.}


\bibitem[\protect\citeauthoryear{Govett, Middlecoff, and Henderson}{Govett
  et~al\mbox{.}}{2010}]%
        {govettGPU}
\bibfield{author}{\bibinfo{person}{Mark Govett}, \bibinfo{person}{Jacques
  Middlecoff}, {and} \bibinfo{person}{Tom Henderson}.}
  \bibinfo{year}{2010}\natexlab{}.
\newblock \showarticletitle{Running the {NIM} next-generation weather model on
  {GPUs}}. In \bibinfo{booktitle}{{\em Proceedings of the 2010 10th IEEE/ACM
  International Conference on Cluster, Cloud and Grid Computing}}. IEEE
  Computer Society, \bibinfo{pages}{792--796}.
\newblock


\bibitem[\protect\citeauthoryear{Govett, Middlecoff, and Henderson}{Govett
  et~al\mbox{.}}{2014}]%
        {GovettDirective}
\bibfield{author}{\bibinfo{person}{Mark Govett}, \bibinfo{person}{Jacques
  Middlecoff}, {and} \bibinfo{person}{Tom Henderson}.}
  \bibinfo{year}{2014}\natexlab{}.
\newblock \showarticletitle{Directive-based Parallelization of the {NIM}
  Weather Model for {GPUs}}. In \bibinfo{booktitle}{{\em Proceedings of the
  First Workshop on Accelerator Programming Using Directives}} {\em
  (\bibinfo{series}{WACCPD '14})}. \bibinfo{publisher}{IEEE Press},
  \bibinfo{address}{Piscataway, NJ, USA}, \bibinfo{pages}{55--61}.
\newblock
\showISBNx{978-1-4799-7023-0}
\showDOI{%
\url{https://doi.org/10.1109/WACCPD.2014.9}}


\bibitem[\protect\citeauthoryear{Govett, Rosinski, Middlecoff, Henderson, Lee,
  MacDonald, Wang, Madden, Schramm, and Duarte}{Govett et~al\mbox{.}}{2017}]%
        {govett2017parallelization}
\bibfield{author}{\bibinfo{person}{Mark Govett}, \bibinfo{person}{Jim
  Rosinski}, \bibinfo{person}{Jacques Middlecoff}, \bibinfo{person}{Tom
  Henderson}, \bibinfo{person}{Jin Lee}, \bibinfo{person}{Alexander MacDonald},
  \bibinfo{person}{Ning Wang}, \bibinfo{person}{Paul Madden},
  \bibinfo{person}{Julie Schramm}, {and} \bibinfo{person}{Antonio Duarte}.}
  \bibinfo{year}{2017}\natexlab{}.
\newblock \showarticletitle{Parallelization and Performance of the {NIM}
  Weather Model on {CPU}, {GPU} and {MIC} Processors}.
\newblock \bibinfo{journal}{{\em Bulletin of the American Meteorological
  Society\/}} (\bibinfo{year}{2017}).
\newblock


\bibitem[\protect\citeauthoryear{Group}{Group}{2012}]%
        {PGIAcc}
\bibfield{author}{\bibinfo{person}{The~Portland Group}.}
  \bibinfo{year}{2012}\natexlab{}.
\newblock \bibinfo{title}{{PGI} Accelerator Compilers with {OpenACC}
  Directives}.
\newblock   (\bibinfo{year}{2012}).
\newblock
\newblock
\shownote{Retrieved 2017-12-22 from The Portland Group:
  \url{http://www.pgroup.com/resources/accel.htm}.}


\bibitem[\protect\citeauthoryear{Harris}{Harris}{2007}]%
        {harris2007optimizing}
\bibfield{author}{\bibinfo{person}{Mark Harris}.}
  \bibinfo{year}{2007}\natexlab{}.
\newblock \showarticletitle{Optimizing {CUDA}}.
\newblock \bibinfo{journal}{{\em {SC07}: High Performance Computing With
  {CUDA}\/}} (\bibinfo{year}{2007}).
\newblock


\bibitem[\protect\citeauthoryear{Intel}{Intel}{2010}]%
        {Xeon5670}
\bibfield{author}{\bibinfo{person}{Intel}.} \bibinfo{year}{2010}\natexlab{}.
\newblock \bibinfo{title}{{Intel Xeon} Processor {X5670}}.
\newblock   (\bibinfo{year}{2010}).
\newblock
\newblock
\shownote{Retrieved 2017-12-22 from Intel:
  \url{http://ark.intel.com/products/47920/Intel-Xeon-Processor-X5670-12M-Cache-2_93-GHz-6_40-GTs-Intel-QPI}.}


\bibitem[\protect\citeauthoryear{Intel}{Intel}{2012}]%
        {XeonE5-2670}
\bibfield{author}{\bibinfo{person}{Intel}.} \bibinfo{year}{2012}\natexlab{}.
\newblock \bibinfo{title}{{Intel Xeon} Processor {E5-2670}}.
\newblock   (\bibinfo{year}{2012}).
\newblock
\newblock
\shownote{Retrieved 2017-12-22 from Intel:
  \url{http://ark.intel.com/products/64595/Intel-Xeon-Processor-E5-2670-20M-Cache-2_60-GHz-8_00-GTs-Intel-QPI}.}


\bibitem[\protect\citeauthoryear{Intel}{Intel}{2016}]%
        {XeonE5-2695v4}
\bibfield{author}{\bibinfo{person}{Intel}.} \bibinfo{year}{2016}\natexlab{}.
\newblock \bibinfo{title}{{Intel Xeon} Processor {E5-2695 v4}}.
\newblock   (\bibinfo{year}{2016}).
\newblock
\newblock
\shownote{Retrieved 2017-12-22 from Intel:
  \url{http://ark.intel.com/products/91316/Intel-Xeon-Processor-E5-2695-v4-45M-Cache-2_10-GHz}.}


\bibitem[\protect\citeauthoryear{Ishida, Muroi, Kawano, and Kitamura}{Ishida
  et~al\mbox{.}}{2010}]%
        {ishida2010development}
\bibfield{author}{\bibinfo{person}{Junichi Ishida}, \bibinfo{person}{Chiashi
  Muroi}, \bibinfo{person}{Kohei Kawano}, {and} \bibinfo{person}{Yuji
  Kitamura}.} \bibinfo{year}{2010}\natexlab{}.
\newblock \showarticletitle{Development of a new nonhydrostatic model {ASUCA}
  at {JMA}}.
\newblock \bibinfo{journal}{{\em CAS/JSC WGNE Research Activities in
  Atmospheric and Oceanic Modelling\/}}  \bibinfo{volume}{40}
  (\bibinfo{year}{2010}), \bibinfo{pages}{0511--0512}.
\newblock


\bibitem[\protect\citeauthoryear{James~Beyer}{James~Beyer}{2016}]%
        {OpenMP45GPU}
\bibfield{author}{\bibinfo{person}{{NVIDIA}~Inc. James~Beyer}.}
  \bibinfo{year}{2016}\natexlab{}.
\newblock \bibinfo{title}{Targeting {GPUs} with {OpenMP} 4.5 Device
  Directives}.
\newblock   (\bibinfo{year}{2016}).
\newblock
\newblock
\shownote{Retrieved 2017-12-22 from {NVIDIA}:
  \url{http://on-demand.gputechconf.com/gtc/2016/presentation/s6510-jeff-larkin-targeting-gpus-openmp.pdf}.}


\bibitem[\protect\citeauthoryear{James C.~Beyer}{James C.~Beyer}{2013}]%
        {CrayAcc}
\bibfield{author}{\bibinfo{person}{Cray~Inc. James C.~Beyer}.}
  \bibinfo{year}{2013}\natexlab{}.
\newblock \bibinfo{title}{The use of {OpenACC} and {OpenMP} Accelerator
  directives with the Cray Compilation Environment ({CCE})}.
\newblock   (\bibinfo{year}{2013}).
\newblock
\newblock
\shownote{Retrieved 2017-12-22 from {GPU} Technology Conference:
  \url{http://on-demand.gputechconf.com/gtc/2013/presentations/S3084-OpenACC-OpenMP-Directives-CCE.pdf}.}


\bibitem[\protect\citeauthoryear{Kwiatkowski}{Kwiatkowski}{2001}]%
        {kwiatkowski2001evaluation}
\bibfield{author}{\bibinfo{person}{Jan Kwiatkowski}.}
  \bibinfo{year}{2001}\natexlab{}.
\newblock \showarticletitle{Evaluation of parallel programs by measurement of
  its granularity}. In \bibinfo{booktitle}{{\em International Conference on
  Parallel Processing and Applied Mathematics}}. Springer,
  \bibinfo{pages}{145--153}.
\newblock


\bibitem[\protect\citeauthoryear{Lapillonne and Fuhrer}{Lapillonne and
  Fuhrer}{2014}]%
        {lapillonne2014using}
\bibfield{author}{\bibinfo{person}{Xavier Lapillonne} {and}
  \bibinfo{person}{Oliver Fuhrer}.} \bibinfo{year}{2014}\natexlab{}.
\newblock \showarticletitle{Using compiler directives to port large scientific
  applications to {GPUs}: An example from atmospheric science}.
\newblock \bibinfo{journal}{{\em Parallel Processing Letters\/}}
  \bibinfo{volume}{24}, \bibinfo{number}{01} (\bibinfo{year}{2014}).
\newblock


\bibitem[\protect\citeauthoryear{Luszczek, Bailey, Dongarra, Kepner, Lucas,
  Rabenseifner, and Takahashi}{Luszczek et~al\mbox{.}}{2006}]%
        {luszczek2006hpc}
\bibfield{author}{\bibinfo{person}{Piotr~R Luszczek}, \bibinfo{person}{David~H
  Bailey}, \bibinfo{person}{Jack~J Dongarra}, \bibinfo{person}{Jeremy Kepner},
  \bibinfo{person}{Robert~F Lucas}, \bibinfo{person}{Rolf Rabenseifner}, {and}
  \bibinfo{person}{Daisuke Takahashi}.} \bibinfo{year}{2006}\natexlab{}.
\newblock \showarticletitle{The {HPC} Challenge ({HPCC}) benchmark suite}. In
  \bibinfo{booktitle}{{\em Proceedings of the 2006 ACM/IEEE conference on
  Supercomputing}}. \bibinfo{pages}{213}.
\newblock


\bibitem[\protect\citeauthoryear{Maruyama, Nomura, Sato, and Matsuoka}{Maruyama
  et~al\mbox{.}}{2011}]%
        {maruyama}
\bibfield{author}{\bibinfo{person}{Naoya Maruyama}, \bibinfo{person}{Tatsuo
  Nomura}, \bibinfo{person}{Kento Sato}, {and} \bibinfo{person}{Satoshi
  Matsuoka}.} \bibinfo{year}{2011}\natexlab{}.
\newblock \showarticletitle{Physis: An Implicitly Parallel Programming Model
  for Stencil Computations on Large-scale {GPU}-accelerated Supercomputers}. In
  \bibinfo{booktitle}{{\em Proceedings of 2011 International Conference for
  High Performance Computing, Networking, Storage and Analysis}} {\em
  (\bibinfo{series}{SC '11})}. \bibinfo{publisher}{ACM}, \bibinfo{address}{New
  York, NY, USA}, Article \bibinfo{articleno}{11},
  \bibinfo{numpages}{12}~pages.
\newblock
\showISBNx{978-1-4503-0771-0}
\showDOI{%
\url{https://doi.org/10.1145/2063384.2063398}}


\bibitem[\protect\citeauthoryear{Michalakes and Vachharajani}{Michalakes and
  Vachharajani}{2008}]%
        {MichalakesV08}
\bibfield{author}{\bibinfo{person}{John Michalakes} {and}
  \bibinfo{person}{Manish Vachharajani}.} \bibinfo{year}{2008}\natexlab{}.
\newblock \showarticletitle{{GPU} acceleration of numerical weather
  prediction}. In \bibinfo{booktitle}{{\em IPDPS}} (2009-06-29).
  \bibinfo{publisher}{IEEE}, \bibinfo{pages}{1--7}.
\newblock
\showURL{%
\url{http://dblp.uni-trier.de/db/conf/ipps/ipdps2008.html#MichalakesV08}}


\bibitem[\protect\citeauthoryear{Mielikainen, Huang, and Huang}{Mielikainen
  et~al\mbox{.}}{2014}]%
        {wrf_on_phi}
\bibfield{author}{\bibinfo{person}{Jarno Mielikainen}, \bibinfo{person}{Bormin
  Huang}, {and} \bibinfo{person}{Allen Huang}.}
  \bibinfo{year}{2014}\natexlab{}.
\newblock \showarticletitle{Using {Intel} {Xeon} {Phi} to accelerate the {WRF}
  {TEMF} planetary boundary layer scheme}. In \bibinfo{booktitle}{{\em SPIE
  Sensing Technology+ Applications}}. International Society for Optics and
  Photonics, \bibinfo{pages}{91240T--91240T}.
\newblock


\bibitem[\protect\citeauthoryear{Mielikainen, Huang, Huang, and
  Goldberg}{Mielikainen et~al\mbox{.}}{2012}]%
        {mielikainen2012gpu}
\bibfield{author}{\bibinfo{person}{Jarno Mielikainen}, \bibinfo{person}{Bormin
  Huang}, \bibinfo{person}{Hung-Lung~Allen Huang}, {and}
  \bibinfo{person}{Mitchell~D Goldberg}.} \bibinfo{year}{2012}\natexlab{}.
\newblock \showarticletitle{{GPU} acceleration of the updated Goddard shortwave
  radiation scheme in the weather research and forecasting ({WRF}) model}.
\newblock \bibinfo{journal}{{\em IEEE Journal of Selected Topics in Applied
  Earth Observations and Remote Sensing\/}} \bibinfo{volume}{5},
  \bibinfo{number}{2} (\bibinfo{year}{2012}), \bibinfo{pages}{555--562}.
\newblock


\bibitem[\protect\citeauthoryear{M{\"u}ller and Aoki}{M{\"u}ller and
  Aoki}{2018}]%
        {MUELLER2017HYBRID}
\bibfield{author}{\bibinfo{person}{Michel M{\"u}ller} {and}
  \bibinfo{person}{Takayuki Aoki}.} \bibinfo{year}{2018}\natexlab{}.
\newblock \showarticletitle{Hybrid Fortran: High Productivity GPU Porting
  Framework Applied to Japanese Weather Prediction Model}. In
  \bibinfo{booktitle}{{\em Accelerator Programming Using Directives}},
  \bibfield{editor}{\bibinfo{person}{Sunita Chandrasekaran} {and}
  \bibinfo{person}{Guido Juckeland}} (Eds.). \bibinfo{publisher}{Springer
  International Publishing}, \bibinfo{address}{Cham}, \bibinfo{pages}{20--41}.
\newblock
\showISBNx{978-3-319-74896-2}


\bibitem[\protect\citeauthoryear{Norman, Mametjanov, and Taylor}{Norman
  et~al\mbox{.}}{2017}]%
        {norman2017exascale}
\bibfield{author}{\bibinfo{person}{Matthew~R Norman}, \bibinfo{person}{Azamat
  Mametjanov}, {and} \bibinfo{person}{Mark Taylor}.}
  \bibinfo{year}{2017}\natexlab{}.
\newblock \bibinfo{title}{Exascale Programming Approaches for the Accelerated
  Model for Climate and Energy}.
\newblock   (\bibinfo{year}{2017}).
\newblock


\bibitem[\protect\citeauthoryear{OpenACC}{OpenACC}{2015}]%
        {openACCSpec}
\bibfield{author}{\bibinfo{person}{OpenACC}.} \bibinfo{year}{2015}\natexlab{}.
\newblock \bibinfo{title}{The {OpenACC} Application Programming Interface
  Version 2.5}.
\newblock   (\bibinfo{year}{2015}).
\newblock
\newblock
\shownote{Retrieved 2017-12-22 from OpenACC.org:
  \url{http://www.openacc.org/sites/default/files/inline-files/OpenACC_2pt5.pdf}.}


\bibitem[\protect\citeauthoryear{Preshing}{Preshing}{2012}]%
        {SingleThreadCPUPerf}
\bibfield{author}{\bibinfo{person}{Jeff Preshing}.}
  \bibinfo{year}{2012}\natexlab{}.
\newblock \bibinfo{title}{A Look Back at Single-Threaded {CPU} Performance}.
\newblock   (\bibinfo{year}{2012}).
\newblock
\newblock
\shownote{Retrieved 2017-12-22:
  \url{http://preshing.com/20120208/a-look-back-at-single-threaded-cpu-performance}.}


\bibitem[\protect\citeauthoryear{Prickett~Morgan}{Prickett~Morgan}{2012}]%
        {k20linpack}
\bibfield{author}{\bibinfo{person}{Timothy Prickett~Morgan}.}
  \bibinfo{year}{2012}\natexlab{}.
\newblock \bibinfo{title}{{NVIDIA} launches not one but two {Kepler2} {GPU}
  coprocessors}.
\newblock   (\bibinfo{year}{2012}).
\newblock
\newblock
\shownote{Retrieved 2017-12-22 from The Register:
  \url{http://www.theregister.co.uk/2012/11/12/nvidia_tesla_k20_k20x_gpu_coprocessors/?page=2}.}


\bibitem[\protect\citeauthoryear{Rendell, Ch~Apm~An, and M{\"u}ller}{Rendell
  et~al\mbox{.}}{2013}]%
        {openmp4:early}
\bibfield{author}{\bibinfo{person}{Alistair~P Rendell}, \bibinfo{person}{B
  Ch~Apm~An}, {and} \bibinfo{person}{Matthias~S M{\"u}ller}.}
  \bibinfo{year}{2013}\natexlab{}.
\newblock \bibinfo{title}{OpenMP in the Era of Low Power Devices and
  Accelerators}.
\newblock   (\bibinfo{year}{2013}).
\newblock


\bibitem[\protect\citeauthoryear{Ruetsch, Phillips, and Fatica}{Ruetsch
  et~al\mbox{.}}{2010}]%
        {ruetsch2010gpu}
\bibfield{author}{\bibinfo{person}{Greg Ruetsch}, \bibinfo{person}{Everett
  Phillips}, {and} \bibinfo{person}{Massimiliano Fatica}.}
  \bibinfo{year}{2010}\natexlab{}.
\newblock \showarticletitle{{GPU} acceleration of the long-wave rapid radiative
  transfer model in {WRF} using {CUDA} {Fortran}}. In \bibinfo{booktitle}{{\em
  Many-Core and Reconfigurable Supercomputing Conference}}.
\newblock


\bibitem[\protect\citeauthoryear{Sakamoto, Ishida, Kawano, Matsubayashi,
  Aranami, Hara, Kusabiraki, Muroi, and Kitamura}{Sakamoto
  et~al\mbox{.}}{2014}]%
        {sakamotodevelopment}
\bibfield{author}{\bibinfo{person}{M Sakamoto}, \bibinfo{person}{J Ishida},
  \bibinfo{person}{K Kawano}, \bibinfo{person}{K Matsubayashi},
  \bibinfo{person}{K Aranami}, \bibinfo{person}{T Hara}, \bibinfo{person}{H
  Kusabiraki}, \bibinfo{person}{C Muroi}, {and} \bibinfo{person}{Y Kitamura}.}
  \bibinfo{year}{2014}\natexlab{}.
\newblock \bibinfo{title}{Development of Yin-Yang Grid Global Model Using a New
  Dynamical Core {ASUCA}}.
\newblock   (\bibinfo{year}{2014}).
\newblock


\bibitem[\protect\citeauthoryear{Shimokawabe, Aoki, Ishida, Kawano, and
  Muroi}{Shimokawabe et~al\mbox{.}}{2011}]%
        {shimokawabe2011145}
\bibfield{author}{\bibinfo{person}{Takashi Shimokawabe},
  \bibinfo{person}{Takayuki Aoki}, \bibinfo{person}{Junichi Ishida},
  \bibinfo{person}{Kohei Kawano}, {and} \bibinfo{person}{Chiashi Muroi}.}
  \bibinfo{year}{2011}\natexlab{}.
\newblock \showarticletitle{145 {TFlops} performance on 3990 {GPUs} of
  {TSUBAME} 2.0 supercomputer for an operational weather prediction}.
\newblock \bibinfo{journal}{{\em Procedia Computer Science\/}}
  \bibinfo{volume}{4} (\bibinfo{year}{2011}), \bibinfo{pages}{1535--1544}.
\newblock


\bibitem[\protect\citeauthoryear{Shimokawabe, Aoki, Muroi, Ishida, Kawano,
  Endo, Nukada, Maruyama, and Matsuoka}{Shimokawabe et~al\mbox{.}}{2010}]%
        {shimokawabe201080}
\bibfield{author}{\bibinfo{person}{Takashi Shimokawabe},
  \bibinfo{person}{Takayuki Aoki}, \bibinfo{person}{Chiashi Muroi},
  \bibinfo{person}{Junichi Ishida}, \bibinfo{person}{Kohei Kawano},
  \bibinfo{person}{Toshio Endo}, \bibinfo{person}{Akira Nukada},
  \bibinfo{person}{Naoya Maruyama}, {and} \bibinfo{person}{Satoshi Matsuoka}.}
  \bibinfo{year}{2010}\natexlab{}.
\newblock \showarticletitle{An 80-fold speedup, 15.0 {TFlops} full {GPU}
  acceleration of non-hydrostatic weather model {ASUCA} production code}. In
  \bibinfo{booktitle}{{\em Proceedings of the 2010 ACM/IEEE International
  Conference for High Performance Computing, Networking, Storage and
  Analysis}}. IEEE Computer Society, \bibinfo{pages}{1--11}.
\newblock


\bibitem[\protect\citeauthoryear{Shimokawabe, Aoki, and Onodera}{Shimokawabe
  et~al\mbox{.}}{2014}]%
        {shimokawabe:2014}
\bibfield{author}{\bibinfo{person}{Takashi Shimokawabe},
  \bibinfo{person}{Takayuki Aoki}, {and} \bibinfo{person}{Naoyuki Onodera}.}
  \bibinfo{year}{2014}\natexlab{}.
\newblock \showarticletitle{High-productivity Framework on {GPU}-rich
  Supercomputers for Operational Weather Prediction Code {ASUCA}}. In
  \bibinfo{booktitle}{{\em Proceedings of the International Conference for High
  Performance Computing, Networking, Storage and Analysis}} {\em
  (\bibinfo{series}{SC '14})}. \bibinfo{publisher}{IEEE Press},
  \bibinfo{address}{Piscataway, NJ, USA}, \bibinfo{pages}{251--261}.
\newblock
\showISBNx{978-1-4799-5500-8}
\showDOI{%
\url{https://doi.org/10.1109/SC.2014.26}}


\bibitem[\protect\citeauthoryear{Sutter}{Sutter}{2005}]%
        {sutter2005free}
\bibfield{author}{\bibinfo{person}{Herb Sutter}.}
  \bibinfo{year}{2005}\natexlab{}.
\newblock \showarticletitle{The free lunch is over: A fundamental turn toward
  concurrency in software}.
\newblock \bibinfo{journal}{{\em Dr. Dobb's journal\/}} \bibinfo{volume}{30},
  \bibinfo{number}{3} (\bibinfo{year}{2005}), \bibinfo{pages}{202--210}.
\newblock


\bibitem[\protect\citeauthoryear{Tezaur, Watkins, and Demeshko}{Tezaur
  et~al\mbox{.}}{[n. d.]}]%
        {tezaurtowards}
\bibfield{author}{\bibinfo{person}{Irina Tezaur}, \bibinfo{person}{Jerry
  Watkins}, {and} \bibinfo{person}{Irina Demeshko}.} \bibinfo{year}{[n.
  d.]}\natexlab{}.
\newblock \showarticletitle{Towards Performance-Portability of the Albany/FELIX
  Land-Ice Solver to New and Emerging Architectures Using Kokkos}.
\newblock  (\bibinfo{year}{[n. d.]}).
\newblock


\bibitem[\protect\citeauthoryear{{The OpenMP Architecture Review Board}}{{The
  OpenMP Architecture Review Board}}{2013}]%
        {openmp4}
\bibfield{author}{\bibinfo{person}{{The OpenMP Architecture Review Board}}.}
  \bibinfo{year}{2013}\natexlab{}.
\newblock \bibinfo{title}{OpenMP Application Program Interface Version 4.0}.
\newblock   (\bibinfo{date}{July} \bibinfo{year}{2013}).
\newblock
\newblock
\shownote{Retrieved 2017-12-22 from OpenMP.org:
  \url{http://www.openmp.org/mp-documents/OpenMP4.0.0.pdf}.}


\bibitem[\protect\citeauthoryear{TOP500}{TOP500}{2016}]%
        {Top500-2016-11}
\bibfield{author}{\bibinfo{person}{TOP500}.} \bibinfo{year}{2016}\natexlab{}.
\newblock \bibinfo{title}{{Top500} List November 2016}.
\newblock   (\bibinfo{year}{2016}).
\newblock
\newblock
\shownote{Retrieved 2017-12-22 from Top500.org:
  \url{https://www.top500.org/lists/2016/11}.}


\bibitem[\protect\citeauthoryear{Underwood, Plimpton, Brightwell, Vaughan, and
  Davis}{Underwood et~al\mbox{.}}{2006}]%
        {underwood2006simple}
\bibfield{author}{\bibinfo{person}{Keith~D Underwood},
  \bibinfo{person}{Steven~J Plimpton}, \bibinfo{person}{Ronald~B Brightwell},
  \bibinfo{person}{Courtenay~T Vaughan}, {and} \bibinfo{person}{Mike Davis}.}
  \bibinfo{year}{2006}\natexlab{}.
\newblock \bibinfo{booktitle}{{\em A Simple Synchronous Distributed-Memory
  Algorithm for the {HPCC} {RandomAccess} Benchmark}}.
\newblock \bibinfo{type}{{T}echnical {R}eport}. \bibinfo{institution}{Sandia
  National Laboratories (SNL-NM), Albuquerque, NM (United States); Sandia
  National Laboratories, Albuquerque, NM}.
\newblock


\bibitem[\protect\citeauthoryear{University}{University}{2017}]%
        {reedbush}
\bibfield{author}{\bibinfo{person}{Tokyo University}.}
  \bibinfo{year}{2017}\natexlab{}.
\newblock \bibinfo{title}{Reedbush Introduction ({in Japanese})}.
\newblock   (\bibinfo{year}{2017}).
\newblock
\newblock
\shownote{Retrieved 2017-12-22 from Tokyo University:
  \url{http://www.cc.u-tokyo.ac.jp/system/reedbush/reedbush_intro.html}.}


\bibitem[\protect\citeauthoryear{Vanderbauwhede and Takemi}{Vanderbauwhede and
  Takemi}{2013}]%
        {vanderbauwhede2013investigation}
\bibfield{author}{\bibinfo{person}{Wim Vanderbauwhede} {and}
  \bibinfo{person}{Tetsuya Takemi}.} \bibinfo{year}{2013}\natexlab{}.
\newblock \showarticletitle{An investigation into the feasibility and benefits
  of {GPU}/multicore acceleration of the weather research and forecasting
  model}. In \bibinfo{booktitle}{{\em High Performance Computing and Simulation
  (HPCS), 2013 International Conference on}}. IEEE, \bibinfo{pages}{482--489}.
\newblock


\bibitem[\protect\citeauthoryear{Wicker and Skamarock}{Wicker and
  Skamarock}{2002}]%
        {wicker2002time}
\bibfield{author}{\bibinfo{person}{Louis~J Wicker} {and}
  \bibinfo{person}{William~C Skamarock}.} \bibinfo{year}{2002}\natexlab{}.
\newblock \showarticletitle{Time-splitting methods for elastic models using
  forward time schemes}.
\newblock \bibinfo{journal}{{\em Monthly weather review\/}}
  \bibinfo{volume}{130}, \bibinfo{number}{8} (\bibinfo{year}{2002}),
  \bibinfo{pages}{2088--2097}.
\newblock


\bibitem[\protect\citeauthoryear{Williams, Waterman, and Patterson}{Williams
  et~al\mbox{.}}{2009}]%
        {roofline}
\bibfield{author}{\bibinfo{person}{Samuel Williams}, \bibinfo{person}{Andrew
  Waterman}, {and} \bibinfo{person}{David Patterson}.}
  \bibinfo{year}{2009}\natexlab{}.
\newblock \showarticletitle{Roofline: An Insightful Visual Performance Model
  for Multicore Architectures}.
\newblock \bibinfo{journal}{{\em Commun. ACM\/}} \bibinfo{volume}{52},
  \bibinfo{number}{4} (\bibinfo{date}{April} \bibinfo{year}{2009}),
  \bibinfo{pages}{65--76}.
\newblock
\showISSN{0001-0782}
\showDOI{%
\url{https://doi.org/10.1145/1498765.1498785}}


\bibitem[\protect\citeauthoryear{Xu, Han, and Dandapanthu}{Xu
  et~al\mbox{.}}{2017}]%
        {p100linpack}
\bibfield{author}{\bibinfo{person}{Rengan Xu}, \bibinfo{person}{Frank Han},
  {and} \bibinfo{person}{Nishanth Dandapanthu}.}
  \bibinfo{year}{2017}\natexlab{}.
\newblock \bibinfo{title}{Application Performance on {P100-PCIe} {GPUs}}.
\newblock   (\bibinfo{year}{2017}).
\newblock
\newblock
\shownote{Retrieved 2017-12-22 from Dell:
  \url{http://en.community.dell.com/techcenter/high-performance-computing/b/general_hpc/archive/2017/03/14/application-performance-on-p100-pcie-gpus}.}


\end{thebibliography}

% \includepdf[pages=-]{data/response_letter.pdf}

\end{document}